\documentclass[twocolumn]{aastex631}

\usepackage{amsmath,amsfonts,amssymb, mathtools}
\usepackage{enumitem}

\newcommand{\dif}{\ensuremath{\mathrm{d}}}
\newcommand{\vect}[1]{\boldsymbol{\mathbf{#1}}}

\shorttitle{}
\shortauthors{Chen et al.}
\graphicspath{{./}{figures/}}

\begin{document}

\title{Impacts and Statistical Mitigation of Missing Data on the 21cm Power Spectrum:
\\ A Case Study with the Hydrogen Epoch of Reionization Array}

\email{$^{\bigstar}$kfchen@mit.edu}

\author[0000-0002-3839-0230]{Kai-Feng Chen$^{\bigstar}$}
\affiliation{MIT Kavli Institute, Massachusetts Institute of Technology, Cambridge, MA 02139, USA}
\affiliation{Department of Physics, Massachusetts Institute of Technology, Cambridge, MA 02139, USA}
\affiliation{Department of Physics and Trottier Space Institute, McGill University, Montreal QC H3A 2T8, Canada}

\author[0000-0001-7716-9312]{Michael J. Wilensky}
\altaffiliation{CITA National Fellow}
\affiliation{Department of Physics and Trottier Space Institute, McGill University, Montreal QC H3A 2T8, Canada}
\affiliation{Jodrell Bank Centre for Astrophysics, University of Manchester, Manchester M13 9PL, UK}

\author[0000-0001-6876-0928]{Adrian Liu}
\affiliation{Department of Physics and Trottier Space Institute, McGill University, Montreal QC H3A 2T8, Canada}

\author[0000-0003-3336-9958]{Joshua S.\ Dillon}
\affiliation{Department of Astronomy, University of California, Berkeley, Berkeley, CA 94720, USA}
\affiliation{Radio Astronomy Laboratory, University of California, Berkeley, Berkeley, CA 94720, USA}

\author[0000-0002-4117-570X]{Jacqueline N. Hewitt}
\affiliation{MIT Kavli Institute, Massachusetts Institute of Technology, Cambridge, MA 02139, USA}
\affiliation{Department of Physics, Massachusetts Institute of Technology, Cambridge, MA 02139, USA}

\author{Tyrone Adams}
\affiliation{South African Radio Astronomy Observatory, Cape Town 7925, South Africa}

\author[0000-0002-4810-666X]{James E. Aguirre}
\affiliation{Department of Physics and Astronomy, University of Pennsylvania, Philadelphia, PA 19104, USA}

\author{Rushelle Baartman}
\affiliation{South African Radio Astronomy Observatory, Cape Town 7925, South Africa}

\author[0000-0001-9428-8233]{Adam P. Beardsley}
\affiliation{Department of Physics, Winona State University, Winona, MN 55987, USA}

\author[0000-0002-2293-9639]{Lindsay M. Berkhout}
\affiliation{Department of Physics and Trottier Space Institute, McGill University, Montreal QC H3A 2T8, Canada}

\author[0000-0002-0916-7443]{Gianni Bernardi}
\affiliation{South African Radio Astronomy Observatory, Cape Town 7925, South Africa}
\affiliation{INAF-Istituto di Radioastronomia, via Gobetti 101, 40129 Bologna, Italy}
\affiliation{Department of Physics and Electronics, Rhodes University, PO Box 94, Grahamstown 6140, South Africa}

\author{Tashalee S. Billings}
\affiliation{Department of Physics and Astronomy, University of Pennsylvania, Philadelphia, PA 19104, USA}

\author[0000-0002-8475-2036]{Judd D. Bowman}
\affiliation{School of Earth and Space Exploration, Arizona State University, Tempe, AZ 85287, USA}

\author[0000-0001-5668-3101]{Philip  Bull}
\affiliation{Jodrell Bank Centre for Astrophysics, University of Manchester, Manchester M13 9PL, UK}
\affiliation{Department of Physics and Astronomy,  University of Western Cape, Cape Town 7535, South Africa}

\author[0000-0002-8465-9341]{Jacob  Burba}
\affiliation{Jodrell Bank Centre for Astrophysics, University of Manchester, Manchester M13 9PL, UK}

\author{Ruby Byrne}
\affiliation{Cahill Center for Astronomy and Astrophysics, California Institute of Technology, Pasadena, CA 91125, USA}

\author{Steven  Carey}
\affiliation{Cavendish Astrophysics, University of Cambridge, Cambridge CB3 0HE, UK}

\author{Samir  Choudhuri}
\affiliation{Centre for Strings, Gravitation and Cosmology, Department of Physics, Indian Institute of Technology Madras, Chennai 600036, India}

\author[0009-0008-2574-3878]{Tyler Cox}
\affiliation{Department of Astronomy, University of California, Berkeley, Berkeley, CA 94720, USA}

\author[0000-0003-3197-2294]{David. R. DeBoer}
\affiliation{Radio Astronomy Laboratory, University of California, Berkeley, Berkeley, CA 94720, USA}

\author{Matt Dexter}
\affiliation{Radio Astronomy Laboratory, University of California, Berkeley, Berkeley, CA 94720, USA}

\author{Nico  Eksteen}
\affiliation{South African Radio Astronomy Observatory, Cape Town 7925, South Africa}

\author{John  Ely}
\affiliation{Cavendish Astrophysics, University of Cambridge, Cambridge CB3 0HE, UK}

\author[0000-0002-0086-7363]{Aaron  Ewall-Wice}
\affiliation{Department of Astronomy, University of California, Berkeley, Berkeley, CA 94720, USA}
\affiliation{Radio Astronomy Laboratory, University of California, Berkeley, Berkeley, CA 94720, USA}

\author[0000-0002-0658-1243]{Steven R. Furlanetto}
\affiliation{Department of Physics and Astronomy, University of California, Los Angeles, CA 90095, USA}

\author{Kingsley  Gale-Sides}
\affiliation{Cavendish Astrophysics, University of Cambridge, Cambridge CB3 0HE, UK}

\author{Hugh  Garsden}
\affiliation{Jodrell Bank Centre for Astrophysics, University of Manchester, Manchester M13 9PL, UK}

\author{Bharat Kumar Gehlot}
\affiliation{School of Earth and Space Exploration, Arizona State University, Tempe, AZ 85287, USA}

\author[0000-0002-1712-737X]{Adélie  Gorce}
\affiliation{Institut d’Astrophysique Spatiale, CNRS, Université Paris-Saclay, 91405 Orsay, France}

\author[0000-0002-0829-167X]{Deepthi  Gorthi}
\affiliation{Department of Astronomy, University of California, Berkeley, Berkeley, CA 94720, USA}

\author{Ziyaad  Halday}
\affiliation{South African Radio Astronomy Observatory, Cape Town 7925, South Africa}

\author[0000-0001-7532-645X]{Bryna J. Hazelton}
\affiliation{Department of Physics, University of Washington, Seattle, WA 98195, USA}
\affiliation{eScience Institute, University of Washington, Seattle, WA 98195, USA}

\author{Jack  Hickish}
\affiliation{Radio Astronomy Laboratory, University of California, Berkeley, Berkeley, CA 94720, USA}

\author[0000-0002-0917-2269]{Daniel C. Jacobs}
\affiliation{School of Earth and Space Exploration, Arizona State University, Tempe, AZ 85287, USA}

\author{Alec  Josaitis}
\affiliation{Cavendish Astrophysics, University of Cambridge, Cambridge CB3 0HE, UK}

\author[0000-0002-8211-1892]{Nicholas S. Kern}
\altaffiliation{NASA Hubble Fellow}
\affiliation{MIT Kavli Institute, Massachusetts Institute of Technology, Cambridge, MA 02139, USA}
\affiliation{Department of Physics, Massachusetts Institute of Technology, Cambridge, MA 02139, USA}

\author[0000-0002-1876-272X]{Joshua  Kerrigan}
\affiliation{Department of Physics, Brown University, Providence, RI 02912, USA}

\author[0000-0003-0953-313X]{Piyanat  Kittiwisit}
\affiliation{Department of Physics and Astronomy,  University of Western Cape, Cape Town 7535, South Africa}

\author[0000-0002-2950-2974]{Matthew  Kolopanis}
\affiliation{School of Earth and Space Exploration, Arizona State University, Tempe, AZ 85287, USA}

\author[0000-0002-4693-0102]{Paul  La~Plante}
\affiliation{Department of Computer Science, University of Nevada, Las Vegas, NV 89154, USA}
\affiliation{Nevada Center for Astrophysics, University of Nevada, Las Vegas, NV 89154, USA}

\author{Adam  Lanman}
\affiliation{MIT Kavli Institute, Massachusetts Institute of Technology, Cambridge, MA 02139, USA}

\author[0000-0001-8108-0986]{Yin-Zhe Ma}
\affiliation{Department of Physics, Stellenbosch University, Matieland 7602, South Africa}

\author{David H.\ E.\ MacMahon}
\affiliation{Radio Astronomy Laboratory, University of California, Berkeley, Berkeley, CA 94720, USA}

\author{Lourence  Malan}
\affiliation{South African Radio Astronomy Observatory, Cape Town 7925, South Africa}

\author{Cresshim  Malgas}
\affiliation{South African Radio Astronomy Observatory, Cape Town 7925, South Africa}

\author{Keith  Malgas}
\affiliation{South African Radio Astronomy Observatory, Cape Town 7925, South Africa}

\author{Bradley  Marero}
\affiliation{South African Radio Astronomy Observatory, Cape Town 7925, South Africa}

\author{Zachary E. Martinot}
\affiliation{Department of Physics and Astronomy, University of Pennsylvania, Philadelphia, PA 19104, USA}

\author{Lisa McBride}
\affiliation{Department of Physics and Trottier Space Institute, McGill University, Montreal QC H3A 2T8, Canada}
\affiliation{Institut d’Astrophysique Spatiale, CNRS, Université Paris-Saclay, 91405 Orsay, France}

\author[0000-0003-3374-1772]{Andrei  Mesinger}
\affiliation{Scuola Normale Superiore, 56126 Pisa, PI, Italy}

\author{Nicel Mohamed-Hinds}
\affiliation{Department of Physics, University of Washington, Seattle, WA 98195, USA}

\author{Mathakane Molewa}
\affiliation{South African Radio Astronomy Observatory, Cape Town 7925, South Africa}

\author[0000-0001-7694-4030]{Miguel F. Morales}
\affiliation{Department of Physics, University of Washington, Seattle, WA 98195, USA}

\author[0000-0003-3059-3823]{Steven G. Murray}
\affiliation{Scuola Normale Superiore, 56126 Pisa, PI, Italy}
\affiliation{School of Earth and Space Exploration, Arizona State University, Tempe, AZ 85287, USA}

\author{Hans Nuwegeld}
\affiliation{South African Radio Astronomy Observatory, Cape Town 7925, South Africa}

\author[0000-0002-5400-8097]{Aaron R. Parsons}
\affiliation{Department of Astronomy, University of California, Berkeley, Berkeley, CA 94720, USA}

\author[0000-0003-0073-5528]{Robert Pascua}
\affiliation{Department of Physics and Trottier Space Institute, McGill University, Montreal QC H3A 2T8, Canada}

\author{Yuxiang  Qin}
\affiliation{Research School of Astronomy and Astrophysics, Australian National University, Canberra, ACT 2611, Australia}

\author{Eleanor Rath}
\affiliation{MIT Kavli Institute, Massachusetts Institute of Technology, Cambridge, MA 02139, USA}
\affiliation{Department of Physics, Massachusetts Institute of Technology, Cambridge, MA 02139, USA}

\author{Nima Razavi-Ghods}
\affiliation{Cavendish Astrophysics, University of Cambridge, Cambridge CB3 0HE, UK}

\author{James Robnett}
\affiliation{National Radio Astronomy Observatory, Socorro, NM 87801, USA}

\author{Mario G. Santos}
\affiliation{South African Radio Astronomy Observatory, Cape Town 7925, South Africa}
\affiliation{Department of Physics and Astronomy,  University of Western Cape, Cape Town 7535, South Africa}

\author{Peter Sims}
\affiliation{School of Earth and Space Exploration, Arizona State University, Tempe, AZ 85287, USA}

\author{Saurabh Singh}
\affiliation{Raman Research Institute, Sadashivanagar, Bangalore 560080, India}

\author{Dara Storer}
\affiliation{Department of Physics, University of Washington, Seattle, WA 98195, USA}

\author{Hilton Swarts}\affiliation{South African Radio Astronomy Observatory, Cape Town 7925, South Africa}

\author{Jianrong Tan}
\affiliation{Department of Physics and Astronomy, University of Pennsylvania, Philadelphia, PA 19104, USA}

\author{Pieter van Wyngaarden}
\affiliation{South African Radio Astronomy Observatory, Cape Town 7925, South Africa}

\author{Haoxuan Zheng}
\affiliation{Department of Physics, Massachusetts Institute of Technology, Cambridge, MA 02139, USA}


\begin{abstract}
The precise characterization and mitigation of systematic effects is one of the biggest roadblocks impeding the detection of the fluctuations of cosmological 21cm signals. Missing data in radio cosmological experiments, often due to radio frequency interference (RFI), \replaced{poses}{pose} a particular challenge to power spectrum analysis as it could lead to the ringing of bright foreground modes in Fourier space, heavily contaminating the cosmological signals. Here we show that the problem of missing data becomes even more arduous in the presence of systematic effects. Using a realistic numerical simulation, we demonstrate that partially flagged data combined with systematic effects can introduce significant foreground ringing. We show that such an effect can be mitigated through inpainting the missing data. We present a rigorous statistical framework that incorporates the process of inpainting missing data into a quadratic estimator of the 21cm power spectrum. Under this framework, the uncertainties associated with our inpainting method and its impact on power spectrum statistics can be understood. These results are applied to the latest Phase II observations taken by the Hydrogen Epoch of Reionization Array, forming a crucial component in power spectrum analyses as we move toward detecting 21cm signals in the ever more noisy RFI environment.
\end{abstract}



\section{Introduction} \label{sec:intro}
The redshifted $21$cm signal from neutral hydrogen contains rich astrophysical and cosmological information across cosmic time \citep[for a review, see][]{Furlanetto2006:Review, Pritchard2012:Review}. Observations of such a signal will open up large volumes of the unexplored universe, allowing us to constrain initial conditions of star and galaxy formation and to better understand the dark sector of our universe \citep{Chang2008:BAO_IM, Morales2010:Review, Bull2015:21cmIM, Mesinger2016:Review}. In particular, radio interferometers have been built or proposed to detect the spatial fluctuation of the $21$cm signal to probe large-scale structures (LSS), the Epoch of Reionization (EoR), and the cosmic dawn. The high spectral resolution that can be achieved in radio astronomy make these experiments especially advantageous for probing small-scale fluctuations along the line of sight. Examples of these experiments include the Donald C. Baker Precision Array for Probing the EoR (PAPER, \citealt{PAPER2010:Overview}), the LOw Frequency Arrray (LOFAR, \citealt{LOFAR2013:Overview}), the New extension in Nançay upgrading LOFAR (NenuFAR, \citealt{NenuFAR2012:Overview}), the Murchison Widefield Array (MWA, \citealt{MWA2013:PhaseI_Overview, MWA2018:PhaseII_Overview}), the (upgraded) Giant Metrewave Radio Telescope (GMRT, \citealt{GMRT2013:21cmLimit, GMRT2017:uGMRT_Overview}), the MeerKAT telescope \citep{MeerKAT2016:MeerKLASS}, the Canadian Hydrogen Intensity Mapping Experiment (CHIME, \citealt{CHIME2022:Overview}), the Hydrogen Epoch of Reionization Array (HERA, \citealt{HERA2017:PhaseI_Overview, Berkhout2024:HERA_PhaseII}), the Canadian Hydrogen Observatory and Radio-transient Detector (CHORD, \citealt{CHORD2019:Overview}), the Owens Valley Long Wavelength Array (OVRO-LWA, \citealt{LWA2019:21cmLimit}), the Hydrogen Intensity and Real-time Analysis eXperiment (HIRAX, \citealt{HIRAX2022:Overview}), and the Square Kilometre Array (SKA, \citealt{SKA2015:HI_IM, SKA2015:EoR}). While none of these experiments have reported a direct detection of the 21cm power spectrum on its own beyond redshift $z\sim1$ so far \citep{Paul2023:MeerKAT_Detection}, many efforts have been put in to derive sensitive upper limits on the 21cm power spectrum at various redshifts \citep{Ghosh2011:GMRT_Limit, GMRT2013:21cmLimit, Parsons2014:PaperLimit, Dillon2014:MWA_Limit, Dillon2015:MWA_Limit, Ewall-Wice2016:Limit_Relection, Beardsley2016:MWA_Limit, Patil2017:LofarLimit, Li2019:MWA_Limit, Barry2019:MWA_Limit, LWA2019:21cmLimit, Gehlot2019:LofarLimit, Kolopanis2019:PaperLimit, Trott2020:MWA_Limit, Mertens2020:LofarLimit, Chakraborty2021:uGMRT_Limit, Garsden2021:LWA_Limit, Yoshiura2021:MWA_Limit, HERA2022:h1c_idr2_limit, HERA2023:h1c_idr3_limit, Wilensky2023:MWA_Limit, Munshi2024:Nenufar_Limit}.

The main challenge to a successful measurement of the 21cm power spectrum lies in mitigating the bright foreground emission. Radio emission from astrophysical foregrounds can be many orders of magnitude brighter than the cosmological signals. Ideally, the spectral smoothness of the foreground emission could allow one to localize the foreground signature in Fourier space (known as the $\textit{foreground wedge}$, \citealt{Datta2010:FGwedge, Parsons2012:delay_spectrum_wedge, Vedantham2012:image_wedge, Trott2012:wedge, Morales2012:wedge, Hazelton2013:wedge, Thyagarajan2013:wedge, Liu2014:EoR_WindowI}), making it possible to either model and subtract the foreground, or filter and avoid the foreground \citep[see][for a review]{Liu2020:Review}. Unfortunately, the lack of a perfect characterization of one's instrument could hinder these foreground mitigation strategies as any unknown spectral variations could lead to a leakage of bright foreground modes out of their intrinsically localized region. Examples of these instrumental systematic effects include calibration errors \citep{Barry2016:CalibrationError, Patil2016:CalibrationError, Ewall-Wice2017:CalibrationError, Bryne2019:CalibrationError, Mouri_Sardarabadi2019:CalibrationError, Dillon2020:redcal}, incomplete or incorrect knowledge of the antenna response \citep{Neben2016:BeamErrorMWA, Joseph2018:BeamError, Ansah-Narh2018:BeamPerturbation, Orosz2019:BeamVariation, Joseph2020:BeamVariation, Choudhuri2021:BeamVariation, Kim2022:BeamPerturbation}, or internal cable reflections, signal chain cross-talk and mutual coupling of antennas \citep{Ewall-Wice2016:Limit_Relection, Kern2019:Relection_Model, Ung2020:MWA_Coupling_Sim, Josaitis2022:MutualCoupling, Rath_Pascua2024:Mutual_Coupling}. 

In addition to systematic effects arising from the instrumental signal processing chain, a growing concern for cosmological experiments in the radio band is the increasing amount of radio frequency interference (RFI). RFI from terrestrial and satellite communications can be observed even by telescopes located in extremely radio-quiet sites \citep{Bowman2010:EDGES_RFI, Offringa2013:LOFAR_RFI, Offringa2015:MWA_RFI, Sokolowski2016:MWA_RFI, Sihlangu2020:MeerKAT_RFI, Lourenco2023:ASCAP_RFI}. The impact of RFI on 21cm power spectrum is two-fold: while most RFI is bright and can be easily identified and masked in the data \citep[e.g.,][]{Offringa2012:AoFlagger, Wilensky2019:SSINS}, residual faint RFI can still introduce power comparable to the cosmological signal across a wide range of Fourier modes \citep{Wilensky2023:MWA_Limit}; secondly, even if all the RFI is perfectly flagged, the gaps in data created by flagged channels are particularly problematic for power spectrum analyses as these discontinuities along the frequency axis may give rise to ringing of bright foreground modes in Fourier space. Owing in part to the continuously deploying satellite constellations \citep{Di_Vruno2023:Starlink_Lofar, Grigg2023:Starlink_SKA}, the ever more noisy RFI environment makes mitigating the impacts of RFI an imminent task in 21cm cosmology.

Here, we focus on dealing with gaps in the data created by flagging RFI. A conservative approach to deal with this problem is by multiplying the data with a taper function that goes to zero near the gaps \citep{Kolopanis2019:PaperLimit} or even to directly exclude an entire integration time that is affected by RFI \citep{Wilensky2023:MWA_Limit}. However, these methods will significantly limit the accessible frequency bandwidth the experiments could probe. Moreover, tapers that are dictated by RFI gap instead of the intrinsic properties of the instrument will, in general, be sub-optimal and result in a reduced sensitivity. Another common strategy to reduce the impact of gaps created by RFI is to average the data across different observations that trace the same cosmological mode. This could be averaging across different baselines through redundant baseline averaging or $uv$-plane gridding, averaging observations of the same patch of the sky from different times \citep{HERA2022:h1c_idr2_limit}, \added{or averaging data with the same frequency separation \citep{Bharadwaj2019:MAPS, Pal2021:TGE, Elahi2024:TGE_on_MWA}}. This way, only channels with RFI flags that are persistent along certain axis require further treatments. In this work, however, we use a realistic simulation to show that in the presence of varying systematic effects, single-baseline delay power spectra with only partially flagged data still exhibit foreground ringing that can be seen at current state-of-the-art sensitivity levels. This is similar to what has been identified in \citet{Offringa2019:RFI, Wilensky2022:flag_systematics} in the context of a gridded power spectrum estimator. 

Numerous other techniques have been proposed to deal with RFI gaps through extracting information from the unflagged data. \citet{Parsons2009:CLEAN} adapted the CLEAN algorithm to operate in the spectral dimension \citep{Hogbom1974:CLEAN, Roberts1987:CLEAN_Time_Series}, proposing an iterative deconvolution method to mitigate incomplete frequency sampling. The deconvolution based method was applied to the PAPER and HERA analyses \citep{Parsons2014:PaperLimit, HERA2022:h1c_idr2_limit, HERA2023:h1c_idr3_limit}. Another camp of methods is to fit the unflagged data with a certain set of basis functions to directly perform the transformation to Fourier space or to $\textit{inpaint}$ the missing frequency channels. These methods have a long history in cosmic microwave background analyses to handle missing sky coverage \citep{de_Oliveira-Costa2006:CMB_GLS, Abrial2008:CMB_Inpainting, Feeney2011:CMB_SkyReconstruction, Starck2013:CMB_Inpainting, Gruetjen2017:CMB_Inpainting}. For instance, \citet{GMRT2013:21cmLimit} adopted a Hermite basis to perform the line-of-sight transformation; \citet{Trott2016:CHIPS} performed a least square spectral analysis (LSSA, \citealt{Vanivek1969:LSSA, Vanivek1971:LSSA}) to fit for Fourier coefficients in MWA data; \citet{Patil2017:LofarLimit, Gehlot2019:LofarLimit, Mertens2020:LofarLimit} also partly utilized the LSSA method for the LOFAR data, while \citet{Barry2019:MWA_Limit} adopted the similar Lomb-Scargle method \citep{Lomb1976, Scargle1982} for the MWA data; \citet{Ewall-Wice2021:DAYNENU} proposed filtering the foreground using the discrete prolate spheroidal sequence (DPSS, \citealt{Slepian1978:DPSS}), which was later applied to the CHIME observations \citep{CHIME2023:21cm_x_QSO_ELG, CHIME2023:21cm_x_Lya}; lastly, many machine learning based methods such as Gaussian Process Regression \citep{Rybicki1992:GPR, Mertens2018:GPR, Offringa2019:RFI, Kern2021:GPR} and Convolutional Neural Networks \citep{Pagano2023:Inpainting} have also gained growing interest in the community and have been actively applied to real data. 

While most of these methods perform reasonably well on data that are not heavily flagged \citep{Chakraborty2022:Inpainting, Pagano2023:Inpainting}, an important question yet to be fully addressed is how these methods impact the statistics of the power spectrum estimator. The non-trivial correlation unavoidably introduced in the Fourier space by incomplete frequency sampling and the uncertainty in these methods itself could be increasingly important as we flag more and more data. \citet{Ewall-Wice2021:DAYNENU} used an empirical covariance matrix to estimate the effect of DPSS foreground filtering on power spectrum window functions. \citet{Kern2021:GPR} investigated the effect of Gaussian Process Regression on power spectrum window functions under the optimal quadratic estimator framework. \citet{Kennedy2023:Gibbs_Sampling, Burba2024:Bayesian21cm} developed a Bayesian Gibbs sampling framework to estimate uncertainties in the recovered power spectra in the presence of flags. In this work, we focus on the statistical impact on 21cm delay spectra introduced by the DPSS inpainting method. DPSS inpainting has been shown in \citet{Pagano2023:Inpainting} to have the smallest error on inpainting narrow RFI gaps, and its linear nature allows us to more easily examine its statistical impact. We use a Bayesian framework to derive the uncertainties associated with the inpainted data. The uncertainties and correlations introduced by data inpainting are then incorporated into a quadratic estimator framework to estimate their statistical impact on the 21cm power spectrum. Our results also take into account the effects of additional analysis steps such as averaging across sidereal days, coherent and incoherent time average. These results are applied to the HERA phase II observations \citep{Berkhout2024:HERA_PhaseII} to construct a set of strategies to handle missing data in the upcoming 21cm power spectrum analysis. 

This paper is organized as follows. In Sec.\,\ref{sec:motivate}, we give an intuition on why non-uniform data sampling can lead to foreground ringing in the presence of systematic effects. This intuition is further verified in Sec.\,\ref{sec:simulation} using a realistic simuation. In Sec.\,\ref{sec:inpainting}, we introduce the DPSS inpainting method and demonstrate its impact on the statistics of the power spectrum estimator through our simulations. These results are applied to data from HERA phase II observations in Sec.\,\ref{sec:hera}. Conclusions are given in Sec.\,\ref{sec:conclusion}.

\section{Motivation} \label{sec:motivate}

To build intuition on how non-uniform data sampling can affect the 21cm power spectrum, consider the interferometric response of two antennas, \textit{i.e.}, the \textit{visibility}, measured at a frequency $\nu$,
\begin{equation}
\label{eq:measurement_eq}
    V(\mathbf{b}_{ij}, \nu) = \int \mathrm{d}\Omega\,I(\hat{\mathbf{s}}, \nu) B_{ij}(\hat{\mathbf{s}}, \nu) \exp\left(-\frac{i2\pi\nu}{c}\mathbf{b}_{ij}\cdot\hat{\mathbf{s}}\right).
\end{equation}
Here, $i$ and $j$ are antenna indices, $I$ is the specific intensity of the sky, $B_{ij}$ is the cross power beam, $\mathbf{b}_{ij}$ is the baseline vector, and $\hat{\mathbf{s}}$ is the unit vector on the sky which we integrate over. A rough estimator of the 21cm power spectrum can then be formed as
\begin{equation}
    \label{eq:delay_spectrum}
    \hat{P}_{21\mathrm{cm}, \alpha} \propto \Tilde{V}^*(\mathbf{b}_{1}, \tau_\alpha)\Tilde{V}(\mathbf{b}_{2}, \tau_\alpha),
\end{equation}
where $\Tilde{V}$ is obtained by taking the \textit{delay transform} \citep{Parsons2009:CLEAN} of the visibility
\begin{equation}
    \Tilde{V}(\mathbf{b}, \tau) = \int \mathrm{d}\nu\, V(\mathbf{b}, \nu) \gamma(\nu) \exp(-i2\pi\nu\tau)
\end{equation}
in which $\gamma(\nu)$ is a tapering function chosen by the analyst. Throughout this work, if not otherwise specified, we choose $\gamma(\nu)$ to be the 4-term Blackman-Harris window \citep{Harris1978, Kolopanis2019:PaperLimit}. In general, we can decompose the delay transformed visibility into
\begin{equation}
    \Tilde{V} = \Tilde{s} + \Tilde{e} + \Tilde{n}
\end{equation}
where $s$ is the signal from foreground emission, $e$ is the desired EoR signal, and $n$ is the thermal noise. Because the foreground emission is spectrally smooth, we expect the foreground signal to be below the noise level beyond a certain delay depending on the baseline length \citep{Parsons2012:delay_spectrum_wedge}. This creates a range in delay space that allows us to more easily search for the EoR signal, known as the \textit{EoR window} \citep{Datta2010:FGwedge, Liu2014:EoR_WindowI}.

If the observation is affected by RFI, causing some of the frequency channels to be flagged, the EoR window can be contaminated. We can write the flagged visibility as $V_\mathrm{flagged}(\nu) = V(\nu) \times w(\nu)$ where $w$ is a sampling function that takes the value of $1$ or $0$. If we then take the delay transform of the flagged visibility, by the convolution theorem, we obtain
\begin{equation}
    \Tilde{V}_\mathrm{flagged}(\tau) = (\Tilde{V}\circledast\Tilde{w})(\tau),
\end{equation}
where we use $\circledast$ to denote the convolution of two functions. As the sampling function $w$ is a sum of multiple top hat functions, its delay transform will be of the form
\begin{equation}
    \Tilde{w}(\tau) = \sum_{i} a_i\operatorname{sinc}(\pi a_i \tau) e^{i2\pi\nu_i\tau},
\end{equation}
where $\nu_i$ and $a_i$ denote the location and width of the flag respectively. Since $\Tilde{w}(\tau)$ has unbounded support, this will cause the foreground $\Tilde{s}_\mathrm{flagged} = \Tilde{s}\circledast\Tilde{w}$ to be no longer limited within a certain delay range, leading to contamination of the EoR window. This impact of missing data on power spectra is well understood in the literature among different contexts. In this work, however, we are focusing on a more subtle effect that arises from averaging over data that might contain flags. 

Consider the example of a drift-scan telescope. As the earth rotates, a drift-scan 21cm experiment can observe repeatedly at the same local sidereal time (LST) to increase sensitivity. The LST-averaged visibility after a long observation period can be written as
\begin{equation}
    V_\mathrm{avg}(\mathbf{b}, \nu) = \displaystyle\frac{\sum_i V_i(\mathbf{b}, \nu)\times w_i(\nu)}{\sum_i w_i(\nu)}
\end{equation}
where $i$ runs over the different sidereal days the visibility is observed on. Here, we will consider the regime where none of the frequency channels are flagged every day, $\textit{i.e.}$, $\sum_i w_i(\nu) > 0$ for all $\nu$. In this regime, if the visibility consists only of foreground sky emission, cosmological signal, and thermal noise, the non-uniform sampling of data will have little impact on the final power spectrum since
\begin{equation}
\begin{aligned}
     V_\mathrm{avg} &= \displaystyle\frac{\sum_i (s + e + n_i)\times w_i}{\sum_i w_i}\\
     &=s + e + \frac{\sum_i n_i w_i}{\sum_i w_i} \\
     \Rightarrow \Tilde{V}_\mathrm{avg} &= \Tilde{s} + \Tilde{e} + \sum_i \Tilde{K}_i\circledast \Tilde{n}_i,
\end{aligned}
\end{equation}
where $K_i \equiv \left(w_i / \sum_j w_j\right)$ and its delay transformed counterpart $\Tilde{K}_i$ would be a sum of sinc-like function. Here, we assume only the thermal noise component $n_i$ is different from night to night\footnote{We note that we also assume that the observation is taken at the exact same grid of LST every night. This could cease to be true in practice due to misaligned discretization in the instrument or choices made by the analysts. The presence of such a slight deviation in nightly observing time can introduce extra correlation in the power spectrum as investigated in \citet{Wilensky2022:flag_systematics}.}. In this case, we see that the kernel $\Tilde{K}_i$ induced by the non-uniform sampling only introduces correlation in the noise spectrum. Since the $\Tilde{K}_i\circledast \Tilde{n}_i$ terms still have zero mean, $\Tilde{V}_\mathrm{avg}$ is still able to give us an unbiased estimate of the cosmological signal in the EoR window. 

In reality, however, every component in the visibility can vary from night to night due to systematic effects. For example, gain uncertainties after calibration, small motions in the instrument, or internal instrument couplings can all introduce tiny night-to-night variations on the measured visibilities. If we quantify these nightly varying systematic effects by a multiplicative bias $(1 + \varepsilon_i)$, the resulting LST-averaged visibility would be
\begin{equation}
\label{eq:systematic_and_flag_real}
\begin{aligned}
     V_\mathrm{avg} &= \displaystyle\frac{\sum_i (1 + \varepsilon_i)(s + e + n_i)\times w_i}{\sum_i w_i}\\
     &=s + e + \frac{\sum_i \varepsilon_i w_i}{\sum_i w_i}(s + e)  + \frac{\sum_i (1+\varepsilon_i)n_i w_i}{\sum_i w_i}\\
     &= s + e + \sum_i \varepsilon_i K_i(s+e) + \sum_i K_i (1+\varepsilon_i)n_i
\end{aligned} 
\end{equation}

Here, even if we assume the multiplicative biases $\varepsilon$ do not depend on frequency, we see that

\begin{equation}
\label{eq:systematic_and_flag_fourier}
\Tilde{V}_\mathrm{avg} = \Tilde{s} + \Tilde{e} + \sum_i \varepsilon_i \Tilde{K}_i\circledast (\Tilde{s}+\Tilde{e}) + \sum_i (1+\varepsilon_i) \Tilde{K}_i\circledast \Tilde{n}_i.
\end{equation}
In particular, the terms proportional to $\Tilde{K}_i\circledast \Tilde{s}$ would lead to contamination of the EoR window due to the convolution of the kernel with the bright foreground component. We stress that such a contamination exists even though for every frequency channel we have measured some amount of good data. Moreover, unlike thermal noise, the fluctuation in the systematic effects $\varepsilon_i$ may not be centered around zero and are usually highly uncertain, making it difficult to remove the contamination simply via averaging the data across some other axes. Quantifying and mitigating this interplay between systematic effects and non-uniform data sampling is precisely the subject of this work.

\begin{figure*}[tbh]
    \centering
    \epsscale{1.175}
    \plotone{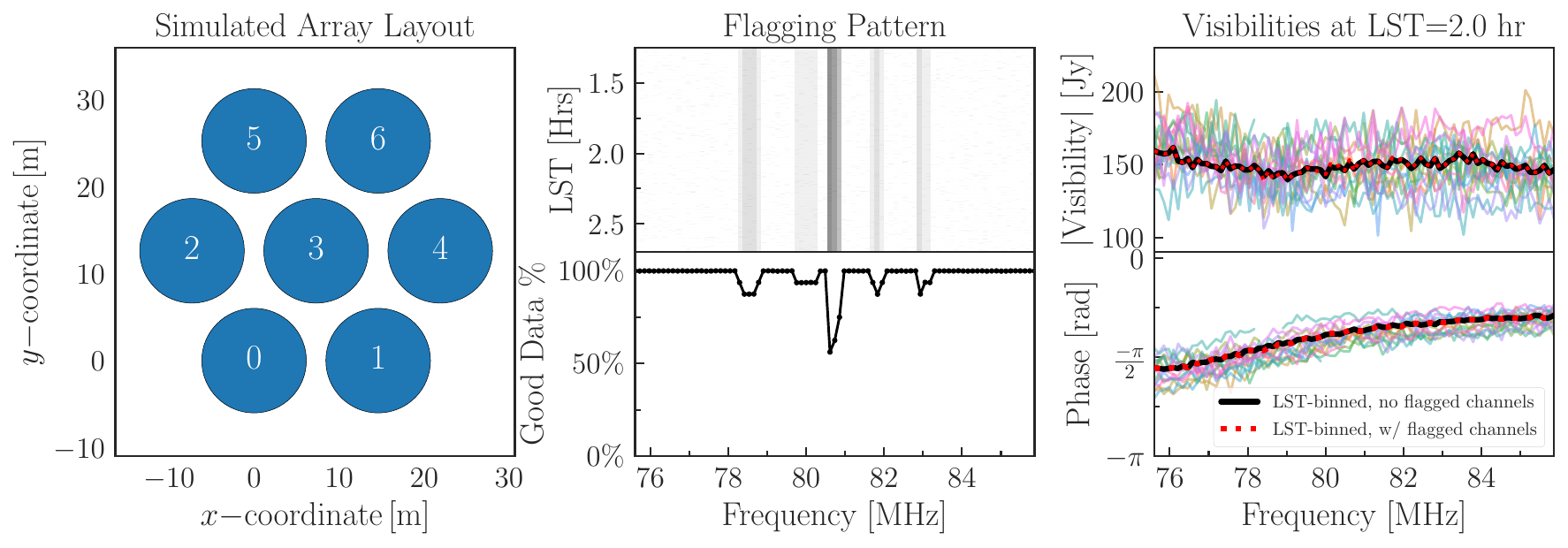}
    \caption{$\textit{Left}$: An example interferometric array layout we simulate in this work. antennas are placed on the vertices and the center of a hexagon with a side of $14.6$ meters. $\textit{Center}$: An example of the simulated flagging patterns we draw in this work. The top panel shows the flagging patterns for the baseline formed by antenna $0$ and $1$ summed over all nights of observation. Darker color indicates a higher amount of flags. The lower panel shows the overall percentage of good data as a function of frequency. $\textit{Right}$: Absolute value of the nightly visibilities at a particular local sidereal time (color lines) and the sidereal-day-averaged visibility with (solid black) or without (dotted black) flags. \label{fig:vis_sim}}

\end{figure*}

\section{Simulation} \label{sec:simulation}
As an illustration, we construct a simple interferometric array similar to the Hydrogen Epoch of Reionization Array (HERA) and simulate its response to foreground sky emission with various instrumental systematic effects. We construct a seven-element interferometric array with the antennas located on the vertices and the center of a hexagon with a side of 14.6 meters. The array layout can be seen in the leftmost panel of \autoref{fig:vis_sim}. Starting from Eq.\,\eqref{eq:measurement_eq}, we simulate the visibility response from point sources in the GLEAM catalog \citep{Hurley-Walker2017:GLEAMI, Hurley-Walker2019:GLEAMII} and the diffuse emission given by the Global Sky Model \citep{de_Oliveira-Costa2008:GSM, Zheng2017:GSM}. We assume each antenna has an Airy beam profile 
\begin{equation}
    B\left(\hat{\mathbf{s}}(\theta, \phi); \nu\right) = \left[\frac{2J_1(2\pi\nu a \sin\theta/c)}{2\pi\nu a \sin\theta/c}\right]^2
\end{equation}
where $J_1$ is the Bessel function of the first kind, $\theta$ is the zenith angle, and $a$ is the aperture radius which we set to be six meters \citep{Neben2016:HERA_Beam, HERA2017:PhaseI_Overview}. Here, we simulate the array to observe the sky every 10 seconds over an overlapping 1.5 hour period across 16 nights. The visibility is sampled in 120 kHz-wide frequency channels to imitate the response of a HERA-like digital back end \citep{Berkhout2024:HERA_PhaseII}. To simulate potential night-to-night fluctuations due to systematic effects and investigate their interplay with RFI flags, we consider the following additional ingredients to our simulation:
\begin{enumerate}
    \item Antenna gain uncertainties: Calibrating the instrument and solving for the antenna's gain solutions is a vital task in 21cm cosmology. While many calibration strategies exist in the literature \citep{Mitchell2008:Calibration, Yatawatta2008:SAGE_Calibration, Liu2010:redcal, Sullivan2012:Calibration, Sievers2017:Calibration, Li2018:Calibration, Armel_Mbou_Sob2019:Calibration, Dillon2020:redcal, Sims2022:Calibration, Byrne2023:Calibration, Cox2023:nucal}, no calibration method is perfect. Here, we consider a small uncertainty in the observed visibility $V^\mathrm{obs}$ compared to the underlying true visibility $V^\mathrm{true}$ by setting
    \begin{equation}
        V^\mathrm{obs}_{ij} = g_i g^*_j V^\mathrm{true}_{ij}
    \end{equation}
    where $i$ and $j$ are antenna indices and $g_j$ is the antenna-based complex gain uncertainty. We model $g_j \equiv 1 + (a + bi)$ with $a$ and $b$ randomly drawn from a uniform distribution between $-0.05$ and $0.05$ for each antenna at each night to simulate a $\sim 5\%$ uncertainty in gain solutions. 

    \item Beam perturbations: Accurate descriptions of the antenna primary beam response is another key to a precise measurement of 21cm power spectrum. However, the antenna response can be subject to small changes due to perturbations to the instrument itself. Here, we consider another systematic effect arises from random motions in the feed. This is modeled with a generalized antenna primary power beam profile \citep{Orosz2019:BeamVariation}
    \begin{equation}
        B_{i}(\hat{\mathbf{s}}(\theta, \phi); \nu) = \left[\frac{2 J_1\left(2\pi\nu a \sin\theta'/ c\right)}{2\pi\nu a \sin\theta'/c}\right]^2,
    \end{equation}
    where the perturbed pointing $\theta'$ is defined as
    \begin{equation}
    \begin{aligned}
        &\sin\theta'(\theta, \phi, x_i, y_i) \equiv  \\ &\displaystyle\sqrt{\left(\sin\theta\cos\phi - \frac{x_i}{z}\right)^2 + \left(\sin\theta\sin\phi - \frac{y_i}{z}\right)^2}
    \end{aligned}
    \end{equation}
    in which $x_i, y_i$ represent a small perturbation in the feed position and $z$ is the height of the feed. Here, we assume $z = 4.5\mathrm{m}$ and draw $x_i$ and $y_i$ from a Gaussian distribution with zero mean and $\sigma = 2\mathrm{cm}$ \citep{HERA_Beam_Movement} for each antenna at each night of observation. The cross power beam in Eq.\,\eqref{eq:measurement_eq} is then obtained as $B_{ij} = \sqrt{B_{i} B_{j}}$.
    
    \item Instrument coupling: Another known concern for radio interferometers is the potential coupling among antennas. These couplings can occur internally in the analog signal chain due to impedance mismatches or in the field through reflections from one antenna to another. Because of the delay in signal propagation, these instrument couplings will cause the foreground signal to leak into higher delay ranges. Here, we adopt a model presented in \citet{Josaitis2022:MutualCoupling} and \citet{Rath_Pascua2024:Mutual_Coupling} to simulate this effect. The coupled visibility $V^\mathrm{cpl}_{ij}$ is modeled as
    \begin{equation}
    \begin{aligned}
        V^\mathrm{cpl}_{ij} = V^\mathrm{true}_{ij} & -\sum_{k \neq i} \Gamma\frac{V_{kj}^\mathrm{true}}{\left|\mathbf{b}_{ik}\right|/\lambda} e^{+i 2 \pi \nu \left|\mathbf{b}_{ik}\right| / c}  \\
        & +\sum_{k \neq j} \Gamma^*\frac{V_{ik}^\mathrm{true}}{\left|\mathbf{b}_{kj}\right|/\lambda} e^{-i 2 \pi \nu \left|\mathbf{b}_{kj}\right| / c},
    \end{aligned}
    \end{equation}
    where the coupling coefficient $\Gamma \equiv a + bi$ is drawn nightly by randomly choosing $a, b \in [-0.01, 0.01]$

    \item Thermal Noise: After modeling the signal and the systematic effects, we add a noise realization to each baseline for each time integration and frequency channel according to the radiometer equation and the auto-correlation visibilities \citep{Tan2021:HERA_ErrorBar}
    \begin{equation}
    \label{eq:noise}
        \sigma_{ij}^\mathrm{rms}(\nu, t) = \sqrt{\frac{V_{ii}(\nu, t) V_{jj}(\nu, t)}{\Delta\nu\Delta t}}.
    \end{equation}    
    Here $\sigma_{ij}^\mathrm{rms}$ denotes the standard deviation of the thermal noise for the visibility measured by baseline $\vect{b}_{ij}$, $V_{ii}$ is the auto-correlation visibility measured by antenna $i$, $\Delta \nu$ is the correlator channel width, and $\Delta t$ is the correlator integration time.

    \item RFI Flags: We create two types of flags for each of the antennas in our simulation during each night of observations. Firstly, a channel at a given time and frequency is randomly flagged according to a binomial distribution with a probability of $0.001$. These flags mimic the effect from irregular short-timescale RFI sources such as airplane reflections. Secondly, we create RFI flags for a few frequency channels across an entire night. For each night, we identify $n\in\{0, 1, 2\}$ frequency channels that are common to all antennas to flag. We start by picking the flagged frequency channels with a uniform weighting, and the channels that have been flagged the night before will be more likely to be flagged to simulate periodic RFI sources such as satellite communications or TV stations. For each antenna, the RFI flags are centered at these frequency channels with a randomly chosen width of $0$ (not flagged) to $3$ channels (the central frequency channel and the two neighbouring channels are all flagged) to simulate leakage of bright RFI sources. 
\end{enumerate}

The middle panel of \autoref{fig:vis_sim} shows an example of the flagging patterns for one baseline in our simulation. The nightly variation in the visibilities due to changing systematic effects is shown in the rightmost panel of \autoref{fig:vis_sim}. The colored lines in the top panel are nightly visibilities at a particular local sidereal time and the black lines indicate the averaged visibility with (solid) or without (dotted) flags. Although there exists only a very small difference between the two averaged visibilities, we will soon see that this could still introduce a non-negligible effect on the power spectrum, highlighting the necessity of looking beyond frequency-time waterfall plots in diagnosing systematic effects.


\begin{figure}[tb]
    \centering
    \epsscale{1.175}
    \plotone{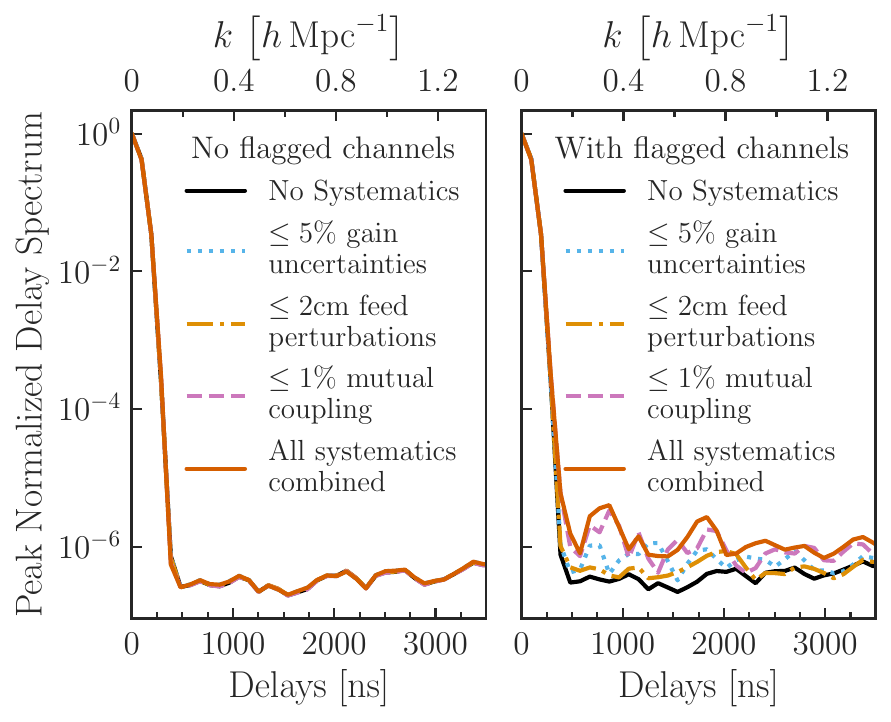}
    \caption{Comparison of power spectra of simulated flagged and unflagged visibilities from a single $14.6$ meter east-west baseline with various systematics effects. For a consistent comparison, we use simulated data with the exact same noise realization and the same realization for all the systematic effects. The only difference between the same line in the left and the right panel is in whether some channels are flagged or not during the sidereal-day average. The flagging pattern that goes into this particular case can be seen in \autoref{fig:vis_sim}.\label{fig:PS_compare}}

\end{figure}
 
\autoref{fig:PS_compare} shows the delay spectra for a single 14.6 meter baseline following Eq.\,\eqref{eq:delay_spectrum}. These delay spectra are formed by first coherently averaging 300 seconds of visibilities after phasing them to a common pointing center. The delay spectra within the 1.5 hour window of observations are then averaged incoherently to further increase the sensitivity. In \autoref{fig:PS_compare}, the different color lines show the delay spectrum from the visibility with different systematics considered. These systematic effects on themselves do not have any significant impact on the power spectrum, as can be seen in the left panel of \autoref{fig:PS_compare}. However, once we include flags during the sidereal day averaging, this introduces discontinuities in the visibility which raise the noise floor by almost an order-of-magnitude. More quantitatively, going back to eq.\,\eqref{eq:systematic_and_flag_real} and \eqref{eq:systematic_and_flag_fourier}, this demonstrates that even if we are in a regime where systematic effect is below the noise level in the EoR window ($\bar{\varepsilon}\Tilde{s} \ll \Tilde{n}$ at high delays), the ringing of bright foreground modes due to RFI flags can cause these systematic effects to be more prominent.

\begin{figure}[tbh]
    \centering
    \epsscale{1.175}
    \plotone{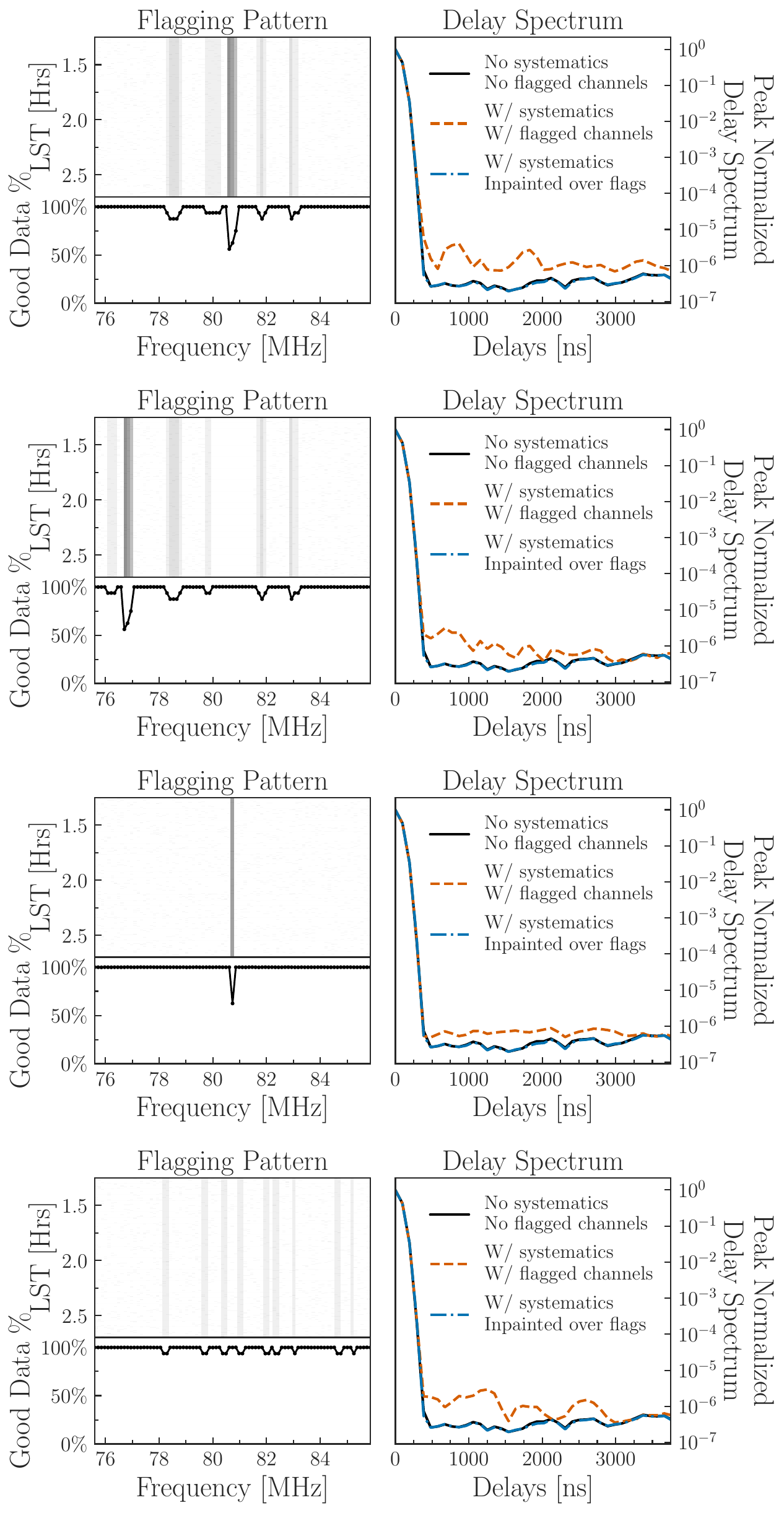}
    \caption{Power spectra with or without data inpainting under different simulated flagging patterns. Similar to Fig\,\ref{fig:vis_sim} and \ref{fig:PS_compare}, each row shows the flagging patterns (left) and the corresponding power spectra (right) for one 14.6 meter east-west baseline integrated over 16 nights. For a consistent comparison, the noise realization and the systematic effects are the same across the four different columns, the only difference is in the flagging patterns. \label{fig:PS_list}}

\end{figure}

The flagging pattern that goes into the right panel of \autoref{fig:PS_compare} is given in \autoref{fig:vis_sim}. While in this example, one of the channels is flagged around half of the time, causing a more significant discontinuity in the averaged visibility, foreground ringing can occur even when only a tiny fraction of the data are flagged. \autoref{fig:PS_list} shows the level of foreground ringing under a variety of scenarios. In all these cases, we only change the flagging patterns as shown in each of the panel on the left-hand side. For a consistent comparison, the noise realization and the systematic effects are kept the same. The bottom panel of Fig\,\ref{fig:PS_list} shows an example in which a tiny fraction ($\lessapprox 10\%$) of flagged data across the frequency range \replaced{is}{are} still enough to cause contamination to the EoR window. Fig\,\ref{fig:PS_list} also shows that the significance of this effect depends on both the width of the flags and the location of the flags. In the second panel from the top, we move the large gap from the first panel at around $80$MHz to the edge of the band, the leakage of bright foreground modes to the EoR window is significantly reduce due to the tapering function we applied. Similar effects can be seen in the third panel where we replace the wide gap in the first panel with a gap that is only a single channel wide. While carefully choosing a frequency band to estimate the power spectrum can help reduce the leakage of bright foreground modes, a mitigation strategy for these gaps in the data is required if we are to probe 21cm power spectra at arbitrary redshifts of interest.

\section{Data Inpainting} \label{sec:inpainting}
The power spectra derived from our simulation in Sec.\,\ref{sec:motivate} show us that even with channels that are flagged less than 10\% of the time, one cannot simply average the remaining good data in the presence of nightly varying systematic effects. Certain mitigation strategies are still needed to avoid the low-level ringing of bright foreground modes. A common method to deal with missing data due to RFI is to fill in the data based on our best guess of the signal, known as data inpainting \citep{Parsons2009:CLEAN, GMRT2013:21cmLimit, Trott2016:CHIPS, Barry2019:MWA_Limit, Ewall-Wice2021:DAYNENU}. Here, we will focus on a particular linear data inpainting method utilizing the discrete prolate spheroidal sequence (DPSS, \citealt{Slepian1978:DPSS, Ewall-Wice2021:DAYNENU}). Our inpainting procedure is discussed in Sec.\,\ref{subsec:inpaint_method}. In Sec.\,\ref{subsec:inpaint_error} we calculate the uncertainties and correlations in the visibility introduced by the inpainting procedure. The impact of inpainting on the final power spectrum estimator is discussed in Sec.\,\ref{subsec:inpaint_PS_error}.

\subsection{Inpainting Method} \label{subsec:inpaint_method}
As the contamination on the EoR window comes from the ringing of bright foreground modes, in order to mitigate this effect, we would like to fill the gap in our data with our best guess of the signals from foreground combined with systematic effects without introducing extra structures in the high delay modes. This can be achieved by choosing a set of band-limited basis functions $f_i(\nu)$ in frequency space with Fourier dual that is localized in a certain delay range. Such a set of basis functions is known as the discrete prolate spheroidal sequence (DPSS). Here, we follow \citet{Slepian1978:DPSS} and re-derive the properties of DPSS for the reader's convenience. Let $[\nu_0 - W/2, \nu_0 + W/2]$ be the frequency range where one is inpainting data, and $[-T, T]$ be the delay range we want our model to be localized in. We would like to find a sequence $f_i(\nu)$ that maximizes

\begin{equation} \label{eq:spectral_problem}
    \lambda_i \equiv \frac{\int_{-T}^{+T} |\Tilde{f}_i(\tau)|^2\,\mathrm{d}\tau }{\int_{-\infty}^{+\infty} |\Tilde{f}_i(\tau)|^2\,\mathrm{d}\tau}
\end{equation}
where we use $\Tilde{f}_i$ to denote the Fourier dual of $f_i$. Using the Parseval–Plancherel identity, the denominator of Eq.\,\eqref{eq:spectral_problem} can be written as 
\begin{equation}
\begin{aligned}
    \int_{-\infty}^{+\infty} |\Tilde{f}_i(\tau)|^2\,\mathrm{d}\tau &= \int_{-\infty}^{+\infty} |f_i(\nu)|^2\,\mathrm{d}\nu \\
    &= \int_{\nu_0 - W/2}^{\nu_0 + W/2} f_i^*(\nu) f_i(\nu)\,\mathrm{d}\nu,
\end{aligned}
\end{equation}
in which we use the fact that $f$ is band-limited. Meanwhile, the numerator of Eq.\,\eqref{eq:spectral_problem} gives
\begin{equation}
\begin{aligned}
    \int_{-T}^{+T} |\Tilde{f}_i(\tau)|^2\,\mathrm{d}\tau  &= \int_{-\infty}^{+\infty} \Tilde{f}_i^*(\tau) \Tilde{B}(\tau; T)\Tilde{f}_i(\tau) \,\mathrm{d}\tau  \\
    &= \int_{-\infty}^{+\infty} \Tilde{f}_i^*(\tau) \widetilde{B \circledast f_i}(\tau) \,\mathrm{d}\tau \\
    &= \int_{-\infty}^{+\infty} f_i^*(\nu) (B \circledast f_i)(\nu) \,\mathrm{d}\nu
\end{aligned}
\end{equation}
where $\Tilde{B}(\tau; T)$ is a top hat function in delay space with a width of $2T$ and $B$ is the inverse Fourier transform in frequency space 
\begin{equation}
    B = \frac{\sin{2\pi T\nu}}{\pi \nu}.
\end{equation}
In particular, this shows us that a set of solution to Eq.\,\eqref{eq:spectral_problem} can be found by solving for the band-limited eigenfunctions $f_i$ of
\begin{equation}
    \int_{\nu_0 - W/2}^{\nu_0 + W/2} \frac{\sin{2\pi T(\nu - \nu')}}{\pi (\nu-\nu')} f_i(\nu')\,\mathrm{d}\nu' = \lambda_i f_i(\nu).
\end{equation}
In the discretized version, our desired basis can be obtained by solving for the eigenvectors to the matrix 
\begin{equation}
    B_{ij} = \frac{\sin{2\pi T(\nu_i - \nu_j)}}{\pi (\nu_i - \nu_j)},
\end{equation}
known as the prolate matrix. The eigenvectors to the prolate matrix are precisely the discrete prolate spheroidal sequence (DPSS). While these eigenvectors form a complete set of basis, depending on their eigenvalue, not all of them has a highly concentrated power in our desired delay range. Following Eq.\,\eqref{eq:spectral_problem}, the eigenvalues $\lambda_i$ is strictly between $0$ and $1$. It has been shown that the eigenvalues are either close to $1$ or to $0$ with only a few in the transition zone \citep{Karnik2020:DPSS_Bounds}. In this work, we choose $T = 500\,\mathrm{ns}$ and select $f_i$ with $\lambda_i \geq 10^{-12}$ to form a set of basis functions to fit for the foreground signal in the flagged channels. A lower eigenvalue cut here ensures that we have a more complete basis for signal that is localized in the $[-T, +T]$ delay range. While this also gives us some eigenvectors with some structure in the high delay EoR window, the number of these eigenvectors are fairly limited. \added{We note that ideally, the value $T$ should be chosen at the horizon scale $\tau_\mathrm{H}\equiv |\vect{b}|/c$ for each baseline $\vect{b}$. This allows the chosen DPSS basis to capture all the bright foreground modes without affecting the EoR window. For the shortest $14.6$ meter baseline in our array, $\tau_\mathrm{H} \sim 50\mathrm{ns}$. In reality, one often chooses a $T$ that is slightly larger than $\tau_\mathrm{H}$ to also capture any foreground leakage beyond the wedge due to systematic effects. In this work, we choose $T=500\,\mathrm{ns}$ as it has been shown in data obtained by the Phase II HERA observations that foreground can leak to such a delay due to mutual coupling of antennas \citep{Rath_Pascua2024:Mutual_Coupling}.}

Once a set of basis $\{f_i\}_{i=1}^N$ is chosen, we can obtain our best guess of the foreground structure parameterized by these basis functions through the unflagged data. This is achieved by finding a set of coefficients $\mathbf{b}$ that solves the linear system 
\begin{equation}
    \mathbf{v}_\mathrm{obs} = \mathbf{W}\left(\mathbf{A}\mathbf{b} + \mathbf{n}\right)
\end{equation}
where $\mathbf{v}_\mathrm{obs}$ is the observed data, $A_{ij} = f_j(\nu_i)$ is the \textit{design matrix}, and $\mathbf{W}$ is a diagonal matrix that is $1$ for unflagged channels and $0$ otherwise. The maximum likelihood estimator of $\mathbf{b}$ is then
\begin{equation}\label{eq:b_ML}
    {\mathbf{\hat{b}}} = (\mathbf{A}^\dagger\mathbf{W}^\dagger\mathbf{N}^{-1}\mathbf{W}\mathbf{A})^{+} \mathbf{A}^\dagger\mathbf{W}^\dagger\mathbf{N}^{-1}\mathbf{W} \mathbf{v}_\mathrm{obs}
\end{equation}
where $M^+$ denotes the Moore–Penrose pseudo-inverse of a matrix $M$ and $\mathbf{N}\equiv\langle \mathbf{n} \mathbf{n}^\dagger\rangle$ is the noise covariance matrix. Although we do not know the noise properties for the channels that are flagged, only the combination $\mathbf{W}^\dagger\mathbf{N}^{-1}\mathbf{W} \equiv \mathbf{N}_\mathrm{u}^{-1}$ appears in Eq.\,\eqref{eq:b_ML}. Here, we assume the noise at different frequency channels are independent with each other and write $\mathbf{N}_\mathrm{u}^{-1}$ as a diagonal matrix as follows
\begin{equation}
\label{eq:N_unflag}
\left(\mathbf{N}_\mathrm{u}^{-1}\right)_{kk} = 
    \begin{cases}
        &1/\sigma^\mathrm{rms}(\nu_k)^2\quad\textrm{if }\nu_k\textrm{ is unflagged} \\
        &\\
        &0\quad\textrm{otherwise,}
    \end{cases}
\end{equation}
where $\sigma^\mathrm{rms}$ is given in Eq.\,\eqref{eq:noise}. Thus, the estimator $\mathbf{\hat{b}}$ depends solely on the information from unflagged frequency channels. 

With these best-fitted DPSS coefficients, we can inpaint the flagged channels with the predicted foreground model. The entire inpainted visibility can thus be written as
\begin{equation}
\label{eq:vobs_to_vinp}
    \begin{aligned}
        &\mathbf{v}_\mathrm{inp} = \mathbf{W}\mathbf{v}_\mathrm{obs} + (\mathbf{I} - \mathbf{W})\mathbf{A}\mathbf{\hat{b}} \\
        =& \left[\mathbf{W} + (\mathbf{I} - \mathbf{W})\mathbf{A}(\mathbf{A}^\dagger\mathbf{N}_\mathrm{u}^{-1}\mathbf{A})^{+} \mathbf{A}^\dagger\mathbf{N}_\mathrm{u}^{-1}\right]\mathbf{v}_\mathrm{obs}
    \end{aligned}
\end{equation}
For simplicity, we will denote the inpainting operator as $\mathcal{O}_\mathrm{inp}$ from here on. We note that the inpainted visibility is obtained with a completely linear operator on the observed visibility vector at a given time instance. The dashed blue lines in \autoref{fig:PS_list} show the resulting power spectra if we inpaint over the flagged channels each night. Even though the simulated visibilities contained nightly varying systematic effects, we can see that our inpainting procedure is still able to reduce the discontinuities in the data and achieve the expected sensitivity. 

\subsection{Uncertainties in the Inpainted Data}  \label{subsec:inpaint_error}
While the empirical success of using inpainting to mitigate missing data due to RFI is demonstrated in \autoref{fig:PS_list} and in the previous literature, one concern is how these data inpainting methods can affect the statistics of the 21cm power spectrum as more and more channels are flagged. The uncertainties we choose to propagate are a subtle balance between formal statistical and practical considerations to which we devote some pedagogical discussion. Readers interested in the results can jump directly to Eq.\,\eqref{eq:modify_noise_cov}. 

A typical error propagation technique when applying a linear operation to some data is to consider a frequentist thought experiment, where one repeatedly inpaints realizations of $\mathbf{v}_\mathrm{obs}$ with a fixed underlying signal, to obtain realizations of $\mathbf{v}_\mathrm{inp}$. This allows us to compute the covariance which will simply be $\mathcal{O}_\mathrm{inp}\mathbf{N}_\mathrm{u}\mathcal{O}_\mathrm{inp}^\dagger$ with $\mathcal{O}_\mathrm{inp}$ defined in Eq.\eqref{eq:vobs_to_vinp}\footnote{While strictly speaking, $\mathbf{N}_\mathrm{u}^{-1}$ as defined by Eq.\,\eqref{eq:N_unflag} has no inverse, the infinite variance in the flagged channels in $\mathbf{N}_\mathrm{u}$ does not affect our calculations as it is always projected out.}. To examine this form of the covariance, let $\mathbf{P}_\mathrm{f}$ and $\mathbf{P}_\mathrm{u}$ be the projection operator on the flagged and unflagged channels respectively. Then, on the unflagged channels, the uncertainties are
\begin{equation}
\mathbf{P}_\mathrm{u}\mathcal{O}_\mathrm{inp}\mathbf{N}_\mathrm{u}\mathcal{O}_\mathrm{inp}^\dagger\mathbf{P}_\mathrm{u}^\dagger = \mathbf{P}_\mathrm{u}\mathbf{N}_\mathrm{u}\mathbf{P}_\mathrm{u}^\dagger\equiv \mathbf{N}_\mathrm{u}'.
\end{equation}
On the other hand, on the flagged channels, let $\mathcal{O'}_\mathrm{inp}\equiv \mathbf{P}_\mathrm{f}\mathbf{A}(\mathbf{A}^\dagger\mathbf{N}_\mathrm{u}^{-1}\mathbf{A})^{+} \mathbf{A}^\dagger\mathbf{N}_\mathrm{u}^{-1}$ be the inpainting operation that maps the observed data $\mathbf{v}_\mathrm{obs}$ into a model which we use to inpaint the flagged channels. The uncertainties on the flagged channels are
\begin{equation}
\mathbf{P}_\mathrm{f}\mathcal{O}_\mathrm{inp}\mathbf{N}_\mathrm{u}\mathcal{O}_\mathrm{inp}^\dagger\mathbf{P}_\mathrm{f}^\dagger = \mathcal{O'}_\mathrm{inp}\mathbf{N}_\mathrm{u}\mathcal{O'}_\mathrm{inp}^\dagger.
\end{equation}
We see that the uncertainties in the flagged channels correspond to a linear combination of uncertainties in the observed channels. Since there are usually significantly more unflagged data than flagged data, this results in a covariance matrix that often expresses more uncertainty in the unflagged data than in the flagged data. While this result is formally correct under the frequentist assumptions given above, as we propagate such a covariance matrix into our power spectrum estimation framework in Sec.\,\ref{subsec:inpaint_PS_error}, it has the uncomfortable side effect of giving more weight to inpainted solutions in the flagged channels than the actual measured data in the unflagged channels.


If we cast inpainting as an inference problem, there is an alternative Bayesian formulation in terms of the posterior predictive distribution, $P(\mathbf{v}_\mathrm{inp}' | \mathbf{v}_\mathrm{obs}, \mathbf{N}, \mathbf{A})$, that does not have this undesirable property. Here, $\mathbf{v}_\mathrm{inp}'$ is the hypothetical unobserved RFI-free visibility in the flagged channels \textit{including the thermal noise that would corrupt it}, and $\mathbf{v}_\mathrm{obs}$ is the observed visibility\footnote{We note that unlike in the previous subsection where $\mathbf{v}_\mathrm{inp}$ is a vector with dimension $N_\mathrm{freq}$, $\vect{v}_\mathrm{inp}'$ now has dimension $N_\mathrm{freq, flagged}$. This is because by construction, we only inpaint data and predict a solution for the visibility in the flagged channel. Therefore, in the Bayesian approach, we are only interested in the probability distribution of the underlying visibility in the unobserved channels given the observed data.}. As before, $\mathbf{N}$ is the full frequency-frequency noise covariance and $\mathbf{A}$ is the design matrix of a given basis that maps a set of coefficients $\mathbf{b}$ to a foreground shape in frequency space. In Bayesian inference, probability is an extension of Boolean logic where propositions may be ascribed an uncertainty rather than a strict binary truth value. The posterior predictive distribution answers the question, ``Given some known prior information, including a model that is assumed to \textit{completely} describe the processes in the data, and some observed data, what is the probability (density) that some unobserved data lie in the (infinitesimal) interval $[\mathbf{v}_\mathrm{inp}',  \mathbf{v}_\mathrm{inp}' + \mathrm{d}\mathbf{v}_\mathrm{inp}']$?" 

The posterior predictive distribution can be calculated by marginalizing over $\mathbf{b}$ as follows
\begin{equation}\label{eq:inp_prob}
\begin{aligned}
    &P(\mathbf{v}_\mathrm{inp}' | \mathbf{v}_\mathrm{obs}, \mathbf{N}, \mathbf{A})\\ =& \int\mathrm{d}\mathbf{b}\,P(\mathbf{v}_\mathrm{inp}' | \mathbf{b}, \mathbf{N}_\mathrm{f}', \mathbf{A}) P(\mathbf{b} | \mathbf{v}_\mathrm{obs}, \mathbf{N}_\mathrm{u}, \mathbf{A}),
\end{aligned}
\end{equation}
where we use $\mathbf{N}_\mathrm{u}$ to denote the noise covariance of the observed data and $\mathbf{N}_\mathrm{f}' \equiv \mathbf{P}_\mathrm{f}\mathbf{N}\mathbf{P}_\mathrm{u}^\dagger$ for the noise variance of the hypothetical RFI-free data in the flagged channels. While $\mathbf{N}_\mathrm{u}$ can be estimated using the auto-correlations as in Eq.\,\eqref{eq:N_unflag}, we do not know $\mathbf{N}_\mathrm{f}'$, \textit{i.e.}, the noise properties in the flagged channels, a priori. However, since the auto-correlations are very smooth, we can safely assume that we can interpolate the auto-correlations over the flagged channels and infer $\mathbf{N}_\mathrm{f}'$ with very low uncertainty. 

The two factors in the integrand of Eq.\,\eqref{eq:inp_prob} are the intrinsic uncertainties in the flagged channels, conditioning on a underlying signal predicted by $\vect{b}$, and the uncertainties in the signal model $\vect{b}$ itself. The two terms can be written as 
\begin{equation}
\begin{aligned}
       &P(\mathbf{v}_\mathrm{inp}' | \mathbf{b}, \mathbf{N}'_\mathrm{f}, \mathbf{A}) \\
       \propto& \exp{\left[-(\mathbf{v}_\mathrm{inp}' - \mathbf{P}_\mathrm{f}\mathbf{A}\mathbf{b})^\dagger \mathbf{N}'^{-1}_\mathrm{f}(\mathbf{v}_\mathrm{inp}' - \mathbf{P}_\mathrm{f}\mathbf{A}\mathbf{b})\right]},  
\end{aligned}
\end{equation}
and assuming a flat prior on $\mathbf{b}$, 
\begin{equation}
\begin{aligned}
    &P(\mathbf{b} | \mathbf{v}_\mathrm{obs}, \mathbf{N}_\mathrm{u}, \mathbf{A}) \propto P(\mathbf{v}_\mathrm{obs} | \mathbf{b}, \mathbf{N}_\mathrm{u}, \mathbf{A})P(\mathbf{b}) \\
    \propto&\exp{\left[-(\mathbf{v}_\mathrm{obs} - \mathbf{A}\mathbf{b})^\dagger \mathbf{N}_\mathrm{u}^{-1}(\mathbf{v}_\mathrm{obs} - \mathbf{A}\mathbf{b})\right]}.
\end{aligned}
\end{equation}
Note that the maximum likelihood estimator $\mathbf{\hat{b}}$ given in Eq.\,\eqref{eq:b_ML} is where $\partial P(\mathbf{b} | \mathbf{v}_\mathrm{obs}, \mathbf{N}_\mathrm{u}, \mathbf{A}) / \partial \mathbf{b} = 0$. We show in Appendix\,\ref{appendix:derivation} that integrating Eq.\,\eqref{eq:inp_prob} gives 
\begin{equation}
\label{eq:inp_likelihood}
\begin{aligned}
    &P(\mathbf{v}_\mathrm{inp}' | \mathbf{v}_\mathrm{obs}, \mathbf{N}, \mathbf{A})\propto \\
    &\exp\left[-(\mathbf{v}_\mathrm{inp}' - \mathbf{P}_\mathrm{f}\mathbf{A}\mathbf{\hat{b}})^\dagger\mathbf{N}'^{-1}_\mathrm{inp}(\mathbf{v}_\mathrm{inp}' - \mathbf{P}_\mathrm{f}\mathbf{A}\mathbf{\hat{b}})\right] 
\end{aligned}
\end{equation}
where
\begin{equation}
    \vect{N}'_\mathrm{inp} \equiv (\mathbf{N}'_\mathrm{f} + \mathcal{O}_\mathrm{inp}' \mathbf{N}_u\mathcal{O}_\mathrm{inp}'^\dagger).
\end{equation}
We therefore see that the uncertainties in the inpainted channels indeed come from two sources: 1) The intrinsic noise uncertainties in these channels described by $\mathbf{N}'_\mathrm{f}$; 2) the posterior uncertainties associated with inferring $\mathbf{v}_\mathrm{inp}'$ from $\mathbf{v}_\mathrm{obs}$. We note that the second term here is exactly the same as the uncertainties on the flagged channels from the frequentist approach. Since each element of $\mathbf{N}'_\mathrm{f}$ is similar to that of $\mathbf{N}_\mathrm{u}$, the combination of the frequentist uncertainties with $\mathbf{N}'_\mathrm{f}$ means that the inpainted data will have larger uncertainties than the unflagged data in our new formalism.

%
%
%
%

To generalize this to a full frequency-frequency covariance for inpainted visibility $\vect{v}_\mathrm{inp}$, we propose the following modification to the frequentist covariance
\begin{equation}
\label{eq:modify_noise_cov}
    \vect{N}_\mathrm{inp} \equiv (\mathbf{N}_\mathrm{f} + \mathcal{O}_\mathrm{inp} \mathbf{N}_u\mathcal{O}_\mathrm{inp}^\dagger).
\end{equation}
This form of the covariance matrix has several desirable qualitative features. It has the full posterior predictive distribution's uncertainty in addition to the correlations between the inpainting solution and any noise affecting the data used for inference. It does not give more weight to the (highly incomplete) model predictions in the flagged channels. In the absence of any flagging it reduces to the standard error propagation procedure, which is to just propagate the thermal noise variances.

We note that in a fully Bayesian treatment, we would propose a model that fully describes the observed data to the best of our ability, and there would be no essential need to consider the relationship between the flagged and unflagged data. One would summarize the unknown values of the model parameters using the posterior distribution conditioning on the observed data and pass this forward to power spectrum estimation \citep{Kennedy2023:Gibbs_Sampling, Burba2024:Bayesian21cm}. For inpainting, however, we choose a model only to smooth out spectral irregularities from RFI flags in order to generate power spectrum under the quadratic estimator framework (see Sec.\,\ref{subsec:inpaint_PS_error}). Our model is deliberately chosen to be incomplete in a very critical way: it has essentially no structure beyond some delay by construction. We therefore do not choose the fully Bayesian approach but explore this alternative approach by framing the process of inpainting into an inference problem and extract the correlation between the inpainting solution and the observed data. In Sec.\,\ref{subsec:inpaint_PS_error}, we will utilize the result derived here to incorporate the effect of inpainting into our power spectrum quadratic estimator.


\begin{figure}[tb]
    \centering
    \epsscale{1.175}
    \plotone{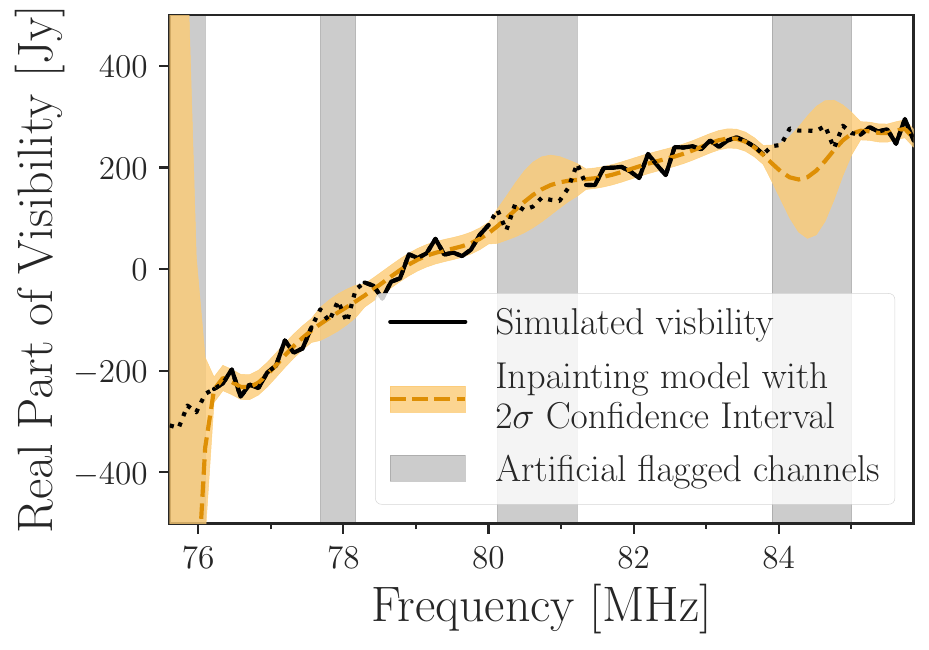}
    \caption{Uncertainties in the inpainted visibilities for various types of flagged channels. The solid black line is the simulated visibility at a single time instance at a single night from a single 14.6 meter baseline. We flag the simulated visibility with two 5-channel wide gaps on the left and two 10-channel wide gaps on the right (shaded gray region). The dashed orange line shows the best-fit inpainted model with the 2$\sigma$ uncertainties given in the shaded region. This can be compared with the dotted black line which represents the underlying true simulated visibility in the flagged region.  \label{fig:inp_err}}
\end{figure}

\autoref{fig:inp_err} shows the performance of our inpainting method across flags with different width at different location of the band. The solid black line is the simulated visibility at a single time instance at a single night from a single 14.6 meter baseline. We flag the simulated visibility with two 5-channel wide gaps on the left and two 10-channel wide gaps on the right. The dashed orange line shows the best-fit inpainted model $\mathbf{A}\mathbf{\hat{b}}$. The uncertainties indicated by the shaded region are given by $\sigma(\nu_i) = \sqrt{(\mathbf{N}_\mathrm{f} + \mathcal{O}_\mathrm{inp}' \mathbf{N}_\mathrm{u}\mathcal{O}'^\dagger_\mathrm{inp})_{ii}/2}$, where the factor of $2$ comes from the fact that we are looking only at the uncertainties in the real part of the visibility. Compared to the true underlying visibility (solid black line), we can clearly see that the inpainting models are less constrained across larger gaps or gaps near the edge of the spectral window. While the best-fit inpainting model can be highly biased in these cases, we stress that the final goal of inpainting is to smoothly connect signal in visibility space to enable us to generate power spectrum free from the bright foreground ringing. Thus, the success of the inpainting method needs to be examined in power spectrum space, which we will discuss next.

\subsection{Impact on Power Spectrum Covariance} \label{subsec:inpaint_PS_error}
While the inpainted visibility might not perfectly recover the true data, the primary goal of inpainting is merely to smoothly connect gaps in visibility space without introducing the bright foreground ringing in the power spectrum estimator. Therefore, the most important quantity to examine is the statistical impact of inpainting on the power spectrum estimator. Here, we focus on the delay power spectra derived under the quadratic estimator (QE, \citealt{Liu2011:QE, Parsons2014:PaperLimit}) formalism. Here, we give a brief summary on the delay power spectra estimator and the evaluation of its statistical properties. 

A per-baseline quadratic estimator $\mathbf{\hat{P}}$ of the $\alpha$-th band-power of the 21cm delay power spectrum can be written as
\begin{equation}
    \hat{P}_\alpha \equiv \mathbf{v}^\dagger \mathbf{E}_\alpha \mathbf{v},
\end{equation}
where $\mathbf{v}$ is the visibility vector, and $\mathbf{E}$ is chosen by the analyst. For simplicity, we consider an estimator that performs a straightforward discrete fourier transform (DFT).
\begin{equation}
    \mathbf{E}_\alpha \equiv M_{\alpha\beta}\mathbf{R}^\dagger \mathbf{Q}^{\mathrm{DFT},\beta} \mathbf{R}
\end{equation}
where $Q_{ij}^{\mathrm{DFT},\beta} = e^{2\pi i \tau_\beta (\nu_i - \nu_j)}$ does the delay transform. Here, we choose $R_{ij} = \gamma(\nu_i)\delta_{ij}$ to be just a tapering matrix, and $M_{\alpha\beta}$ is the normalization factor which will be specified later. 

The expectation value of the estimator is
\begin{equation}
    \langle\hat{P}_\alpha\rangle = \operatorname{tr}\left[\mathbf{E}_\alpha \mathbf{C}\right],
\end{equation}
where $\mathbf{C} \equiv \langle \vect{v}\vect{v}^\dagger\rangle$ is the data covariance matrix. In general, we can decompose the data covariance into
\begin{equation}
    \mathbf{C} = \mathbf{C}_\mathrm{sig} + \mathbf{N},
\end{equation}
where $\mathbf{N}$ is the noise covariance and $\mathbf{C}_\mathrm{sig}$ represents the signal covariance from both the foreground sky emission and the cosmological 21cm signal. Using Eq.\,\eqref{eq:measurement_eq}, the signal covariance can be written as
\begin{equation}
\label{eq:sig_cov}
    \begin{aligned}
        &\left(\mathbf{C}_\mathrm{sig}\right)_{ij} = \langle V_\mathbf{b}(\nu_i)V^*_\mathbf{b}(\nu_j)\rangle \\
        =&\int \dif\eta\dif^2\mathbf{u}\,P(\mathbf{u}, \eta) \times \Big[ \\
        &\quad\Tilde{B}(\mathbf{u}_\mathbf{b} - \mathbf{u}, \nu_i)\Tilde{B}^*(\mathbf{u}_\mathbf{b} - \mathbf{u}, \nu_j)e^{i2\pi\eta(\nu_i - \nu_j)}\Big] \\
        \approx& \int \dif\eta\,P(\mathbf{u}_\mathbf{b}, \eta)e^{i2\pi\eta(\nu_i - \nu_j)}  \times \Big[ \\
        &\quad\int\dif^2\mathbf{u}\,\Tilde{B}(\mathbf{u}_\mathbf{b} - \mathbf{u}, \nu_i)\Tilde{B}^*(\mathbf{u}_\mathbf{b} - \mathbf{u}, \nu_j) \Big]\\
        \approx& \sum_\alpha \bar{P}_\alpha \Delta\eta e^{i2\pi\eta_\alpha(\nu_i - \nu_j)} \int \dif^2\mathbf{\theta} B(\mathbf{\theta}, \nu_i) B^*(\mathbf{\theta}, \nu_j) \\
        \equiv& \sum_\alpha \bar{P}_\alpha \frac{\partial \mathbf{C}_\mathrm{sig}}{\partial P_\alpha},
    \end{aligned}
\end{equation}
where $\eta$ and $\mathbf{u}$ is the Fourier dual of frequency $\nu$ and spatial coordinate $\vect{\theta}$ respectively, $P(\mathbf{u}, \eta)$ is the true signal power spectrum, $\bar{P}_\alpha$ represents the averaged band-power within a narrow bin of $\eta$, and $\Tilde{B}(\mathbf{u}, \nu_i)$ is the spatial Fourier transform of the antenna primary beam profile which we assume to be relatively compact as a function of $\mathbf{u}$. Therefore, 
\begin{equation}
\begin{aligned}
\label{eq:window_function}
        \langle\hat{P}_\alpha\rangle &= \sum_\beta \operatorname{tr}\left[\mathbf{E}_\alpha \frac{\partial \mathbf{C}_\mathrm{sig}}{\partial P_\beta}\right] \bar{P}_\beta + \operatorname{tr}\left[\mathbf{E}_\alpha \mathbf{N}\right] \\
        & \equiv \sum_\beta W_{\alpha\beta} \bar{P}_\beta + b_\alpha,
\end{aligned}
\end{equation}
Here, $W_{\alpha\beta}$, often called the window function \citep{Liu2011:QE, Gorce2023:Window_Function}, represents the connection between the power spectrum estimator and the underlying true power spectrum. $b_\alpha$ is the noise bias in our estimator which can either be subtracted by the analyst or can be avoided by forming an estimator using visibilities from different times (a technique known as $\textit{time interleaving}$, \citealt{Tan2021:HERA_ErrorBar}). The normalization matrix $M_{\alpha\beta}$ can now be chosen such that the window function is power-conserving, \textit{i.e.},
\begin{equation}
    \sum_\beta W_{\alpha\beta} = 1. 
\end{equation}

Meanwhile, the covariance of the power spectrum estimator can be obtained as \citep{Tan2021:HERA_ErrorBar}
\begin{equation}
\begin{aligned}
\label{eq:sigma_N}
    \Sigma_{\alpha\beta} \equiv& \langle \hat{P}_\alpha\hat{P}_\beta^\dagger\rangle  - \langle \hat{P}_\alpha\rangle\langle\hat{P}_\beta^\dagger\rangle\\
    =& \operatorname{tr}\left[
    \mathbf{E}_\alpha \mathbf{C}_\mathrm{sig}\mathbf{E}_{\beta} \mathbf{N} + \mathbf{E}_\alpha \mathbf{N} \mathbf{E}_{\beta}\mathbf{C}_\mathrm{sig}  +
    \mathbf{E}_\alpha \mathbf{N} \mathbf{E}_{\beta} \mathbf{N}
\right],
\end{aligned}
\end{equation}
where we have ignored the contribution from cosmic variance. We note that the power spectrum covariance here includes the signal-noise cross terms that are important in the signal-dominated regime \citep{Kolopanis2019:PaperLimit, Tan2021:HERA_ErrorBar}. 


To include the impact of data inpainting, we can substitute the data covariance matrix to the covariance of the inpainted visibility $\mathbf{v}_\mathrm{inp} = \mathcal{O}_\mathbf{inp}\mathbf{v}_\mathrm{obs}$. Naively, the covariance of the inpainted visibility is therefore
\begin{equation}
\begin{aligned}
        \vect{C}^\mathrm{inp} &= \mathcal{O}_\mathrm{inp}\langle\mathbf{v}_\mathrm{obs}\mathbf{v}_\mathrm{obs}^\dagger\rangle\mathcal{O}_\mathrm{inp}^\dagger \\
        &=\mathcal{O}_\mathrm{inp}\vect{C}_\mathrm{sig}\mathcal{O}_\mathrm{inp}^\dagger + \mathcal{O}_\mathrm{inp}\vect{N}\mathcal{O}_\mathrm{inp}^\dagger 
\end{aligned}
\end{equation}
However, as discussed in Sec.\,\ref{subsec:inpaint_error}, the term $\mathcal{O}_\mathrm{inp}\vect{N}\mathcal{O}_\mathrm{inp}^\dagger$ only captures the uncertainties in the inpainted model due to the noise in the unflagged channels, but does not fully capture the intrinsic uncertainties associated with the flagged channels if we were to observe them. In other words, even if the inpainting model can be determined perfectly, there should still be a term corresponds to the uncertainties due to thermal noise so that inpainting does not artificially increase our sensitivity. Therefore, we model the data covariance after inpainting as 
\begin{equation}
\label{eq:inp_cov}
         \vect{C}^\mathrm{inp} = \mathcal{O}_\mathrm{inp}\vect{C}_\mathrm{sig}\mathcal{O}_\mathrm{inp}^\dagger + \mathcal{O}_\mathrm{inp}\vect{N}_\mathrm{u}\mathcal{O}_\mathrm{inp}^\dagger + \vect{N}_f. 
\end{equation}
Here, we define $\vect{C}_\mathrm{sig}^\mathrm{inp} \equiv \mathcal{O}_\mathrm{inp}\vect{C}_\mathrm{sig}\mathcal{O}_\mathrm{inp}^\dagger$ and $\vect{N}^\mathrm{inp} \equiv \mathcal{O}_\mathrm{inp}\vect{N}_\mathrm{u}\mathcal{O}_\mathrm{inp}^\dagger + \vect{N}_f$. We note that Eq.\,\eqref{eq:inp_cov} denotes the covariance of inpainted visibility at a single time instance and a single night of observation. In practice, we coherently average the inpainted visibility $\vect{v}_\mathrm{inp}^\mathrm{tavg}$ across different sidereal days and within a 300-second window before making the power spectrum. Since the flagging pattern can be different from time-to-time, inpainting is thus a time-varying operation. The covariance for the time-averaged visibility vector is then
\begin{equation}
\label{eq:tavg_cov}
    \begin{aligned}
        \operatorname{Cov}(\mathbf{v}^\mathrm{tavg}_\mathrm{inp}, \mathbf{v}^\mathrm{tavg}_\mathrm{inp}) =& \frac{1}{n^2_\mathrm{night}n^2_{\mathrm{coherent}}}\Big[\sum_{ij} \mathcal{O}_{\mathrm{inp}, i}\vect{C}_\mathrm{sig}\mathcal{O}_{\mathrm{inp}, j}^\dagger \\
        &+ \sum_i \mathcal{O}_{\mathrm{inp}, i}\vect{N}_{\mathrm{u}, i}\mathcal{O}_{\mathrm{inp}, i}^\dagger + \vect{N}_{f, i} \Big],
    \end{aligned}
\end{equation}
where $n_\mathrm{night}$ is the number of nights we observe, $n_{\mathrm{coherent}}$ is the amount of data samples within a $300$-second window, and the index $i$ and $j$ run through all the times and days we average over.

\begin{figure}[tb]
    \centering
    \epsscale{1.175}
    \plotone{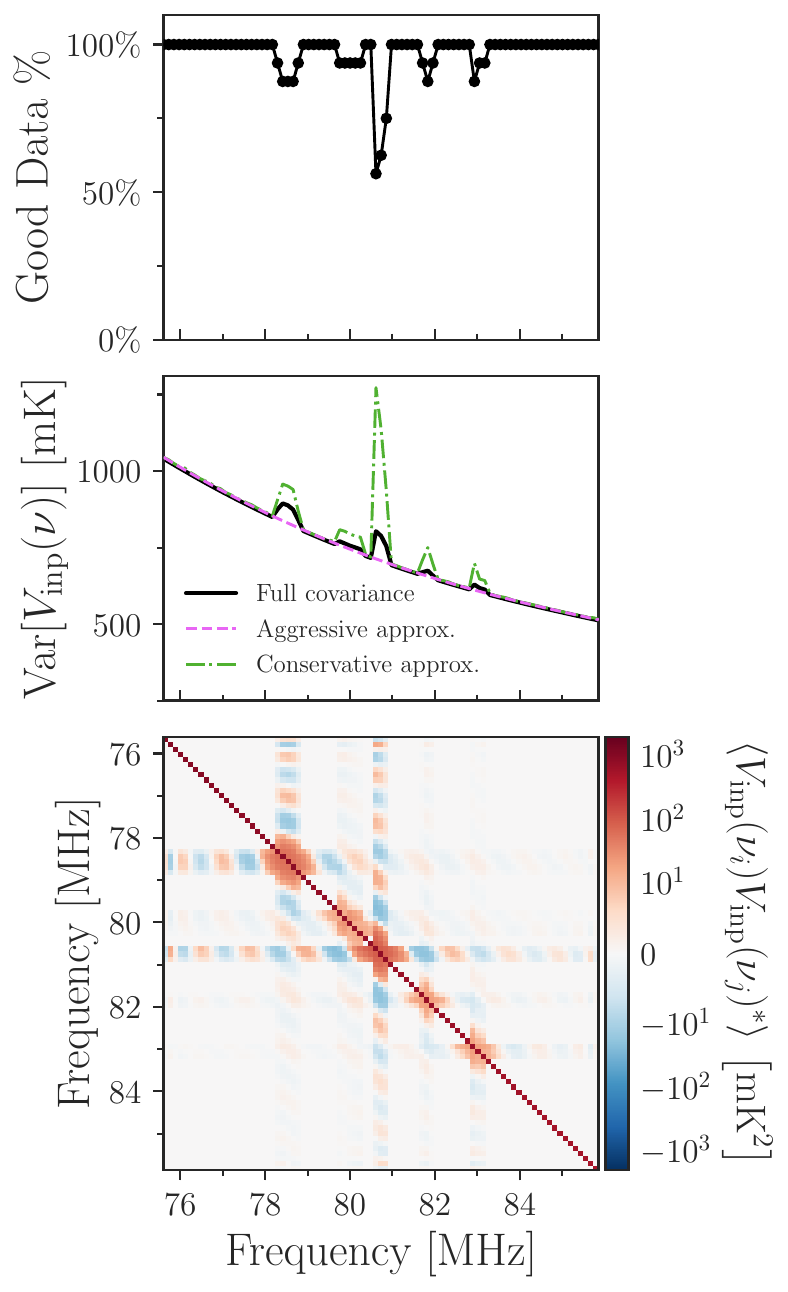}
    \caption{Statistics of inpainted data in the visibility domain in our simulation. \textit{Top}: Overall percentage of good data in the simulation as a function of frequency. The results presented here is for a single 14.6 meter baseline at a single time instance after coherently averaging across sidereal days and a 300-second window. \textit{Center}: Estimates of the noise variance in the inpainted visibility from a full covariance treatment (solid black) and two different approximations (dashed and dash-dotted). \textit{Bottom}: Frequency-frequency noise covariance matrix including the effect of inpainting obtained from the last two terms in Eq.\,\eqref{eq:tavg_cov}. \label{fig:vis_cov}}
\end{figure}

\autoref{fig:vis_cov} shows the noise (co)variance of the inpainted visibility calculated through the last two terms in Eq.\,\eqref{eq:tavg_cov}. Here, the visibility is from a single 14.6 meter baseline in our simulation after coherently averaging across sidereal days and a 300-second window. The percentage of unflagged data as a function of frequency can be seen in the top panel \autoref{fig:vis_cov}. The variance and the covariance of the inpainted visibility are given as the solid black line in the middle panel and the bottom panel respectively. We can see that at the frequency range where there is no flagged channel (i.e., 100\% good data), the noise covariance is completely diagonal and simply follows $\vect{N}_\mathrm{u}$. Meanwhile, inpainting introduces correlation between the flagged channel and the unflagged channel which can be clearly seen in the bottom panel of Fig\,\ref{fig:vis_cov}. 

\begin{figure}[tb]
    \centering
    \epsscale{1.175}
    \plotone{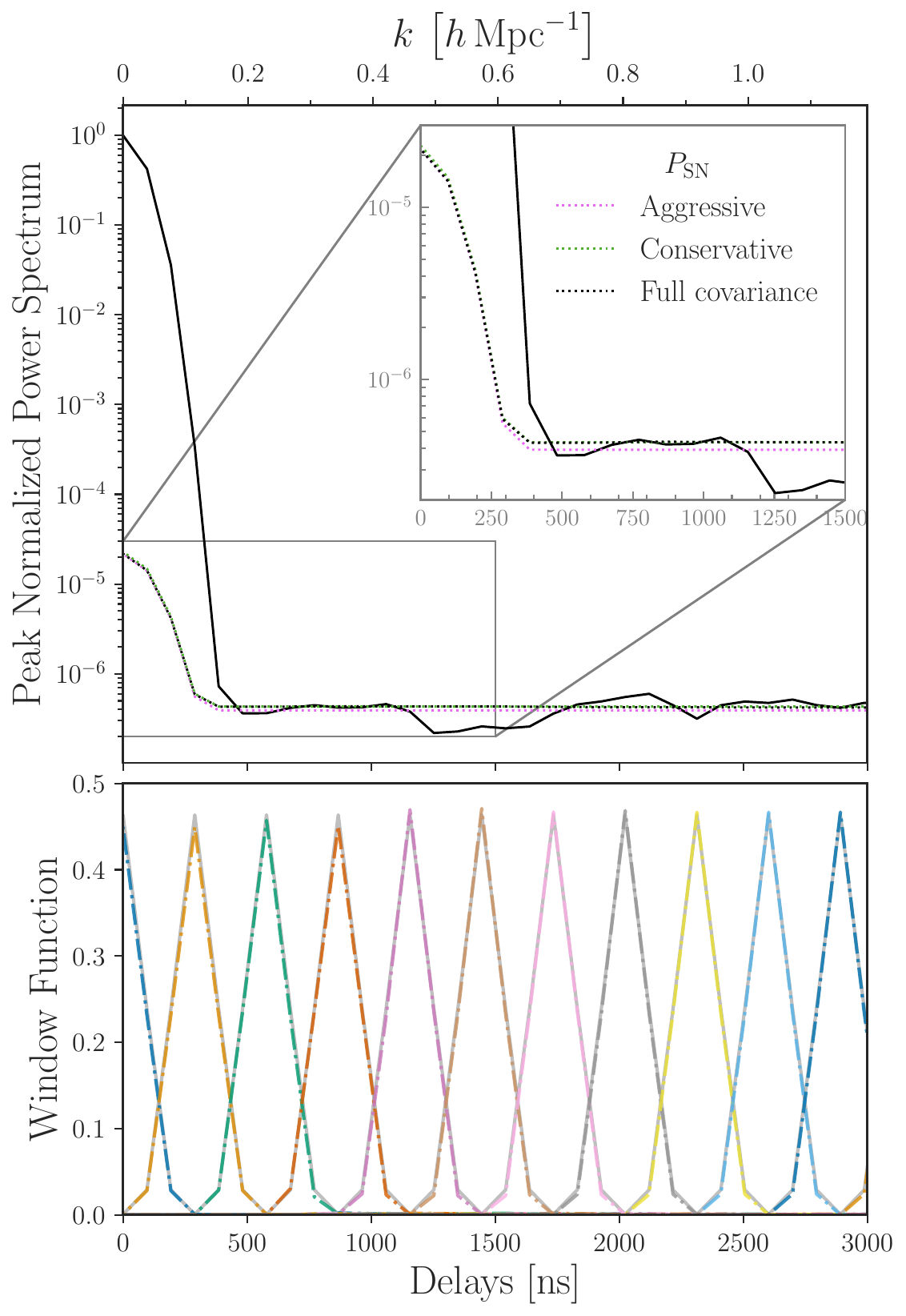}
    \caption{Delay power spectrum and power spectrum window function from an inpainted visibility in our simulation. The delay spectrum is obtained with the inpainted visibility of a single 14.6 meter baseline after coherently averaging across sidereal days and a 300-second window. The delay spectrum is further averaged incoherently within a 1.5-hour window. \textit{Top}: Delay power spectrum (solid black) and error estimates (dotted lines). The three dotted lines correspond to power spectrum error-bars calculated with different inpainted noise covariance matrices as in \autoref{fig:vis_cov}. \textit{Bottom}: Power spectrum window function obtained through Eq.\,\eqref{eq:window_function}. The grey lines show the window functions if we ignore the effect of inpainting while the color dashed lines include the impact of inpainting using Eq.\,\eqref{eq:inp_window_function}. \label{fig:PS_noise}}
\end{figure}

Unfortunately, it might not always be computationally feasible to propagate the full covariance matrix. One simplification one could make is to ignore the off-diagonal correlation and develop an approximation for the variances. For completely unflagged channels, the variances are estimated with the radiometer equation combined with the antenna's auto-correlation
\begin{equation}
\label{eq:noise_estimate}
    \sigma_{ij}^2(\nu) = {\frac{V_{ii} V_{jj}}{N_\mathrm{sample}\Delta\nu\Delta t}},
\end{equation}
where $N_\mathrm{sample}$ traces the amount of data we coherently combined, in this case, $n_\mathrm{night}\times n_{\mathrm{coherent}}$. For channels with flagged data, there are intuitively two ways to assign $N_\mathrm{sample}$. One is to assume we have perfectly predicted the signal in the flagged channels through inpainting, and these flagged channels contain as much information as other unflagged channels. This is the same as assuming the noise covariance is just $\mathbf{N}_\mathrm{u} + \mathbf{N}_\mathrm{f}$, and ignore the uncertainties in inpainting. The variance predicted by this approach is shown as the dashed magenta curve in the middle panel of \autoref{fig:vis_cov}. A more conservative approach is to assume that the flagged channels contain no information at all even after inpainting. This can be achieved by modifying the $N_\mathrm{sample}$ parameter and assume flagged channels do not contribute to $N_\mathrm{sample}$. The variance calculated from this approach is given in dash-dotted green. In reality, inpainted channels do contain some information inferred from the neighbouring channels. We see that the variance calculated by propagating the full covariance matrix lies in between the two approximations. 

\autoref{fig:PS_noise} shows the delay power spectrum from the inpainted visibility and various power spectrum statistics discussed above. The solid black line in the top panel shows the power spectrum for the same inpainted visibility presented in \autoref{fig:vis_cov}. The three dotted lines show the error in the power spectrum estimator obtained by using three different forms of noise covariance for the inpainted visibility in Eq.\,\eqref{eq:sigma_N}. Here, the error power spectrum $P_\mathrm{SN}$ is defined to be the square root of the diagonal terms of the full power spectrum covariance \citep{Tan2021:HERA_ErrorBar}, \textit{i.e.},
\begin{equation}
    P_\mathrm{SN}(k_\alpha) \equiv \sqrt{\Sigma_{\alpha\alpha}}\,,
\end{equation}
where $\Sigma_{\alpha\beta}$ is given in Eq.\,\eqref{eq:sigma_N}. The dotted black line is obtained by using the full noise covariance given by Eq.\,\eqref{eq:tavg_cov}, while the two colored lines are from the two approximations discussed above. We see that while the variance of inpainted visibility in frequency space is bounded by our two approximations as shown in \autoref{fig:vis_cov}, the off-diagonal terms in the full frequency-frequency covariance matrix make the resulting uncertainties in power spectrum space higher than either one of our approximations. In the case where not a significant amount of data \replaced{is}{are} flagged, we see that the conservative approach (dotted green, which assumes \replaced{inpainting data does}{inpainted data do} not contribute to the $N_\mathrm{sample}$ in the noise estimate in Eq.\,\eqref{eq:noise_estimate}) gives a reasonable approximation to the error obtained by propagating the full covariance matrix. Scenarios where this ceases to be true will be discussed in Sec.\,\ref{subsec:hera_inpaint}.

The lower panel of \autoref{fig:PS_noise} shows the window functions for the delay power spectrum estimator at different delay bins. The solid grey lines in the background are the window functions obtained without including the effect of inpainting, \textit{i.e.}, using Eq.\,\eqref{eq:window_function} and \eqref{eq:sig_cov}. To include the effect of inpainting, following Eq.\,\eqref{eq:tavg_cov}, the window functions become
\begin{equation}
\label{eq:inp_window_function}
    W^\mathrm{inp}_{\alpha\beta} = \frac{1}{n^2_\mathrm{night}n^2_{\mathrm{coherent}}}\sum_{ij}\operatorname{tr}\left[\mathbf{E}_\alpha \mathcal{O}_{\mathrm{inp}, i}\frac{\partial \mathbf{C}_\mathrm{sig}}{\partial P_\beta}\mathcal{O}^\dagger_{\mathrm{inp}, j}\right]
\end{equation}
where the index $i$ and $j$ run through all the times and days we average over. The resulting window functions are shown as the colored lines in the lower panel of \autoref{fig:PS_noise}. Here, the difference between including and not including a proper treatment of the effect of inpainting is small as not a large percentage of data \replaced{is}{are} flagged. The effect of inpainting on window functions will be further examined in Sec.\,\ref{subsec:hera_inpaint}.

\begin{figure*}[tbh]
    \centering
    \epsscale{1.175}
    \plotone{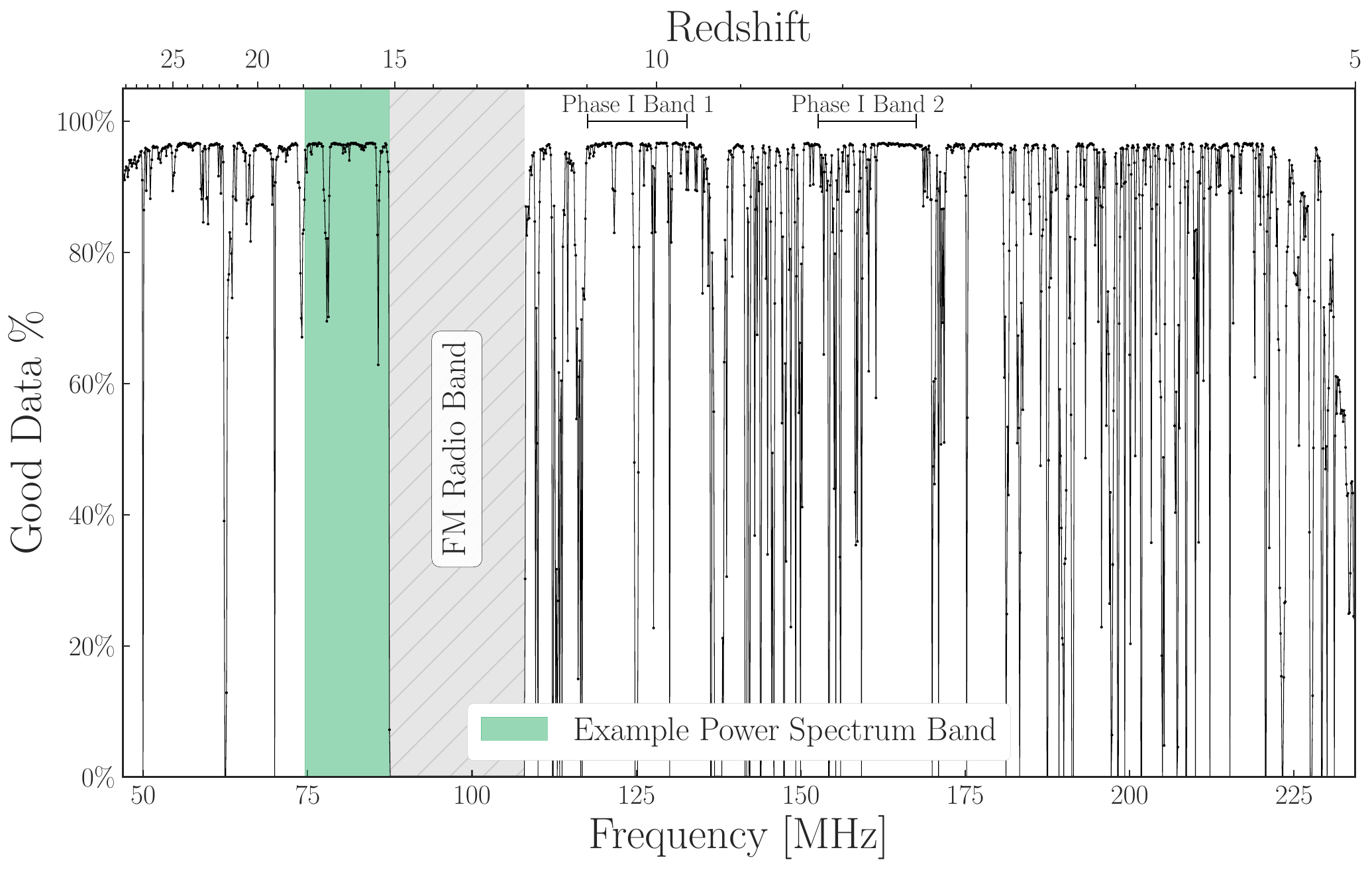}
    \caption{Percentage of good data for the shortest east-west baseline group across 14 nights of observations with the HERA Phase II instruments. The upgraded Vivaldi feed extends our observation range to roughly from 50 to 250 MHz. The grey region is excluded due to heavy contamination from FM radio. A $12$MHz-wide frequency range where we use to derive sample power spectra is marked in green. \label{fig:h6c_flags}}
\end{figure*}

\begin{figure}[tbh]
    \centering
    \epsscale{1.175}
    \plotone{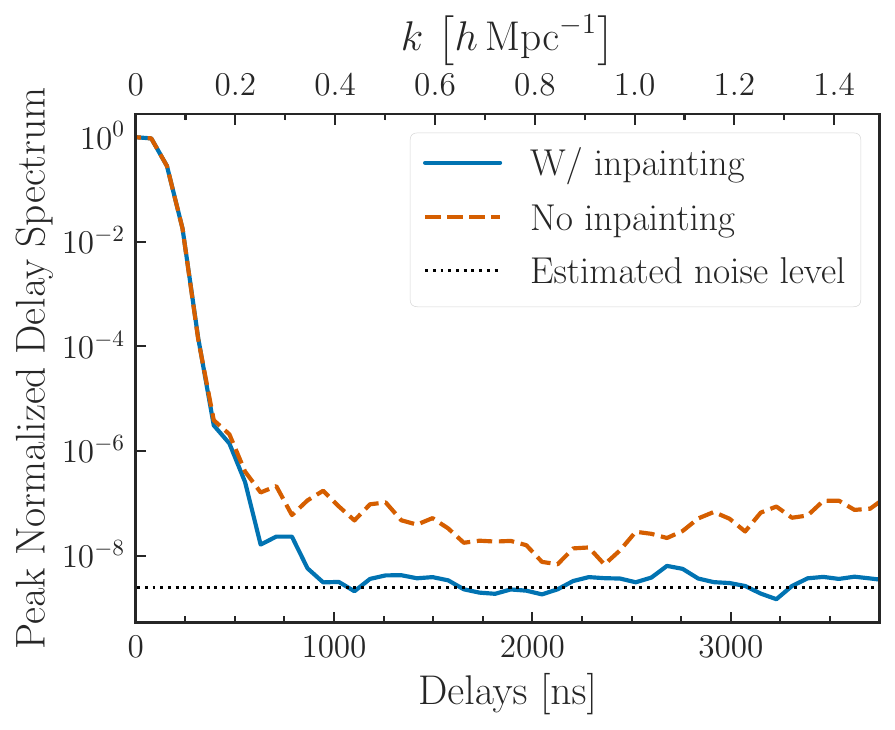}
    \caption{Delay power spectra from the H6C Phase II data with (solid blue) or without (dashed orange) nightly inpainting. The power spectra are derived from 14 nights of data observed by all the 14.6 meter east-west baselines with a 300-second coherent time average and an one-hour incoherent time average. The frequency range where the power spectra is derived and the RFI flagging pattern in the data can be seen in the green band in \autoref{fig:h6c_flags}. The expected sensitivity for \replaced{this data}{these data} obtained through the radiometer equation is given in the dotted line. \label{fig:H6C_PS}}
\end{figure}

\section{Application to HERA} \label{sec:hera}
Equipped with the intuition and statistical tools from the previous sections, in this section we discuss the application to the Phase II data obtained by the Hydrogen Epoch of Reionization Array (HERA). HERA is a drift-scan EoR experiment with 350 parabolic 14-meter dishes located in the Karoo desert in South Africa. HERA Phase II involves an upgrade to the Vivaldi feeds that allows for observations in a wider frequency range from 50-250 MHz. In Sec.\,\ref{subsec:hera_data}, we introduce a small set of HERA Phase II observations taken during October 2022. Detailed strategies for inpainting and results are given in Sec.\,\ref{subsec:hera_inpaint}.

\subsection{Data and Flags} \label{subsec:hera_data}
The first season of scientific observation for HERA Phase II began in October, 2022. A preliminary analysis of the first 14 nights of high-quality observations was performed. While over 170 antennas were constructed at the time, only $\sim90$ were producing good data passing various stringent antenna metrics. 

The data presented here \replaced{has}{have} passed through multiple stages of analyses in order to properly calibrate the data and detect RFI. The calibration strategy adopted here is similar to the HERA Phase I analyses \citep{HERA2022:h1c_idr2_limit, HERA2023:h1c_idr3_limit}. The RFI flagging pipeline has been upgraded in the Phase II analysis to better capture time-varying broad-band RFI that are likely the results of lightning. The details of the RFI identification routine is documented in the HERA Team Memos \citep{H6C_IDR1, H6C_IDR2, H6C_IDR3}, here we give a brief summary of the RFI flagging process.

In general, RFI in the HERA data are identified through a combination of different metrics to isolate the outliers. For the data presented here, two types of flags are created. The first is an antenna flag where we flag bad antenna polarization across the band. These flags are created on a nightly basis, first for every two integration times, and later harmonized so that we continuously flag an antenna across a large period if it is flagged for a significant percentage of times within. The second type of flags is an array-wide RFI mask. These flags are created nightly by identifying misbehaving time and frequency channels in the array-averaged auto-correlation functions and cross-correlation functions. Such a mask can capture low-level RFI as they can only be seen after we average down the data to increase sensitivity. 

\autoref{fig:h6c_flags} shows the amount of good data for the shortest east-west baseline group across 14 nights of observations. For the remainder of this section, we focus on the data below the FM radio band which was previously inaccessible by the Phase I instruments. \replaced{The low-band data offers}{The low-band data offer} a unique opportunity to observe deep into the cosmic dawn but also presents significant challenges due to the brighter foreground contamination. Mitigating any potential leakage of bright foreground modes is therefore even more crucial at this frequency range. 

\begin{figure*}[tbh]
    \centering
    \epsscale{1.175}
    \plotone{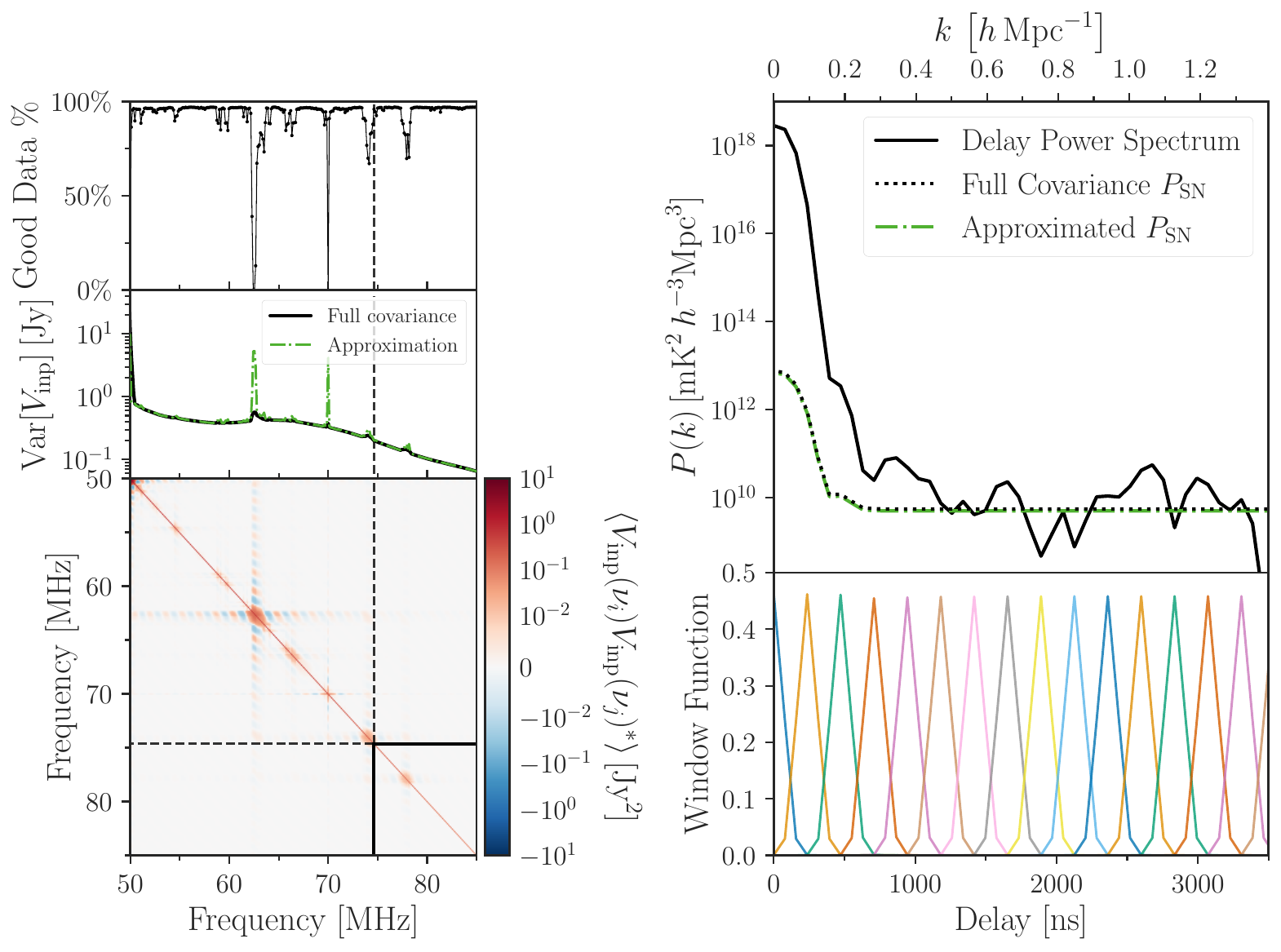}
    \caption{Statistical properties of nightly inpainted visibility from the HERA Phase II data. Similar to \autoref{fig:H6C_PS}, the results shown here are derived from data observed by all the 14.6 meter east-west baselines in the array. The visibility is coherently averaged across 14 nights of observations and a 300-second window. \textit{Left}: Statistical properties in the frequency space including flagging ratio and (co)variance of the inpainted visibility data. \textit{Right}: Statistical properties in the Fourier space including delay power spectrum, power spectrum error estimate, and the window function. \label{fig:H6C_PS_Stats}}
\end{figure*}

\begin{figure}[tbh]
    \centering
    \epsscale{1.175}
    \plotone{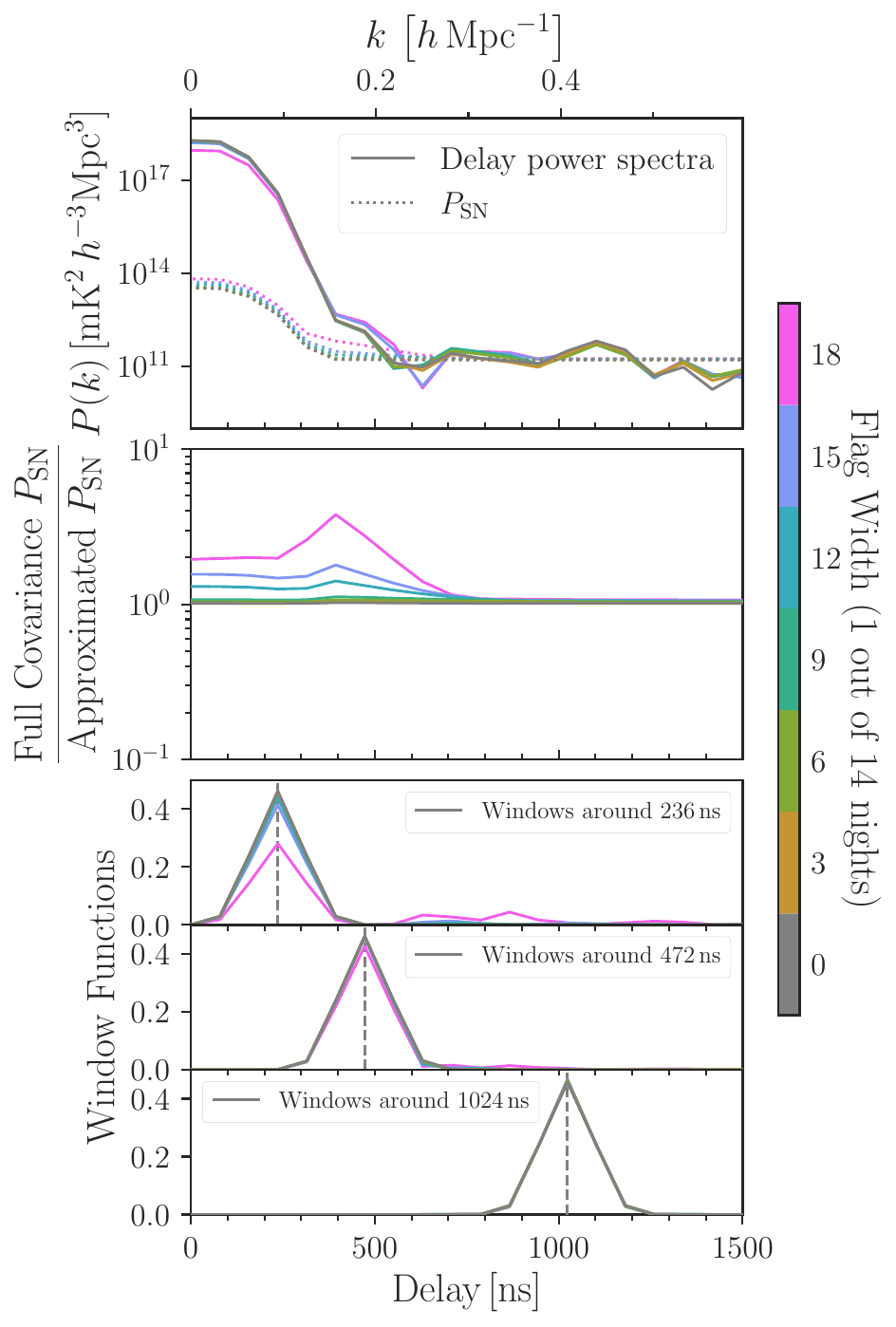}
    \caption{Changes in power spectrum statistics as a function of flagged channel width. Here, we artificially flag channels in 1 of the 14 nights in the observed HERA Phase II data and inpaint over them to investigate the impact of inpainting over gaps with different width. The flagged channels are placed contiguously at the center of the frequency range ($\sim81$MHz) to maximize the effect of inpainting on the power spectrum statistics.  \textit{From top to bottom}: Delay power spectra (solid line) and error estimates (dotted line); Ratio between the power spectrum error estimates derived from the full inpainting covariance versus those derived from the conservative approximation described in Sec.\,\ref{subsec:inpaint_PS_error}; Window functions around different delay bins.  \label{fig:H6C_simulated_gap_1night}}
\end{figure}

\begin{figure}[tbh]
    \centering
    \epsscale{1.175}
    \plotone{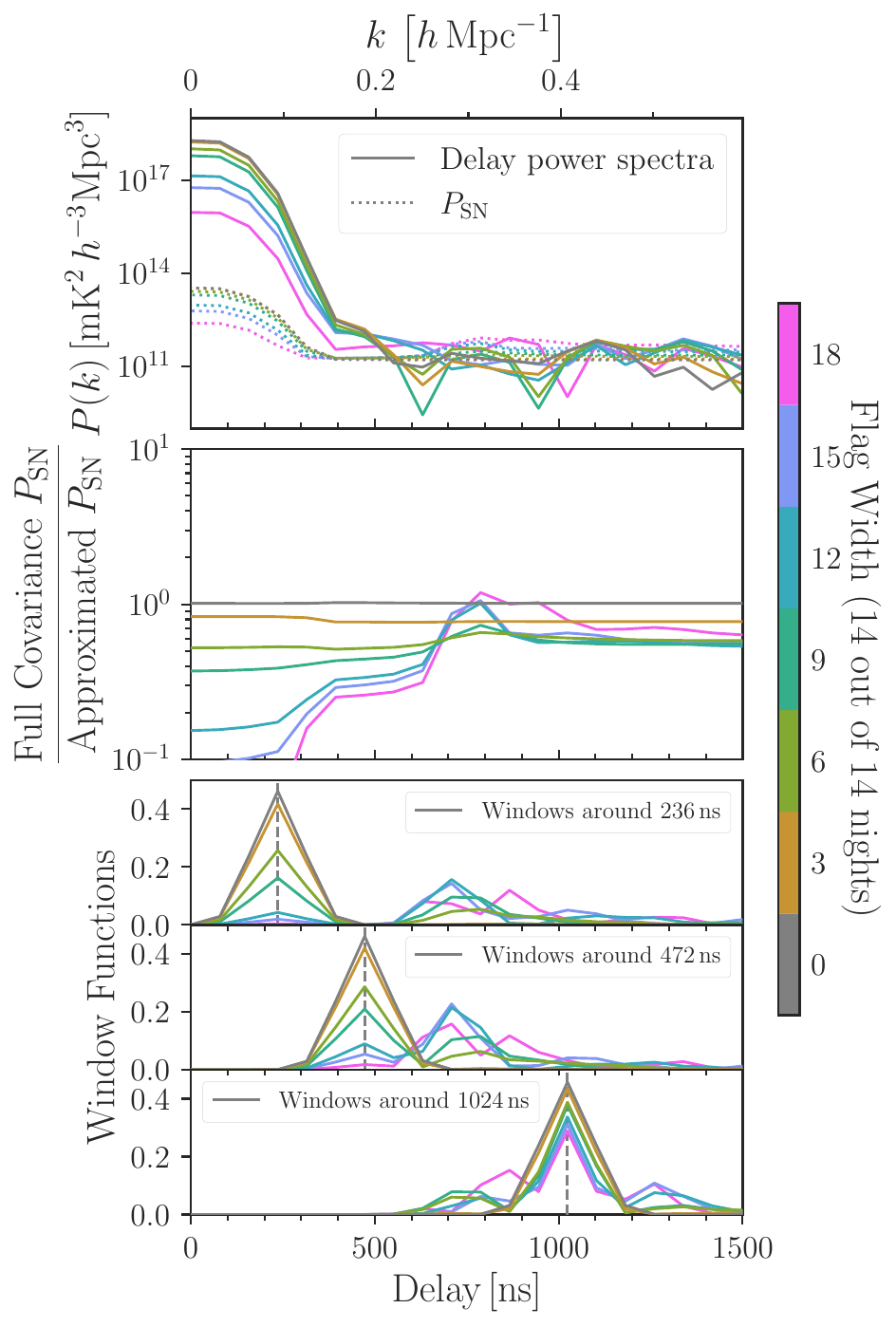}
    \caption{Same as \autoref{fig:H6C_simulated_gap_1night}, but with artificially flagged channels across all 14 nights in the observed HERA Phase II data.  \label{fig:H6C_simulated_gap_14night}}
\end{figure}

\subsection{Inpainting Strategies and Results} \label{subsec:hera_inpaint}

Following the methodology outlined in Sec.\,\ref{subsec:inpaint_method}, \replaced{the HERA data is}{the HERA data are} inpainted with a DPSS basis localized within $\pm500$ nanoseconds in Fourier space. \added{As we discussed in Sec.\,\ref{subsec:inpaint_method}, we choose $500$ nanoseconds so that the DPSS basis not just captures the bright foreground modes within the wedge, but also any potential foreground leakage due to systematic effects. While this does mean that we are inpainting into the EoR window for some baselines, the formalism developed in Sec.\,\ref{sec:inpainting} makes sure that we are able to accurately take into account the impact of inpainting in these delay ranges.} To reduce the uncertainties in inpainting, we fit the DPSS basis to all the data points below the FM radio band \citep{Ewall-Wice2021:DAYNENU, Kern2021:GPR}. Following this inpainting step, we select a $12$MHz frequency window (marked in green in \autoref{fig:h6c_flags}) to examine the power spectrum. 

To demonstrate the necessity of data inpainting, we calculate the delay power spectra within this frequency window with or without inpainting in \autoref{fig:H6C_PS}. The power spectra are derived from 14 nights of data observed by all the 14.6 meter east-west baselines with a 300-second coherent time average and an 1.5-hour incoherent time average. Similar to what we have already seen from the simulation, the delay spectrum from the inpainted visibility (solid blue) agrees well at high delay with the expected sensitivity obtained through the radiometer equation (dotted black). Meanwhile, even though every frequency channel within this band-of-interest has at least 60\% of good data, we see that the resulting power spectrum (dashed orange) from directly averaging the visibilities across sidereal day exhibits significant leakage of bright foreground modes. The result without inpainting is more than one order-of-magnitude away from the expected noise floor. This demonstrates two points: (1) Nightly varying systematic effects exists in the HERA Phase II data and (2) Inpainting over gaps due to RFI on a nightly basis is necessary to reduce the interplay between flags and systematic effects. We note that we also observe deviation from the expected noise level at lower delay ($\leq 1000\mathrm{ns}$) even when inpainting is performed. This is due to other systematic effects such as instrument coupling \citep{Kern2019:Relection_Model, Josaitis2022:MutualCoupling, Rath_Pascua2024:Mutual_Coupling} and can be mitigated with additional analysis steps \citep{Kern2019:Relection_Model, Garsden2024:FRF, Pascua2024:FRF}. These additional systematic mitigation strategies are not employed here for simplicity. 

\autoref{fig:H6C_PS_Stats} shows the power spectrum from the nightly inpainted HERA Phase II data together with all the statistical quantities discussed in Sec.\,\ref{subsec:inpaint_error} and \ref{subsec:inpaint_PS_error}. Similar to \autoref{fig:vis_cov}, the panels on the left-hand side show the properties of the inpainted visibility in frequency space. The lower-left panel shows the frequency-frequency noise covariance matrix after inpainting following Eq.\,\eqref{eq:tavg_cov} while the variance is shown as the solid black line in the middle panel. An approximation to the variance from the full covariance matrix is given in the dash-dotted green line. This is the \textit{conservative} approximation shown in \autoref{fig:vis_cov} as it has been shown to better approximate the power spectrum error bar in \autoref{fig:PS_noise}. We note that for the conservative approximation, we assign $N_\mathrm{sample}$ in Eq.\,\eqref{eq:noise_estimate} using only the number of unflagged channels. Here, for channels that are completely flagged across all nights, we take $N_\mathrm{sample}$ to be the smallest positive number across the band to avoid dividing by zero in Eq.\,\eqref{eq:noise_estimate}.

The right two panels of \autoref{fig:H6C_PS_Stats} show the resulting power spectrum and various power spectrum statistics for the nightly inpainted HERA Phase II data. We note that as we inpaint over a larger frequency range than we derive power spectra. All the power spectrum statistics shown here are derived with the covariance matrix projected to the frequency subspace where we calculate the power spectrum estimator. We see that in the presence of only a small fraction of flagged data, the approximated $P_\mathrm{SN}$ agrees well with $P_\mathrm{SN}$ derived from the full frequency-frequency covariance, and the effect of inpainting on the window function is also negligible. 

However, as we inpaint over more data or over larger gaps, our proposed approximation might not be able to faithfully capture the impact of inpainting on power spectrum statistics. To determine the cases where the effect of inpainting is no longer statistically negligible, we artificially inject flags of different width into the observed data. To maximize the effect of missing data, we position the flag at the center of the frequency band where we derive the power spectrum ($\sim 81$MHz). \autoref{fig:H6C_simulated_gap_1night} and \ref{fig:H6C_simulated_gap_14night} show the resulting changes in power spectrum statistics as a function of flagged channel width. In \autoref{fig:H6C_simulated_gap_1night}, we introduce flags with different width in one out of the 14 nights of data while flags are introduced every night in \autoref{fig:H6C_simulated_gap_14night}. 

The top panel of in \autoref{fig:H6C_simulated_gap_1night} and \ref{fig:H6C_simulated_gap_14night} show the delay power spectra (solid lines) and error estimates (dotted lines) derived from the full covariance of the inpainted visibility. \added{We see that as the width of the injected flag increases, especially when flagging over all 14 nights, the amplitude of the power spectrum decreases at lower delays. This is because of the correlation introduced by inpainting. Inpainting over large gaps causes the low-delay modes to correlate strongly with high-delay modes. The power at lower delays is thus diluted as we are measuring a weighted average of the intrinsic power at low and high delays. It is therefore important to carefully take into account the power spectrum window function when interpreting data with a significant amount of flags.} 

Meanwhile, the ratio between the full-covariance error estimates and the approximation is given in the second panel of \autoref{fig:H6C_simulated_gap_1night} and \autoref{fig:H6C_simulated_gap_14night}. In the case where only one out of the 14 nights is flagged, our approximated error estimates always under-estimate the error. This is because even though the approximation agrees well with the diagonal terms in the full covariance matrix, the off-diagonal terms make the error in the full-covariance approach higher. The two approaches do agree well at high delays and the differences at low delays are within $50\%$ for flags that are narrower than $15$-channel wide. On the other hand, when all 14 nights are flagged, our approximation actually over-estimate the noise level. This is because as $N_\mathrm{sample}$ goes to zero, the estimated variance in the conservative approximation becomes extremely large\footnote{Formally speaking, the estimated variance goes to infinity as $N_\mathrm{sample}$ goes to zero. To regularize this behavior, we use the minimal non-zero value of $N_\mathrm{sample}$ across the frequency range where we inpaint instead of zero to estimate the noise variance at the channels that are completely flagged.}. This can be seen from our approximation for some of the channels near $\sim60$MHz in \autoref{fig:H6C_PS_Stats}. In the full covariance matrix, these completely flagged channels do not have infinite uncertainties due to the sufficient amount of information from neighboring channels where we infer the best-fit inpainting model. 

The bottom three panels of \autoref{fig:H6C_simulated_gap_1night} and \ref{fig:H6C_simulated_gap_14night} show the window functions with the effect of inpainting over these artificial flags. We see that in the case where the flagged channel only appears in one of the night, the effect of inpainting on the window function is negligible especially at high delays. When all 14 nights are flagged, however, we see that a flag that spans more than five channels can introduce significant correlation in Fourier space even at high delays, making it crucial to correctly propagate the effect of inpainting into power spectrum statistics. 

\section{Conclusion} \label{sec:conclusion}
Power spectrum analyses in 21cm cosmology are particularly sensitive to missing data or small discontinuities in the data along the frequency axis. These spectral structures can lead to ringing of bright foreground modes in the Fourier space, contaminating a large area of modes that are, in theory, free of foreground signals. 

In this work, we have identified a new source of discontinuities. This arises from averaging measurements that contain varying systematic effects and flagging patterns. Using a realistic simulation, we have shown (\autoref{fig:PS_compare}) that time-varying flagging patterns can couple bright foreground modes with systematic effects in Fourier space, making systematic effects that are otherwise below the noise level prominent in the 21cm power spectrum. This effect can be seen even when each frequency channel in the averaged visibility contains more than $80\%$ of unflagged data (\autoref{fig:PS_list}). 

We have demonstrated that this discontinuity can be suppressed by inpainting data prior to averaging. However, as we inpaint over more and more data, it is crucial to correctly estimate the uncertainties associated with these methods. In this work, we have chosen to focus on inpainting via the discrete prolate spheroidal sequence (DPSS, \citealt{Slepian1978:DPSS, Ewall-Wice2021:DAYNENU}). Thanks to the linear nature of our inpainting method, we have developed a framework to incorporate the inpainting operation into the power spectrum quadratic estimator. This allows us to accurately quantify the effect of data inpainting on important statistical quantities such as power spectrum error bar and window function (\autoref{fig:PS_noise}). 

We have applied the statistical tools developed in this work to inspect a small set of data obtained by the HERA Phase II instrument. To increase computational feasibility in future analyses, we have proposed and tested methods that can well approximate the power spectrum statistics. By artificially introducing flags into the real data, we have quantified the regime where we can safely adopt our proposed approximation and neglect the effect of inpainting (\autoref{fig:H6C_simulated_gap_1night} and \ref{fig:H6C_simulated_gap_14night}). Our results provide a framework that allows us to carefully study gappy data in the ever more noisy RFI environment as we move toward detecting 21cm power spectrum.

\section*{Acknowledgments}
\added{We thank an anonymous referee for valuable comments and suggestions.} This material is based upon work supported by the National Science Foundation under Grant Nos. 1636646 and 1836019 and institutional support from the HERA collaboration partners. This research is funded by the Gordon and Betty Moore Foundation through grant GBMF5215 to the Massachusetts Institute of Technology. HERA is hosted by the South African Radio Astronomy Observatory, which is a facility of the National Research Foundation, an agency of the Department of Science and Innovation.

K.-F.C. acknowledges support from the Mitacs Globalink Research Award, Taiwan Think Global Education Trust Scholarship, and the Taiwan Ministry of Education's Government Scholarship to Study Abroad. A. Liu acknowledges support from the Trottier Space Institute, the New Frontiers in Research Fund Exploration grant program, the Canadian Institute for Advanced Research (CIFAR) Azrieli Global Scholars program, a Natural Sciences and Engineering Research Council of Canada (NSERC) Discovery Grant and a Discovery Launch Supplement, the Sloan Research Fellowship, and the William Dawson Scholarship at McGill. This result is part of a project that has received funding from the European Research Council (ERC) under the European Union's Horizon 2020 research and innovation programme (Grant agreement No. 948764; MJW, PB, JB).

This paper makes use of publicly accessible software developed for the HERA Collaboration\footnote{\url{https://github.com/Hera-Team}} and software built by both HERA members and collaborators\footnote{\url{https://github.com/RadioAstronomySoftwareGroup}}, especially \texttt{pyuvdata} \citep{Hazelton2017:pyuvdata}. Numerical calculations are performed through the \texttt{Python} packages \texttt{numpy} \citep{numpy} and \texttt{scipy} \citep{scipy}. Many of the cosmological and astrophysical calculations in this work rely on the routines wrapped up in \texttt{colossus} \citep{Colossus} and \texttt{astropy} \citep{astropy}. Plots are made available thanks to \texttt{matplotlib} \citep{matplotlib} and \texttt{seaborn} \citep{seaborn}.

\bibliography{ref}

\begin{thebibliography}{}
\expandafter\ifx\csname natexlab\endcsname\relax\def\natexlab#1{#1}\fi
\providecommand{\url}[1]{\href{#1}{#1}}
\providecommand{\dodoi}[1]{doi:~\href{http://doi.org/#1}{\nolinkurl{#1}}}
\providecommand{\doeprint}[1]{\href{http://ascl.net/#1}{\nolinkurl{http://ascl.net/#1}}}
\providecommand{\doarXiv}[1]{\href{https://arxiv.org/abs/#1}{\nolinkurl{https://arxiv.org/abs/#1}}}

\bibitem[{{Abdurashidova} {et~al.}(2022){Abdurashidova}, {Aguirre},
  {Alexander}, {Ali}, {Balfour}, {Beardsley}, {Bernardi}, {Billings}, {Bowman},
  {Bradley}, {Bull}, {Burba}, {Carey}, {Carilli}, {Cheng}, {DeBoer}, {Dexter},
  {de Lera Acedo}, {Dibblee-Barkman}, {Dillon}, {Ely}, {Ewall-Wice}, {Fagnoni},
  {Fritz}, {Furlanetto}, {Gale-Sides}, {Glendenning}, {Gorthi}, {Greig},
  {Grobbelaar}, {Halday}, {Hazelton}, {Hewitt}, {Hickish}, {Jacobs}, {Julius},
  {Kern}, {Kerrigan}, {Kittiwisit}, {Kohn}, {Kolopanis}, {Lanman}, {La Plante},
  {Lekalake}, {Lewis}, {Liu}, {MacMahon}, {Malan}, {Malgas}, {Maree},
  {Martinot}, {Matsetela}, {Mesinger}, {Molewa}, {Morales}, {Mosiane},
  {Murray}, {Neben}, {Nikolic}, {Nunhokee}, {Parsons}, {Patra}, {Pascua},
  {Pieterse}, {Pober}, {Razavi-Ghods}, {Ringuette}, {Robnett}, {Rosie}, {Sims},
  {Singh}, {Smith}, {Syce}, {Thyagarajan}, {Williams}, {Zheng}, \& {HERA
  Collaboration}}]{HERA2022:h1c_idr2_limit}
{Abdurashidova}, Z., {Aguirre}, J.~E., {Alexander}, P., {et~al.} 2022, \apj,
  925, 221, \dodoi{10.3847/1538-4357/ac1c78}

\bibitem[{{Abrial} {et~al.}(2008){Abrial}, {Moudden}, {Starck}, {Fadili},
  {Delabrouille}, \& {Nguyen}}]{Abrial2008:CMB_Inpainting}
{Abrial}, P., {Moudden}, Y., {Starck}, J.~L., {et~al.} 2008, Statistical
  Methodology, 5, 289, \dodoi{10.1016/j.stamet.2007.11.005}

\bibitem[{{Amiri} {et~al.}(2023){Amiri}, {Bandura}, {Chen}, {Deng}, {Dobbs},
  {Fandino}, {Foreman}, {Halpern}, {Hill}, {Hinshaw}, {H{\"o}fer}, {Kania},
  {Landecker}, {MacEachern}, {Masui}, {Mena-Parra}, {Milutinovic},
  {Mirhosseini}, {Newburgh}, {Ordog}, {Pen}, {Pinsonneault-Marotte}, {Polzin},
  {Reda}, {Renard}, {Shaw}, {Siegel}, {Singh}, {Vanderlinde}, {Wang}, {Wiebe},
  {Wulf}, \& {CHIME Collaboration}}]{CHIME2023:21cm_x_QSO_ELG}
{Amiri}, M., {Bandura}, K., {Chen}, T., {et~al.} 2023, \apj, 947, 16,
  \dodoi{10.3847/1538-4357/acb13f}

\bibitem[{{Ansah-Narh} {et~al.}(2018){Ansah-Narh}, {Abdalla}, {Smirnov},
  {Asad}, \& {Shaw}}]{Ansah-Narh2018:BeamPerturbation}
{Ansah-Narh}, T., {Abdalla}, F.~B., {Smirnov}, O.~M., {Asad}, K.~M.~B., \&
  {Shaw}, J.~R. 2018, \mnras, 481, 2694, \dodoi{10.1093/mnras/sty2433}

\bibitem[{{Armel Mbou Sob} {et~al.}(2019){Armel Mbou Sob}, {Landman Bester},
  {Smirnov}, {Kenyon}, \& {Grobler}}]{Armel_Mbou_Sob2019:Calibration}
{Armel Mbou Sob}, U., {Landman Bester}, H., {Smirnov}, O., {Kenyon}, J., \&
  {Grobler}, T. 2019, arXiv e-prints, arXiv:1910.08136,
  \dodoi{10.48550/arXiv.1910.08136}

\bibitem[{{Astropy Collaboration} {et~al.}(2022){Astropy Collaboration},
  {Price-Whelan}, {Lim}, {Earl}, {Starkman}, {Bradley}, {Shupe}, {Patil},
  {Corrales}, {Brasseur}, {N{\"o}the}, {Donath}, {Tollerud}, {Morris},
  {Ginsburg}, {Vaher}, {Weaver}, {Tocknell}, {Jamieson}, {van Kerkwijk},
  {Robitaille}, {Merry}, {Bachetti}, {G{\"u}nther}, {Aldcroft},
  {Alvarado-Montes}, {Archibald}, {B{\'o}di}, {Bapat}, {Barentsen},
  {Baz{\'a}n}, {Biswas}, {Boquien}, {Burke}, {Cara}, {Cara}, {Conroy},
  {Conseil}, {Craig}, {Cross}, {Cruz}, {D'Eugenio}, {Dencheva}, {Devillepoix},
  {Dietrich}, {Eigenbrot}, {Erben}, {Ferreira}, {Foreman-Mackey}, {Fox},
  {Freij}, {Garg}, {Geda}, {Glattly}, {Gondhalekar}, {Gordon}, {Grant},
  {Greenfield}, {Groener}, {Guest}, {Gurovich}, {Handberg}, {Hart},
  {Hatfield-Dodds}, {Homeier}, {Hosseinzadeh}, {Jenness}, {Jones}, {Joseph},
  {Kalmbach}, {Karamehmetoglu}, {Ka{\l}uszy{\'n}ski}, {Kelley}, {Kern},
  {Kerzendorf}, {Koch}, {Kulumani}, {Lee}, {Ly}, {Ma}, {MacBride}, {Maljaars},
  {Muna}, {Murphy}, {Norman}, {O'Steen}, {Oman}, {Pacifici}, {Pascual},
  {Pascual-Granado}, {Patil}, {Perren}, {Pickering}, {Rastogi}, {Roulston},
  {Ryan}, {Rykoff}, {Sabater}, {Sakurikar}, {Salgado}, {Sanghi}, {Saunders},
  {Savchenko}, {Schwardt}, {Seifert-Eckert}, {Shih}, {Jain}, {Shukla}, {Sick},
  {Simpson}, {Singanamalla}, {Singer}, {Singhal}, {Sinha}, {Sip{\H{o}}cz},
  {Spitler}, {Stansby}, {Streicher}, {{\v{S}}umak}, {Swinbank}, {Taranu},
  {Tewary}, {Tremblay}, {de Val-Borro}, {Van Kooten}, {Vasovi{\'c}}, {Verma},
  {de Miranda Cardoso}, {Williams}, {Wilson}, {Winkel}, {Wood-Vasey}, {Xue},
  {Yoachim}, {Zhang}, {Zonca}, \& {Astropy Project Contributors}}]{astropy}
{Astropy Collaboration}, {Price-Whelan}, A.~M., {Lim}, P.~L., {et~al.} 2022,
  \apj, 935, 167, \dodoi{10.3847/1538-4357/ac7c74}

\bibitem[{{Barry} {et~al.}(2016){Barry}, {Hazelton}, {Sullivan}, {Morales}, \&
  {Pober}}]{Barry2016:CalibrationError}
{Barry}, N., {Hazelton}, B., {Sullivan}, I., {Morales}, M.~F., \& {Pober},
  J.~C. 2016, \mnras, 461, 3135, \dodoi{10.1093/mnras/stw1380}

\bibitem[{{Barry} {et~al.}(2019){Barry}, {Wilensky}, {Trott}, {Pindor},
  {Beardsley}, {Hazelton}, {Sullivan}, {Morales}, {Pober}, {Line}, {Greig},
  {Byrne}, {Lanman}, {Li}, {Jordan}, {Joseph}, {McKinley}, {Rahimi},
  {Yoshiura}, {Bowman}, {Gaensler}, {Hewitt}, {Jacobs}, {Mitchell}, {Udaya
  Shankar}, {Sethi}, {Subrahmanyan}, {Tingay}, {Webster}, \&
  {Wyithe}}]{Barry2019:MWA_Limit}
{Barry}, N., {Wilensky}, M., {Trott}, C.~M., {et~al.} 2019, \apj, 884, 1,
  \dodoi{10.3847/1538-4357/ab40a8}

\bibitem[{{Beardsley} {et~al.}(2016){Beardsley}, {Hazelton}, {Sullivan},
  {Carroll}, {Barry}, {Rahimi}, {Pindor}, {Trott}, {Line}, {Jacobs}, {Morales},
  {Pober}, {Bernardi}, {Bowman}, {Busch}, {Briggs}, {Cappallo}, {Corey}, {de
  Oliveira-Costa}, {Dillon}, {Emrich}, {Ewall-Wice}, {Feng}, {Gaensler},
  {Goeke}, {Greenhill}, {Hewitt}, {Hurley-Walker}, {Johnston-Hollitt},
  {Kaplan}, {Kasper}, {Kim}, {Kratzenberg}, {Lenc}, {Loeb}, {Lonsdale},
  {Lynch}, {McKinley}, {McWhirter}, {Mitchell}, {Morgan}, {Neben},
  {Thyagarajan}, {Oberoi}, {Offringa}, {Ord}, {Paul}, {Prabu}, {Procopio},
  {Riding}, {Rogers}, {Roshi}, {Udaya Shankar}, {Sethi}, {Srivani},
  {Subrahmanyan}, {Tegmark}, {Tingay}, {Waterson}, {Wayth}, {Webster},
  {Whitney}, {Williams}, {Williams}, {Wu}, \&
  {Wyithe}}]{Beardsley2016:MWA_Limit}
{Beardsley}, A.~P., {Hazelton}, B.~J., {Sullivan}, I.~S., {et~al.} 2016, \apj,
  833, 102, \dodoi{10.3847/1538-4357/833/1/102}

\bibitem[{{Berkhout} {et~al.}(2024){Berkhout}, {Jacobs}, {Abdurashidova},
  {Adams}, {Aguirre}, {Alexander}, {Ali}, {Baartman}, {Balfour}, {Beardsley},
  {Bernardi}, {Billings}, {Bowman}, {Bradley}, {Bull}, {Burba}, {Carey},
  {Carilli}, {Chen}, {Cheng}, {Choudhuri}, {DeBoer}, {de Lera Acedo}, {Dexter},
  {Dillon}, {Dynes}, {Eksteen}, {Ely}, {Ewall-Wice}, {Fagnoni}, {Fritz},
  {Furlanetto}, {Gale-Sides}, {Garsden}, {Gehlot}, {Ghosh}, {Glendenning},
  {Gorce}, {Gorthi}, {Greig}, {Grobbelaar}, {Halday}, {Hazelton}, {Hewitt},
  {Hickish}, {Huang}, {Josaitis}, {Julius}, {Kariseb}, {Kern}, {Kerrigan},
  {Kim}, {Kittiwisit}, {Kohn}, {Kolopanis}, {Lanman}, {La Plante}, {Liu},
  {Loots}, {Ma}, {MacMahon}, {Malan}, {Malgas}, {Malgas}, {Marero}, {Martinot},
  {Mesinger}, {Molewa}, {Morales}, {Mosiane}, {Murray}, {Neben}, {Nikolic},
  {Devi Nunhokee}, {Nuwegeld}, {Parsons}, {Pascua}, {Patra}, {Pieterse}, {Qin},
  {Rath}, {Razavi-Ghods}, {Riley}, {Robnett}, {Rosie}, {Santos}, {Sims},
  {Singh}, {Storer}, {Swarts}, {Tan}, {Thyagarajan}, {van Wyngaarden},
  {Williams}, {Zheng}, \& {Xu}}]{Berkhout2024:HERA_PhaseII}
{Berkhout}, L.~M., {Jacobs}, D.~C., {Abdurashidova}, Z., {et~al.} 2024, arXiv
  e-prints, arXiv:2401.04304, \dodoi{10.48550/arXiv.2401.04304}

\bibitem[{{Bharadwaj} {et~al.}(2019){Bharadwaj}, {Pal}, {Choudhuri}, \&
  {Dutta}}]{Bharadwaj2019:MAPS}
{Bharadwaj}, S., {Pal}, S., {Choudhuri}, S., \& {Dutta}, P. 2019, \mnras, 483,
  5694, \dodoi{10.1093/mnras/sty3501}

\bibitem[{{Bowman} \& {Rogers}(2010)}]{Bowman2010:EDGES_RFI}
{Bowman}, J., \& {Rogers}, A.~E.~E. 2010, in RFI Mitigation Workshop, 30,
  \dodoi{10.22323/1.107.0030}

\bibitem[{{Bull} {et~al.}(2015){Bull}, {Ferreira}, {Patel}, \&
  {Santos}}]{Bull2015:21cmIM}
{Bull}, P., {Ferreira}, P.~G., {Patel}, P., \& {Santos}, M.~G. 2015, \apj, 803,
  21, \dodoi{10.1088/0004-637X/803/1/21}

\bibitem[{{Burba} {et~al.}(2024){Burba}, {Bull}, {Wilensky}, {Kennedy},
  {Garsden}, \& {Glasscock}}]{Burba2024:Bayesian21cm}
{Burba}, J., {Bull}, P., {Wilensky}, M.~J., {et~al.} 2024, arXiv e-prints,
  arXiv:2403.13767, \dodoi{10.48550/arXiv.2403.13767}

\bibitem[{{Byrne}(2023)}]{Byrne2023:Calibration}
{Byrne}, R. 2023, \apj, 943, 117, \dodoi{10.3847/1538-4357/acac95}

\bibitem[{{Byrne} {et~al.}(2019){Byrne}, {Morales}, {Hazelton}, {Li}, {Barry},
  {Beardsley}, {Joseph}, {Pober}, {Sullivan}, \&
  {Trott}}]{Bryne2019:CalibrationError}
{Byrne}, R., {Morales}, M.~F., {Hazelton}, B., {et~al.} 2019, \apj, 875, 70,
  \dodoi{10.3847/1538-4357/ab107d}

\bibitem[{{Chakraborty} {et~al.}(2022){Chakraborty}, {Datta}, \&
  {Mazumder}}]{Chakraborty2022:Inpainting}
{Chakraborty}, A., {Datta}, A., \& {Mazumder}, A. 2022, \apj, 929, 104,
  \dodoi{10.3847/1538-4357/ac5cc5}

\bibitem[{{Chakraborty} {et~al.}(2021){Chakraborty}, {Datta}, {Roy},
  {Bharadwaj}, {Choudhury}, {Datta}, {Pal}, {Choudhury}, {Choudhuri}, {Dutta},
  \& {Sarkar}}]{Chakraborty2021:uGMRT_Limit}
{Chakraborty}, A., {Datta}, A., {Roy}, N., {et~al.} 2021, \apjl, 907, L7,
  \dodoi{10.3847/2041-8213/abd17a}

\bibitem[{{Chang} {et~al.}(2008){Chang}, {Pen}, {Peterson}, \&
  {McDonald}}]{Chang2008:BAO_IM}
{Chang}, T.-C., {Pen}, U.-L., {Peterson}, J.~B., \& {McDonald}, P. 2008, \prl,
  100, 091303, \dodoi{10.1103/PhysRevLett.100.091303}

\bibitem[{{CHIME Collaboration} {et~al.}(2022){CHIME Collaboration}, {Amiri},
  {Bandura}, {Boskovic}, {Chen}, {Cliche}, {Deng}, {Denman}, {Dobbs},
  {Fandino}, {Foreman}, {Halpern}, {Hanna}, {Hill}, {Hinshaw}, {H{\"o}fer},
  {Kania}, {Klages}, {Landecker}, {MacEachern}, {Masui}, {Mena-Parra},
  {Milutinovic}, {Mirhosseini}, {Newburgh}, {Nitsche}, {Ordog}, {Pen},
  {Pinsonneault-Marotte}, {Polzin}, {Reda}, {Renard}, {Shaw}, {Siegel},
  {Singh}, {Smegal}, {Tretyakov}, {van Gassen}, {Vanderlinde}, {Wang}, {Wiebe},
  {Willis}, \& {Wulf}}]{CHIME2022:Overview}
{CHIME Collaboration}, {Amiri}, M., {Bandura}, K., {et~al.} 2022, \apjs, 261,
  29, \dodoi{10.3847/1538-4365/ac6fd9}

\bibitem[{{CHIME Collaboration} {et~al.}(2023){CHIME Collaboration}, {Amiri},
  {Bandura}, {Chakraborty}, {Dobbs}, {Fandino}, {Foreman}, {Gan}, {Halpern},
  {Hill}, {Hinshaw}, {H{\"o}fer}, {Landecker}, {Li}, {MacEachern}, {Masui},
  {Mena-Parra}, {Milutinovic}, {Mirhosseini}, {Newburgh}, {Ordog}, {Paul},
  {Pen}, {Pinsonneault-Marotte}, {Reda}, {Shaw}, {Siegel}, {Vanderlinde},
  {Wang}, {Wiebe}, \& {Wulf}}]{CHIME2023:21cm_x_Lya}
---. 2023, arXiv e-prints, arXiv:2309.04404, \dodoi{10.48550/arXiv.2309.04404}

\bibitem[{{Choudhuri} {et~al.}(2021){Choudhuri}, {Bull}, \&
  {Garsden}}]{Choudhuri2021:BeamVariation}
{Choudhuri}, S., {Bull}, P., \& {Garsden}, H. 2021, \mnras, 506, 2066,
  \dodoi{10.1093/mnras/stab1795}

\bibitem[{{Cox} {et~al.}(2023){Cox}, {Parsons}, {Dillon}, {Ewall-Wice}, \&
  {Pascua}}]{Cox2023:nucal}
{Cox}, T.~A., {Parsons}, A.~R., {Dillon}, J.~S., {Ewall-Wice}, A., \& {Pascua},
  R. 2023, arXiv e-prints, arXiv:2311.01422, \dodoi{10.48550/arXiv.2311.01422}

\bibitem[{{Crichton} {et~al.}(2022){Crichton}, {Aich}, {Amara}, {Bandura},
  {Bassett}, {Bengaly}, {Berner}, {Bhatporia}, {Bucher}, {Chang}, {Chiang},
  {Cliche}, {Crichton}, {Dave}, {De Villiers}, {Dobbs}, {Ewall-Wice}, {Eyono},
  {Finlay}, {Gaddam}, {Ganga}, {Gayley}, {Gerodias}, {Gibbon}, {Gumba},
  {Gupta}, {Harris}, {Heilgendorff}, {Hilton}, {Hincks}, {Hitz}, {Jalilvand},
  {Julie}, {Kader}, {Kania}, {Karagiannis}, {Karastergiou}, {Kesebonye},
  {Kittiwisit}, {Kneib}, {Knowles}, {Kuhn}, {Kunz}, {Maartens}, {MacKay},
  {MacPherson}, {Monstein}, {Moodley}, {Mugundhan}, {Naidoo}, {Naidu},
  {Newburgh}, {Nistane}, {Di Nitto}, {{\"O}l{\c{c}}ek}, {Pan}, {Paul},
  {Peterson}, {Pieters}, {Pieterse}, {Pillay}, {Polish}, {Randrianjanahary},
  {Refregier}, {Renard}, {Retana-Montenegro}, {Rout}, {Russeeawon}, {Sadr},
  {Saliwanchik}, {Sampath}, {Sanghavi}, {Santos}, {Sengate}, {Shaw}, {Sievers},
  {Smirnov}, {Smith}, {Sob}, {Srianand}, {Stronkhorst}, {Sunder},
  {Tartakovsky}, {Taylor}, {Timbie}, {Tolley}, {Townsend}, {Tyndall},
  {Ungerer}, {van Dyk}, {van Vuuren}, {Vanderlinde}, {Viant}, {Walters},
  {Wang}, {Weltman}, {Woudt}, {Wulf}, {Zavyalov}, \&
  {Zhang}}]{HIRAX2022:Overview}
{Crichton}, D., {Aich}, M., {Amara}, A., {et~al.} 2022, Journal of Astronomical
  Telescopes, Instruments, and Systems, 8, 011019,
  \dodoi{10.1117/1.JATIS.8.1.011019}

\bibitem[{{Datta} {et~al.}(2010){Datta}, {Bowman}, \&
  {Carilli}}]{Datta2010:FGwedge}
{Datta}, A., {Bowman}, J.~D., \& {Carilli}, C.~L. 2010, \apj, 724, 526,
  \dodoi{10.1088/0004-637X/724/1/526}

\bibitem[{{de Oliveira-Costa} \&
  {Tegmark}(2006)}]{de_Oliveira-Costa2006:CMB_GLS}
{de Oliveira-Costa}, A., \& {Tegmark}, M. 2006, \prd, 74, 023005,
  \dodoi{10.1103/PhysRevD.74.023005}

\bibitem[{{de Oliveira-Costa} {et~al.}(2008){de Oliveira-Costa}, {Tegmark},
  {Gaensler}, {Jonas}, {Landecker}, \& {Reich}}]{de_Oliveira-Costa2008:GSM}
{de Oliveira-Costa}, A., {Tegmark}, M., {Gaensler}, B.~M., {et~al.} 2008,
  \mnras, 388, 247, \dodoi{10.1111/j.1365-2966.2008.13376.x}

\bibitem[{{DeBoer} {et~al.}(2017){DeBoer}, {Parsons}, {Aguirre}, {Alexander},
  {Ali}, {Beardsley}, {Bernardi}, {Bowman}, {Bradley}, {Carilli}, {Cheng}, {de
  Lera Acedo}, {Dillon}, {Ewall-Wice}, {Fadana}, {Fagnoni}, {Fritz},
  {Furlanetto}, {Glendenning}, {Greig}, {Grobbelaar}, {Hazelton}, {Hewitt},
  {Hickish}, {Jacobs}, {Julius}, {Kariseb}, {Kohn}, {Lekalake}, {Liu}, {Loots},
  {MacMahon}, {Malan}, {Malgas}, {Maree}, {Martinot}, {Mathison}, {Matsetela},
  {Mesinger}, {Morales}, {Neben}, {Patra}, {Pieterse}, {Pober}, {Razavi-Ghods},
  {Ringuette}, {Robnett}, {Rosie}, {Sell}, {Smith}, {Syce}, {Tegmark},
  {Thyagarajan}, {Williams}, \& {Zheng}}]{HERA2017:PhaseI_Overview}
{DeBoer}, D.~R., {Parsons}, A.~R., {Aguirre}, J.~E., {et~al.} 2017, \pasp, 129,
  045001, \dodoi{10.1088/1538-3873/129/974/045001}

\bibitem[{{Di Vruno} {et~al.}(2023){Di Vruno}, {Winkel}, {Bassa}, {J{\'o}zsa},
  {Brentjens}, {Jessner}, \& {Garrington}}]{Di_Vruno2023:Starlink_Lofar}
{Di Vruno}, F., {Winkel}, B., {Bassa}, C.~G., {et~al.} 2023, \aap, 676, A75,
  \dodoi{10.1051/0004-6361/202346374}

\bibitem[{{Diemer}(2018)}]{Colossus}
{Diemer}, B. 2018, \apjs, 239, 35, \dodoi{10.3847/1538-4365/aaee8c}

\bibitem[{{Dillon} \& {Murray}(2023)}]{H6C_IDR2}
{Dillon}, J., \& {Murray}, S. 2023, {H6C Internal Data Release 2.2}, Tech.
  rep., HERA Analysis Team

\bibitem[{{Dillon} {et~al.}(2024){Dillon}, {Murray}, {Cox}, \&
  {Martinot}}]{H6C_IDR3}
{Dillon}, J., {Murray}, S., {Cox}, T.~A., \& {Martinot}, Z.~E. 2024, {H6C
  Internal Data Release 2.3}, Tech. rep., HERA Analysis Team

\bibitem[{{Dillon} {et~al.}(2014){Dillon}, {Liu}, {Williams}, {Hewitt},
  {Tegmark}, {Morgan}, {Levine}, {Morales}, {Tingay}, {Bernardi}, {Bowman},
  {Briggs}, {Cappallo}, {Emrich}, {Mitchell}, {Oberoi}, {Prabu}, {Wayth}, \&
  {Webster}}]{Dillon2014:MWA_Limit}
{Dillon}, J.~S., {Liu}, A., {Williams}, C.~L., {et~al.} 2014, \prd, 89, 023002,
  \dodoi{10.1103/PhysRevD.89.023002}

\bibitem[{{Dillon} {et~al.}(2015){Dillon}, {Neben}, {Hewitt}, {Tegmark},
  {Barry}, {Beardsley}, {Bowman}, {Briggs}, {Carroll}, {de Oliveira-Costa},
  {Ewall-Wice}, {Feng}, {Greenhill}, {Hazelton}, {Hernquist}, {Hurley-Walker},
  {Jacobs}, {Kim}, {Kittiwisit}, {Lenc}, {Line}, {Loeb}, {McKinley},
  {Mitchell}, {Morales}, {Offringa}, {Paul}, {Pindor}, {Pober}, {Procopio},
  {Riding}, {Sethi}, {Shankar}, {Subrahmanyan}, {Sullivan}, {Thyagarajan},
  {Tingay}, {Trott}, {Wayth}, {Webster}, {Wyithe}, {Bernardi}, {Cappallo},
  {Deshpande}, {Johnston-Hollitt}, {Kaplan}, {Lonsdale}, {McWhirter}, {Morgan},
  {Oberoi}, {Ord}, {Prabu}, {Srivani}, {Williams}, \&
  {Williams}}]{Dillon2015:MWA_Limit}
{Dillon}, J.~S., {Neben}, A.~R., {Hewitt}, J.~N., {et~al.} 2015, \prd, 91,
  123011, \dodoi{10.1103/PhysRevD.91.123011}

\bibitem[{{Dillon} {et~al.}(2020){Dillon}, {Lee}, {Ali}, {Parsons}, {Orosz},
  {Nunhokee}, {La Plante}, {Beardsley}, {Kern}, {Abdurashidova}, {Aguirre},
  {Alexander}, {Balfour}, {Bernardi}, {Billings}, {Bowman}, {Bradley}, {Bull},
  {Burba}, {Carey}, {Carilli}, {Cheng}, {DeBoer}, {Dexter}, {de Lera Acedo},
  {Ely}, {Ewall-Wice}, {Fagnoni}, {Fritz}, {Furlanetto}, {Gale-Sides},
  {Glendenning}, {Gorthi}, {Greig}, {Grobbelaar}, {Halday}, {Hazelton},
  {Hewitt}, {Hickish}, {Jacobs}, {Julius}, {Kerrigan}, {Kittiwisit}, {Kohn},
  {Kolopanis}, {Lanman}, {Lekalake}, {Lewis}, {Liu}, {Ma}, {MacMahon}, {Malan},
  {Malgas}, {Maree}, {Martinot}, {Matsetela}, {Mesinger}, {Molewa}, {Morales},
  {Mosiane}, {Murray}, {Neben}, {Nikolic}, {Pascua}, {Patra}, {Pieterse},
  {Pober}, {Razavi-Ghods}, {Ringuette}, {Robnett}, {Rosie}, {Santos}, {Sims},
  {Smith}, {Syce}, {Tegmark}, {Thyagarajan}, {Williams}, \&
  {Zheng}}]{Dillon2020:redcal}
{Dillon}, J.~S., {Lee}, M., {Ali}, Z.~S., {et~al.} 2020, \mnras, 499, 5840,
  \dodoi{10.1093/mnras/staa3001}

\bibitem[{{Eastwood} {et~al.}(2019){Eastwood}, {Anderson}, {Monroe},
  {Hallinan}, {Catha}, {Dowell}, {Garsden}, {Greenhill}, {Hicks}, {Kocz},
  {Price}, {Schinzel}, {Vedantham}, \& {Wang}}]{LWA2019:21cmLimit}
{Eastwood}, M.~W., {Anderson}, M.~M., {Monroe}, R.~M., {et~al.} 2019, \aj, 158,
  84, \dodoi{10.3847/1538-3881/ab2629}

\bibitem[{{Elahi} {et~al.}(2024){Elahi}, {Bharadwaj}, {Chatterjee}, {Sarkar},
  {Choudhuri}, {Sethi}, \& {Patwa}}]{Elahi2024:TGE_on_MWA}
{Elahi}, K. M.~A., {Bharadwaj}, S., {Chatterjee}, S., {et~al.} 2024, arXiv
  e-prints, arXiv:2410.11380, \dodoi{10.48550/arXiv.2410.11380}

\bibitem[{{Ewall-Wice} {et~al.}(2017){Ewall-Wice}, {Dillon}, {Liu}, \&
  {Hewitt}}]{Ewall-Wice2017:CalibrationError}
{Ewall-Wice}, A., {Dillon}, J.~S., {Liu}, A., \& {Hewitt}, J. 2017, \mnras,
  470, 1849, \dodoi{10.1093/mnras/stx1221}

\bibitem[{{Ewall-Wice} {et~al.}(2016){Ewall-Wice}, {Dillon}, {Hewitt}, {Loeb},
  {Mesinger}, {Neben}, {Offringa}, {Tegmark}, {Barry}, {Beardsley}, {Bernardi},
  {Bowman}, {Briggs}, {Cappallo}, {Carroll}, {Corey}, {de Oliveira-Costa},
  {Emrich}, {Feng}, {Gaensler}, {Goeke}, {Greenhill}, {Hazelton},
  {Hurley-Walker}, {Johnston-Hollitt}, {Jacobs}, {Kaplan}, {Kasper}, {Kim},
  {Kratzenberg}, {Lenc}, {Line}, {Lonsdale}, {Lynch}, {McKinley}, {McWhirter},
  {Mitchell}, {Morales}, {Morgan}, {Thyagarajan}, {Oberoi}, {Ord}, {Paul},
  {Pindor}, {Pober}, {Prabu}, {Procopio}, {Riding}, {Rogers}, {Roshi},
  {Shankar}, {Sethi}, {Srivani}, {Subrahmanyan}, {Sullivan}, {Tingay}, {Trott},
  {Waterson}, {Wayth}, {Webster}, {Whitney}, {Williams}, {Williams}, {Wu}, \&
  {Wyithe}}]{Ewall-Wice2016:Limit_Relection}
{Ewall-Wice}, A., {Dillon}, J.~S., {Hewitt}, J.~N., {et~al.} 2016, \mnras, 460,
  4320, \dodoi{10.1093/mnras/stw1022}

\bibitem[{{Ewall-Wice} {et~al.}(2021){Ewall-Wice}, {Kern}, {Dillon}, {Liu},
  {Parsons}, {Singh}, {Lanman}, {La Plante}, {Fagnoni}, {Acedo}, {DeBoer},
  {Nunhokee}, {Bull}, {Chang}, {Lazio}, {Aguirre}, \&
  {Weinberg}}]{Ewall-Wice2021:DAYNENU}
{Ewall-Wice}, A., {Kern}, N., {Dillon}, J.~S., {et~al.} 2021, \mnras, 500,
  5195, \dodoi{10.1093/mnras/staa3293}

\bibitem[{{Feeney} {et~al.}(2011){Feeney}, {Peiris}, \&
  {Pontzen}}]{Feeney2011:CMB_SkyReconstruction}
{Feeney}, S.~M., {Peiris}, H.~V., \& {Pontzen}, A. 2011, \prd, 84, 103002,
  \dodoi{10.1103/PhysRevD.84.103002}

\bibitem[{{Furlanetto} {et~al.}(2006){Furlanetto}, {Oh}, \&
  {Briggs}}]{Furlanetto2006:Review}
{Furlanetto}, S.~R., {Oh}, S.~P., \& {Briggs}, F.~H. 2006, \physrep, 433, 181,
  \dodoi{10.1016/j.physrep.2006.08.002}

\bibitem[{{Garsden} {et~al.}(2021){Garsden}, {Greenhill}, {Bernardi},
  {Fialkov}, {Price}, {Mitchell}, {Dowell}, {Spinelli}, \&
  {Schinzel}}]{Garsden2021:LWA_Limit}
{Garsden}, H., {Greenhill}, L., {Bernardi}, G., {et~al.} 2021, \mnras, 506,
  5802, \dodoi{10.1093/mnras/stab1671}

\bibitem[{{Garsden} {et~al.}(2024){Garsden}, {Bull}, {Wilensky},
  {Abdurashidova}, {Adams}, {Aguirre}, {Alexander}, {Ali}, {Baartman},
  {Balfour}, {Beardsley}, {Berkhout}, {Bernardi}, {Billings}, {Bowman},
  {Bradley}, {Burba}, {Carey}, {Carilli}, {Chen}, {Cheng}, {Choudhuri},
  {DeBoer}, {de Lera Acedo}, {Dexter}, {Dillon}, {Dynes}, {Eksteen}, {Ely},
  {Ewall-Wice}, {Fagnoni}, {Fritz}, {Furlanetto}, {Gale-Sides}, {Gehlot},
  {Ghosh}, {Glendenning}, {Gorce}, {Gorthi}, {Greig}, {Grobbelaar}, {Halday},
  {Hazelton}, {Hewitt}, {Hickish}, {Huang}, {Jacobs}, {Josaitis}, {Julius},
  {Kariseb}, {Kern}, {Kerrigan}, {Kim}, {Kittiwisit}, {Kohn}, {Kolopanis},
  {Lanman}, {La Plante}, {Liu}, {Loots}, {Ma}, {MacMahon}, {Malan}, {Malgas},
  {Malgas}, {Marero}, {Martinot}, {Mesinger}, {Molewa}, {Morales}, {Mosiane},
  {Murray}, {Neben}, {Nikolic}, {Devi Nunhokee}, {Nuwegeld}, {Parsons},
  {Pascua}, {Patra}, {Pieterse}, {Qin}, {Rath}, {Razavi-Ghods}, {Riley},
  {Robnett}, {Rosie}, {Santos}, {Sims}, {Singh}, {Storer}, {Swarts}, {Tan},
  {Thyagarajan}, {van Wyngaarden}, {Williams}, {Xu}, \&
  {Zheng}}]{Garsden2024:FRF}
{Garsden}, H., {Bull}, P., {Wilensky}, M., {et~al.} 2024, arXiv e-prints,
  arXiv:2402.08659, \dodoi{10.48550/arXiv.2402.08659}

\bibitem[{{Gehlot} {et~al.}(2019){Gehlot}, {Mertens}, {Koopmans}, {Brentjens},
  {Zaroubi}, {Ciardi}, {Ghosh}, {Hatef}, {Iliev}, {Jeli{\'c}}, {}, {Kooistra},
  {Krause}, {Mellema}, {Mevius}, {Mitra}, {Offringa}, {Pandey}, {Sardarabadi},
  {Schaye}, {Silva}, {Vedantham}, \& {Yatawatta}}]{Gehlot2019:LofarLimit}
{Gehlot}, B.~K., {Mertens}, F.~G., {Koopmans}, L.~V.~E., {et~al.} 2019, \mnras,
  488, 4271, \dodoi{10.1093/mnras/stz1937}

\bibitem[{{Ghosh} {et~al.}(2011){Ghosh}, {Bharadwaj}, {Ali}, \&
  {Chengalur}}]{Ghosh2011:GMRT_Limit}
{Ghosh}, A., {Bharadwaj}, S., {Ali}, S.~S., \& {Chengalur}, J.~N. 2011, \mnras,
  411, 2426, \dodoi{10.1111/j.1365-2966.2010.17853.x}

\bibitem[{{Gorce} {et~al.}(2023){Gorce}, {Ganjam}, {Liu}, {Murray},
  {Abdurashidova}, {Adams}, {Aguirre}, {Alexander}, {Ali}, {Baartman},
  {Balfour}, {Beardsley}, {Bernardi}, {Billings}, {Bowman}, {Bradley}, {Bull},
  {Burba}, {Carey}, {Carilli}, {Cheng}, {DeBoer}, {Acedo}, {Dexter}, {Dillon},
  {Eksteen}, {Ely}, {Ewall-Wice}, {Fagnoni}, {Fritz}, {Furlanetto},
  {Gale-Sides}, {Glendenning}, {Gorthi}, {Greig}, {Grobbelaar}, {Halday},
  {Hazelton}, {Hewitt}, {Hickish}, {Jacobs}, {Julius}, {Kariseb}, {Kern},
  {Kerrigan}, {Kittiwisit}, {Kohn}, {Kolopanis}, {Lanman}, {La Plante},
  {Loots}, {MacMahon}, {Malan}, {Malgas}, {Malgas}, {Marero}, {Martinot},
  {Mesinger}, {Molewa}, {Morales}, {Mosiane}, {Neben}, {Nikolic}, {Nuwegeld},
  {Parsons}, {Patra}, {Pieterse}, {Pober}, {Razavi-Ghods}, {Robnett}, {Rosie},
  {Sims}, {Swarts}, {Thyagarajan}, {van Wyngaarden}, {Williams}, \&
  {Zheng}}]{Gorce2023:Window_Function}
{Gorce}, A., {Ganjam}, S., {Liu}, A., {et~al.} 2023, \mnras, 520, 375,
  \dodoi{10.1093/mnras/stad090}

\bibitem[{{Grigg} {et~al.}(2023){Grigg}, {Tingay}, {Sokolowski}, {Wayth},
  {Indermuehle}, \& {Prabu}}]{Grigg2023:Starlink_SKA}
{Grigg}, D., {Tingay}, S.~J., {Sokolowski}, M., {et~al.} 2023, \aap, 678, L6,
  \dodoi{10.1051/0004-6361/202347654}

\bibitem[{{Gruetjen} {et~al.}(2017){Gruetjen}, {Fergusson}, {Liguori}, \&
  {Shellard}}]{Gruetjen2017:CMB_Inpainting}
{Gruetjen}, H.~F., {Fergusson}, J.~R., {Liguori}, M., \& {Shellard}, E.~P.~S.
  2017, \prd, 95, 043532, \dodoi{10.1103/PhysRevD.95.043532}

\bibitem[{{Gupta} {et~al.}(2017){Gupta}, {Ajithkumar}, {Kale}, {Nayak},
  {Sabhapathy}, {Sureshkumar}, {Swami}, {Chengalur}, {Ghosh},
  {Ishwara-Chandra}, {Joshi}, {Kanekar}, {Lal}, \&
  {Roy}}]{GMRT2017:uGMRT_Overview}
{Gupta}, Y., {Ajithkumar}, B., {Kale}, H.~S., {et~al.} 2017, Current Science,
  113, 707, \dodoi{10.18520/cs/v113/i04/707-714}

\bibitem[{{Harris}(1978)}]{Harris1978}
{Harris}, F.~J. 1978, IEEE Proceedings, 66, 51

\bibitem[{{Hazelton} {et~al.}(2017){Hazelton}, {Jacobs}, {Pober}, \&
  {Beardsley}}]{Hazelton2017:pyuvdata}
{Hazelton}, B.~J., {Jacobs}, D.~C., {Pober}, J.~C., \& {Beardsley}, A.~P. 2017,
  The Journal of Open Source Software, 2, 140, \dodoi{10.21105/joss.00140}

\bibitem[{{Hazelton} {et~al.}(2013){Hazelton}, {Morales}, \&
  {Sullivan}}]{Hazelton2013:wedge}
{Hazelton}, B.~J., {Morales}, M.~F., \& {Sullivan}, I.~S. 2013, \apj, 770, 156,
  \dodoi{10.1088/0004-637X/770/2/156}

\bibitem[{{HERA Collaboration} {et~al.}(2023){HERA Collaboration},
  {Abdurashidova}, {Adams}, {Aguirre}, {Alexander}, {Ali}, {Baartman},
  {Balfour}, {Barkana}, {Beardsley}, {Bernardi}, {Billings}, {Bowman},
  {Bradley}, {Breitman}, {Bull}, {Burba}, {Carey}, {Carilli}, {Cheng},
  {Choudhuri}, {DeBoer}, {de Lera Acedo}, {Dexter}, {Dillon}, {Ely},
  {Ewall-Wice}, {Fagnoni}, {Fialkov}, {Fritz}, {Furlanetto}, {Gale-Sides},
  {Garsden}, {Glendenning}, {Gorce}, {Gorthi}, {Greig}, {Grobbelaar}, {Halday},
  {Hazelton}, {Heimersheim}, {Hewitt}, {Hickish}, {Jacobs}, {Julius}, {Kern},
  {Kerrigan}, {Kittiwisit}, {Kohn}, {Kolopanis}, {Lanman}, {La Plante},
  {Lewis}, {Liu}, {Loots}, {Ma}, {MacMahon}, {Malan}, {Malgas}, {Malgas},
  {Maree}, {Marero}, {Martinot}, {McBride}, {Mesinger}, {Mirocha}, {Molewa},
  {Morales}, {Mosiane}, {Mu{\~n}oz}, {Murray}, {Nagpal}, {Neben}, {Nikolic},
  {Nunhokee}, {Nuwegeld}, {Parsons}, {Pascua}, {Patra}, {Pieterse}, {Qin},
  {Razavi-Ghods}, {Robnett}, {Rosie}, {Santos}, {Sims}, {Singh}, {Smith},
  {Swarts}, {Tan}, {Thyagarajan}, {Wilensky}, {Williams}, {van Wyngaarden}, \&
  {Zheng}}]{HERA2023:h1c_idr3_limit}
{HERA Collaboration}, {Abdurashidova}, Z., {Adams}, T., {et~al.} 2023, \apj,
  945, 124, \dodoi{10.3847/1538-4357/acaf50}

\bibitem[{{H{\"o}gbom}(1974)}]{Hogbom1974:CLEAN}
{H{\"o}gbom}, J.~A. 1974, \aaps, 15, 417

\bibitem[{Hunter(2007)}]{matplotlib}
Hunter, J.~D. 2007, Computing in Science \& Engineering, 9, 90,
  \dodoi{10.1109/MCSE.2007.55}

\bibitem[{{Hurley-Walker} {et~al.}(2017){Hurley-Walker}, {Callingham},
  {Hancock}, {Franzen}, {Hindson}, {Kapi{\'n}ska}, {Morgan}, {Offringa},
  {Wayth}, {Wu}, {Zheng}, {Murphy}, {Bell}, {Dwarakanath}, {For}, {Gaensler},
  {Johnston-Hollitt}, {Lenc}, {Procopio}, {Staveley-Smith}, {Ekers}, {Bowman},
  {Briggs}, {Cappallo}, {Deshpande}, {Greenhill}, {Hazelton}, {Kaplan},
  {Lonsdale}, {McWhirter}, {Mitchell}, {Morales}, {Morgan}, {Oberoi}, {Ord},
  {Prabu}, {Shankar}, {Srivani}, {Subrahmanyan}, {Tingay}, {Webster},
  {Williams}, \& {Williams}}]{Hurley-Walker2017:GLEAMI}
{Hurley-Walker}, N., {Callingham}, J.~R., {Hancock}, P.~J., {et~al.} 2017,
  \mnras, 464, 1146, \dodoi{10.1093/mnras/stw2337}

\bibitem[{{Hurley-Walker} {et~al.}(2019){Hurley-Walker}, {Hancock}, {Franzen},
  {Callingham}, {Offringa}, {Hindson}, {Wu}, {Bell}, {For}, {Gaensler},
  {Johnston-Hollitt}, {Kapi{\'n}ska}, {Morgan}, {Murphy}, {McKinley},
  {Procopio}, {Staveley-Smith}, {Wayth}, \&
  {Zheng}}]{Hurley-Walker2019:GLEAMII}
{Hurley-Walker}, N., {Hancock}, P.~J., {Franzen}, T.~M.~O., {et~al.} 2019,
  \pasa, 36, e047, \dodoi{10.1017/pasa.2019.37}

\bibitem[{{Josaitis} {et~al.}(2022){Josaitis}, {Ewall-Wice}, {Fagnoni}, \& {de
  Lera Acedo}}]{Josaitis2022:MutualCoupling}
{Josaitis}, A.~T., {Ewall-Wice}, A., {Fagnoni}, N., \& {de Lera Acedo}, E.
  2022, \mnras, 514, 1804, \dodoi{10.1093/mnras/stac916}

\bibitem[{{Joseph} {et~al.}(2018){Joseph}, {Trott}, \&
  {Wayth}}]{Joseph2018:BeamError}
{Joseph}, R.~C., {Trott}, C.~M., \& {Wayth}, R.~B. 2018, \aj, 156, 285,
  \dodoi{10.3847/1538-3881/aaec0b}

\bibitem[{{Joseph} {et~al.}(2020){Joseph}, {Trott}, {Wayth}, \&
  {Nasirudin}}]{Joseph2020:BeamVariation}
{Joseph}, R.~C., {Trott}, C.~M., {Wayth}, R.~B., \& {Nasirudin}, A. 2020,
  \mnras, 492, 2017, \dodoi{10.1093/mnras/stz3375}

\bibitem[{{Karnik} {et~al.}(2020){Karnik}, {Romberg}, \&
  {Davenport}}]{Karnik2020:DPSS_Bounds}
{Karnik}, S., {Romberg}, J., \& {Davenport}, M.~A. 2020, arXiv e-prints,
  arXiv:2006.00427, \dodoi{10.48550/arXiv.2006.00427}

\bibitem[{{Kennedy} {et~al.}(2023){Kennedy}, {Bull}, {Wilensky}, {Burba}, \&
  {Choudhuri}}]{Kennedy2023:Gibbs_Sampling}
{Kennedy}, F., {Bull}, P., {Wilensky}, M.~J., {Burba}, J., \& {Choudhuri}, S.
  2023, \apjs, 266, 23, \dodoi{10.3847/1538-4365/acc324}

\bibitem[{{Kern} \& {Liu}(2021)}]{Kern2021:GPR}
{Kern}, N.~S., \& {Liu}, A. 2021, \mnras, 501, 1463,
  \dodoi{10.1093/mnras/staa3736}

\bibitem[{{Kern} {et~al.}(2019){Kern}, {Parsons}, {Dillon}, {Lanman},
  {Fagnoni}, \& {de Lera Acedo}}]{Kern2019:Relection_Model}
{Kern}, N.~S., {Parsons}, A.~R., {Dillon}, J.~S., {et~al.} 2019, arXiv
  e-prints, arXiv:1909.11732, \dodoi{10.48550/arXiv.1909.11732}

\bibitem[{{Kim} {et~al.}(2022){Kim}, {Nhan}, {Hewitt}, {Kern}, {Dillon}, {de
  Lera Acedo}, {Dynes}, {Mahesh}, {Fagnoni}, \&
  {DeBoer}}]{Kim2022:BeamPerturbation}
{Kim}, H., {Nhan}, B.~D., {Hewitt}, J.~N., {et~al.} 2022, \apj, 941, 207,
  \dodoi{10.3847/1538-4357/ac9eaf}

\bibitem[{{Kolopanis} {et~al.}(2019){Kolopanis}, {Jacobs}, {Cheng}, {Parsons},
  {Kohn}, {Pober}, {Aguirre}, {Ali}, {Bernardi}, {Bradley}, {Carilli},
  {DeBoer}, {Dexter}, {Dillon}, {Kerrigan}, {Klima}, {Liu}, {MacMahon},
  {Moore}, {Thyagarajan}, {Nunhokee}, {Walbrugh}, \&
  {Walker}}]{Kolopanis2019:PaperLimit}
{Kolopanis}, M., {Jacobs}, D.~C., {Cheng}, C., {et~al.} 2019, \apj, 883, 133,
  \dodoi{10.3847/1538-4357/ab3e3a}

\bibitem[{{Koopmans} {et~al.}(2015){Koopmans}, {Pritchard}, {Mellema},
  {Aguirre}, {Ahn}, {Barkana}, {van Bemmel}, {Bernardi}, {Bonaldi}, {Briggs},
  {de Bruyn}, {Chang}, {Chapman}, {Chen}, {Ciardi}, {Dayal}, {Ferrara},
  {Fialkov}, {Fiore}, {Ichiki}, {Illiev}, {Inoue}, {Jelic}, {Jones}, {Lazio},
  {Maio}, {Majumdar}, {Mack}, {Mesinger}, {Morales}, {Parsons}, {Pen},
  {Santos}, {Schneider}, {Semelin}, {de Souza}, {Subrahmanyan}, {Takeuchi},
  {Vedantham}, {Wagg}, {Webster}, {Wyithe}, {Datta}, \& {Trott}}]{SKA2015:EoR}
{Koopmans}, L., {Pritchard}, J., {Mellema}, G., {et~al.} 2015, in Advancing
  Astrophysics with the Square Kilometre Array (AASKA14), 1,
  \dodoi{10.22323/1.215.0001}

\bibitem[{{Li} {et~al.}(2018){Li}, {Pober}, {Hazelton}, {Barry}, {Morales},
  {Sullivan}, {Parsons}, {Ali}, {Dillon}, {Beardsley}, {Bowman}, {Briggs},
  {Byrne}, {Carroll}, {Crosse}, {Emrich}, {Ewall-Wice}, {Feng}, {Franzen},
  {Hewitt}, {Horsley}, {Jacobs}, {Johnston-Hollitt}, {Jordan}, {Joseph},
  {Kaplan}, {Kenney}, {Kim}, {Kittiwisit}, {Lanman}, {Line}, {McKinley},
  {Mitchell}, {Murray}, {Neben}, {Offringa}, {Pallot}, {Paul}, {Pindor},
  {Procopio}, {Rahimi}, {Riding}, {Sethi}, {Udaya Shankar}, {Steele},
  {Subrahmanian}, {Tegmark}, {Thyagarajan}, {Tingay}, {Trott}, {Walker},
  {Wayth}, {Webster}, {Williams}, {Wu}, \& {Wyithe}}]{Li2018:Calibration}
{Li}, W., {Pober}, J.~C., {Hazelton}, B.~J., {et~al.} 2018, \apj, 863, 170,
  \dodoi{10.3847/1538-4357/aad3c3}

\bibitem[{{Li} {et~al.}(2019){Li}, {Pober}, {Barry}, {Hazelton}, {Morales},
  {Trott}, {Lanman}, {Wilensky}, {Sullivan}, {Beardsley}, {Booler}, {Bowman},
  {Byrne}, {Crosse}, {Emrich}, {Franzen}, {Hasegawa}, {Horsley},
  {Johnston-Hollitt}, {Jacobs}, {Jordan}, {Joseph}, {Kaneuji}, {Kaplan},
  {Kenney}, {Kubota}, {Line}, {Lynch}, {McKinley}, {Mitchell}, {Murray},
  {Pallot}, {Pindor}, {Rahimi}, {Riding}, {Sleap}, {Steele}, {Takahashi},
  {Tingay}, {Walker}, {Wayth}, {Webster}, {Williams}, {Wu}, {Wyithe},
  {Yoshiura}, \& {Zheng}}]{Li2019:MWA_Limit}
{Li}, W., {Pober}, J.~C., {Barry}, N., {et~al.} 2019, \apj, 887, 141,
  \dodoi{10.3847/1538-4357/ab55e4}

\bibitem[{{Liu} {et~al.}(2014){Liu}, {Parsons}, \&
  {Trott}}]{Liu2014:EoR_WindowI}
{Liu}, A., {Parsons}, A.~R., \& {Trott}, C.~M. 2014, \prd, 90, 023018,
  \dodoi{10.1103/PhysRevD.90.023018}

\bibitem[{{Liu} \& {Shaw}(2020)}]{Liu2020:Review}
{Liu}, A., \& {Shaw}, J.~R. 2020, \pasp, 132, 062001,
  \dodoi{10.1088/1538-3873/ab5bfd}

\bibitem[{{Liu} \& {Tegmark}(2011)}]{Liu2011:QE}
{Liu}, A., \& {Tegmark}, M. 2011, \prd, 83, 103006,
  \dodoi{10.1103/PhysRevD.83.103006}

\bibitem[{{Liu} {et~al.}(2010){Liu}, {Tegmark}, {Morrison}, {Lutomirski}, \&
  {Zaldarriaga}}]{Liu2010:redcal}
{Liu}, A., {Tegmark}, M., {Morrison}, S., {Lutomirski}, A., \& {Zaldarriaga},
  M. 2010, \mnras, 408, 1029, \dodoi{10.1111/j.1365-2966.2010.17174.x}

\bibitem[{{Lomb}(1976)}]{Lomb1976}
{Lomb}, N.~R. 1976, \apss, 39, 447, \dodoi{10.1007/BF00648343}

\bibitem[{{Louren{\c{c}}o} {et~al.}(2023){Louren{\c{c}}o}, {Chippendale},
  {Indermuehle}, {Moss}, {Murphy}, {Galvin}, {Hellbourg}, {Hotan}, {Lenc}, \&
  {Whiting}}]{Lourenco2023:ASCAP_RFI}
{Louren{\c{c}}o}, L., {Chippendale}, A.~P., {Indermuehle}, B., {et~al.} 2023,
  arXiv e-prints, arXiv:2312.14422, \dodoi{10.48550/arXiv.2312.14422}

\bibitem[{{Mertens} {et~al.}(2018){Mertens}, {Ghosh}, \&
  {Koopmans}}]{Mertens2018:GPR}
{Mertens}, F.~G., {Ghosh}, A., \& {Koopmans}, L.~V.~E. 2018, \mnras, 478, 3640,
  \dodoi{10.1093/mnras/sty1207}

\bibitem[{{Mertens} {et~al.}(2020){Mertens}, {Mevius}, {Koopmans}, {Offringa},
  {Mellema}, {Zaroubi}, {Brentjens}, {Gan}, {Gehlot}, {Pand ey}, {Sardarabadi},
  {Vedantham}, {Yatawatta}, {Asad}, {Ciardi}, {Chapman}, {Gazagnes}, {Ghara},
  {Ghosh}, {Giri}, {Iliev}, {Jeli{\'c}}, {Kooistra}, {Mondal}, {Schaye}, \&
  {Silva}}]{Mertens2020:LofarLimit}
{Mertens}, F.~G., {Mevius}, M., {Koopmans}, L.~V.~E., {et~al.} 2020, \mnras,
  493, 1662, \dodoi{10.1093/mnras/staa327}

\bibitem[{Mesinger(2016)}]{Mesinger2016:Review}
Mesinger, A. 2016, Underst. Epoch Cosm. Reionization Challenges Prog., 423,
  \dodoi{10.1007/978-3-319-21957-8}

\bibitem[{Mitchell {et~al.}(2009)Mitchell, Greenhill, Wayth, Sault, Lonsdale,
  Cappallo, Morales, \& Ord}]{Mitchell2008:Calibration}
Mitchell, D.~A., Greenhill, L.~J., Wayth, R.~B., {et~al.} 2009, Radio Sci., 44,
  0A01, \dodoi{10.1029/2009RS004263}

\bibitem[{{Morales} {et~al.}(2012){Morales}, {Hazelton}, {Sullivan}, \&
  {Beardsley}}]{Morales2012:wedge}
{Morales}, M.~F., {Hazelton}, B., {Sullivan}, I., \& {Beardsley}, A. 2012,
  \apj, 752, 137, \dodoi{10.1088/0004-637X/752/2/137}

\bibitem[{{Morales} \& {Wyithe}(2010)}]{Morales2010:Review}
{Morales}, M.~F., \& {Wyithe}, J. S.~B. 2010, \araa, 48, 127,
  \dodoi{10.1146/annurev-astro-081309-130936}

\bibitem[{{Mouri Sardarabadi} \&
  {Koopmans}(2019)}]{Mouri_Sardarabadi2019:CalibrationError}
{Mouri Sardarabadi}, A., \& {Koopmans}, L.~V.~E. 2019, \mnras, 483, 5480,
  \dodoi{10.1093/mnras/sty3444}

\bibitem[{{Munshi} {et~al.}(2024){Munshi}, {Mertens}, {Koopmans}, {Offringa},
  {Semelin}, {Aubert}, {Barkana}, {Bracco}, {Brackenhoff}, {Cecconi},
  {Ceccotti}, {Corbel}, {Fialkov}, {Gehlot}, {Ghara}, {Girard},
  {Grie{\ss}meier}, {H{\"o}fer}, {Hothi}, {M{\'e}riot}, {Mevius}, {Ocvirk},
  {Shaw}, {Theureau}, {Yatawatta}, {Zarka}, \&
  {Zaroubi}}]{Munshi2024:Nenufar_Limit}
{Munshi}, S., {Mertens}, F.~G., {Koopmans}, L.~V.~E., {et~al.} 2024, \aap, 681,
  A62, \dodoi{10.1051/0004-6361/202348329}

\bibitem[{{Murray} {et~al.}(2023){Murray}, {Dillon}, \& {Martinot}}]{H6C_IDR1}
{Murray}, S., {Dillon}, J., \& {Martinot}, Z.~E. 2023, {H6C Internal Data
  Release 2.1}, Tech. rep., HERA Collaboration.
\newblock
  \url{https://reionization.org/manual_uploads/HERA124_H6C_IDR_2_Memo_v3.pdf}

\bibitem[{{Neben} {et~al.}(2016{\natexlab{a}}){Neben}, {Hewitt}, {Bradley},
  {Dillon}, {Bernardi}, {Bowman}, {Briggs}, {Cappallo}, {Corey}, {Deshpande},
  {Goeke}, {Greenhill}, {Hazelton}, {Johnston-Hollitt}, {Kaplan}, {Lonsdale},
  {McWhirter}, {Mitchell}, {Morales}, {Morgan}, {Oberoi}, {Ord}, {Prabu},
  {Udaya Shankar}, {Srivani}, {Subrahmanyan}, {Tingay}, {Wayth}, {Webster},
  {Williams}, \& {Williams}}]{Neben2016:BeamErrorMWA}
{Neben}, A.~R., {Hewitt}, J.~N., {Bradley}, R.~F., {et~al.} 2016{\natexlab{a}},
  \apj, 820, 44, \dodoi{10.3847/0004-637X/820/1/44}

\bibitem[{{Neben} {et~al.}(2016{\natexlab{b}}){Neben}, {Bradley}, {Hewitt},
  {DeBoer}, {Parsons}, {Aguirre}, {Ali}, {Cheng}, {Ewall-Wice}, {Patra},
  {Thyagarajan}, {Bowman}, {Dickenson}, {Dillon}, {Doolittle}, {Egan},
  {Hedrick}, {Jacobs}, {Kohn}, {Klima}, {Moodley}, {Saliwanchik}, {Schaffner},
  {Shelton}, {Taylor}, {Taylor}, {Tegmark}, {Wirt}, \&
  {Zheng}}]{Neben2016:HERA_Beam}
{Neben}, A.~R., {Bradley}, R.~F., {Hewitt}, J.~N., {et~al.} 2016{\natexlab{b}},
  \apj, 826, 199, \dodoi{10.3847/0004-637X/826/2/199}

\bibitem[{{Offringa} {et~al.}(2019){Offringa}, {Mertens}, \&
  {Koopmans}}]{Offringa2019:RFI}
{Offringa}, A.~R., {Mertens}, F., \& {Koopmans}, L.~V.~E. 2019, \mnras, 484,
  2866, \dodoi{10.1093/mnras/stz175}

\bibitem[{{Offringa} {et~al.}(2012){Offringa}, {van de Gronde}, \&
  {Roerdink}}]{Offringa2012:AoFlagger}
{Offringa}, A.~R., {van de Gronde}, J.~J., \& {Roerdink}, J.~B.~T.~M. 2012,
  \aap, 539, A95, \dodoi{10.1051/0004-6361/201118497}

\bibitem[{{Offringa} {et~al.}(2013){Offringa}, {de Bruyn}, {Zaroubi}, {van
  Diepen}, {Martinez-Ruby}, {Labropoulos}, {Brentjens}, {Ciardi}, {Daiboo},
  {Harker}, {Jeli{\'c}}, {Kazemi}, {Koopmans}, {Mellema}, {Pandey}, {Pizzo},
  {Schaye}, {Vedantham}, {Veligatla}, {Wijnholds}, {Yatawatta}, {Zarka},
  {Alexov}, {Anderson}, {Asgekar}, {Avruch}, {Beck}, {Bell}, {Bell}, {Bentum},
  {Bernardi}, {Best}, {Birzan}, {Bonafede}, {Breitling}, {Broderick},
  {Br{\"u}ggen}, {Butcher}, {Conway}, {de Vos}, {Dettmar}, {Eisloeffel},
  {Falcke}, {Fender}, {Frieswijk}, {Gerbers}, {Griessmeier}, {Gunst},
  {Hassall}, {Heald}, {Hessels}, {Hoeft}, {Horneffer}, {Karastergiou},
  {Kondratiev}, {Koopman}, {Kuniyoshi}, {Kuper}, {Maat}, {Mann}, {McKean},
  {Meulman}, {Mevius}, {Mol}, {Nijboer}, {Noordam}, {Norden}, {Paas}, {Pandey},
  {Pizzo}, {Polatidis}, {Rafferty}, {Rawlings}, {Reich}, {R{\"o}ttgering},
  {Schoenmakers}, {Sluman}, {Smirnov}, {Sobey}, {Stappers}, {Steinmetz},
  {Swinbank}, {Tagger}, {Tang}, {Tasse}, {van Ardenne}, {van Cappellen}, {van
  Duin}, {van Haarlem}, {van Leeuwen}, {van Weeren}, {Vermeulen}, {Vocks},
  {Wijers}, {Wise}, \& {Wucknitz}}]{Offringa2013:LOFAR_RFI}
{Offringa}, A.~R., {de Bruyn}, A.~G., {Zaroubi}, S., {et~al.} 2013, \aap, 549,
  A11, \dodoi{10.1051/0004-6361/201220293}

\bibitem[{{Offringa} {et~al.}(2015){Offringa}, {Wayth}, {Hurley-Walker},
  {Kaplan}, {Barry}, {Beardsley}, {Bell}, {Bernardi}, {Bowman}, {Briggs},
  {Callingham}, {Cappallo}, {Carroll}, {Deshpande}, {Dillon}, {Dwarakanath},
  {Ewall-Wice}, {Feng}, {For}, {Gaensler}, {Greenhill}, {Hancock}, {Hazelton},
  {Hewitt}, {Hindson}, {Jacobs}, {Johnston-Hollitt}, {Kapi{\'n}ska}, {Kim},
  {Kittiwisit}, {Lenc}, {Line}, {Loeb}, {Lonsdale}, {McKinley}, {McWhirter},
  {Mitchell}, {Morales}, {Morgan}, {Morgan}, {Neben}, {Oberoi}, {Ord}, {Paul},
  {Pindor}, {Pober}, {Prabu}, {Procopio}, {Riding}, {Udaya Shankar}, {Sethi},
  {Srivani}, {Staveley-Smith}, {Subrahmanyan}, {Sullivan}, {Tegmark},
  {Thyagarajan}, {Tingay}, {Trott}, {Webster}, {Williams}, {Williams}, {Wu},
  {Wyithe}, \& {Zheng}}]{Offringa2015:MWA_RFI}
{Offringa}, A.~R., {Wayth}, R.~B., {Hurley-Walker}, N., {et~al.} 2015, \pasa,
  32, e008, \dodoi{10.1017/pasa.2015.7}

\bibitem[{Oliphant(2006--)}]{numpy}
Oliphant, T. 2006--, {NumPy}: A guide to {NumPy}, USA: Trelgol Publishing.
\newblock \url{http://www.numpy.org/}

\bibitem[{{Orosz} {et~al.}(2019){Orosz}, {Dillon}, {Ewall-Wice}, {Parsons}, \&
  {Thyagarajan}}]{Orosz2019:BeamVariation}
{Orosz}, N., {Dillon}, J.~S., {Ewall-Wice}, A., {Parsons}, A.~R., \&
  {Thyagarajan}, N. 2019, \mnras, 487, 537, \dodoi{10.1093/mnras/stz1287}

\bibitem[{{Paciga} {et~al.}(2013){Paciga}, {Albert}, {Bandura}, {Chang},
  {Gupta}, {Hirata}, {Odegova}, {Pen}, {Peterson}, {Roy}, {Shaw}, {Sigurdson},
  \& {Voytek}}]{GMRT2013:21cmLimit}
{Paciga}, G., {Albert}, J.~G., {Bandura}, K., {et~al.} 2013, \mnras, 433, 639,
  \dodoi{10.1093/mnras/stt753}

\bibitem[{{Pagano} {et~al.}(2023){Pagano}, {Liu}, {Liu}, {Kern}, {Ewall-Wice},
  {Bull}, {Pascua}, {Ravanbakhsh}, {Abdurashidova}, {Adams}, {Aguirre},
  {Alexander}, {Ali}, {Baartman}, {Balfour}, {Beardsley}, {Bernardi},
  {Billings}, {Bowman}, {Bradley}, {Burba}, {Carey}, {Carilli}, {Cheng},
  {DeBoer}, {de Lera Acedo}, {Dexter}, {Dillon}, {Eksteen}, {Ely}, {Fagnoni},
  {Fritz}, {Furlanetto}, {Gale-Sides}, {Glendenning}, {Gorthi}, {Greig},
  {Grobbelaar}, {Halday}, {Hazelton}, {Hewitt}, {Hickish}, {Jacobs}, {Julius},
  {Kariseb}, {Kerrigan}, {Kittiwisit}, {Kohn}, {Kolopanis}, {Lanman}, {La
  Plante}, {Loots}, {MacMahon}, {Malan}, {Malgas}, {Malgas}, {Marero},
  {Martinot}, {Mesinger}, {Molewa}, {Morales}, {Mosiane}, {Neben}, {Nikolic},
  {Nuwegeld}, {Parsons}, {Patra}, {Pieterse}, {Razavi-Ghods}, {Robnett},
  {Rosie}, {Sims}, {Smith}, {Swarts}, {Thyagarajan}, {van Wyngaarden},
  {Williams}, \& {Zheng}}]{Pagano2023:Inpainting}
{Pagano}, M., {Liu}, J., {Liu}, A., {et~al.} 2023, \mnras, 520, 5552,
  \dodoi{10.1093/mnras/stad441}

\bibitem[{{Pal} {et~al.}(2021){Pal}, {Bharadwaj}, {Ghosh}, \&
  {Choudhuri}}]{Pal2021:TGE}
{Pal}, S., {Bharadwaj}, S., {Ghosh}, A., \& {Choudhuri}, S. 2021, \mnras, 501,
  3378, \dodoi{10.1093/mnras/staa3831}

\bibitem[{{Parsons} \& {Backer}(2009)}]{Parsons2009:CLEAN}
{Parsons}, A.~R., \& {Backer}, D.~C. 2009, \aj, 138, 219,
  \dodoi{10.1088/0004-6256/138/1/219}

\bibitem[{{Parsons} {et~al.}(2012){Parsons}, {Pober}, {Aguirre}, {Carilli},
  {Jacobs}, \& {Moore}}]{Parsons2012:delay_spectrum_wedge}
{Parsons}, A.~R., {Pober}, J.~C., {Aguirre}, J.~E., {et~al.} 2012, \apj, 756,
  165, \dodoi{10.1088/0004-637X/756/2/165}

\bibitem[{{Parsons} {et~al.}(2010){Parsons}, {Backer}, {Foster}, {Wright},
  {Bradley}, {Gugliucci}, {Parashare}, {Benoit}, {Aguirre}, {Jacobs},
  {Carilli}, {Herne}, {Lynch}, {Manley}, \& {Werthimer}}]{PAPER2010:Overview}
{Parsons}, A.~R., {Backer}, D.~C., {Foster}, G.~S., {et~al.} 2010, \aj, 139,
  1468, \dodoi{10.1088/0004-6256/139/4/1468}

\bibitem[{{Parsons} {et~al.}(2014){Parsons}, {Liu}, {Aguirre}, {Ali},
  {Bradley}, {Carilli}, {DeBoer}, {Dexter}, {Gugliucci}, {Jacobs}, {Klima},
  {MacMahon}, {Manley}, {Moore}, {Pober}, {Stefan}, \&
  {Walbrugh}}]{Parsons2014:PaperLimit}
{Parsons}, A.~R., {Liu}, A., {Aguirre}, J.~E., {et~al.} 2014, \apj, 788, 106,
  \dodoi{10.1088/0004-637X/788/2/106}

\bibitem[{{Pascua} {et~al.}(2024){Pascua}, {Martinot}, {Liu}, {Aguirre},
  {Kern}, {Dillon}, {Wilensky}, {Fagnoni}, {de Lera Acedo}, \&
  {DeBoer}}]{Pascua2024:FRF}
{Pascua}, R., {Martinot}, Z.~E., {Liu}, A., {et~al.} 2024, arXiv e-prints,
  arXiv:2410.01872, \dodoi{10.48550/arXiv.2410.01872}

\bibitem[{Patil {et~al.}(2017)Patil, Yatawatta, Koopmans, de~Bruyn, Brentjens,
  Zaroubi, Asad, Hatef, Jeli{\'{c}}, Mevius, Offringa, Pandey, Vedantham,
  Abdalla, Brouw, Chapman, Ciardi, Gehlot, Ghosh, Harker, Iliev, Kakiichi,
  Majumdar, Mellema, Silva, Schaye, Vrbanec, \&
  Wijnholds}]{Patil2017:LofarLimit}
Patil, A., Yatawatta, S., Koopmans, L., {et~al.} 2017, \apj, 838, 65,
  \dodoi{10.3847/1538-4357/aa63e7}

\bibitem[{{Patil} {et~al.}(2016){Patil}, {Yatawatta}, {Zaroubi}, {Koopmans},
  {de Bruyn}, {Jeli{\'c}}, {Ciardi}, {Iliev}, {Mevius}, {Pandey}, \&
  {Gehlot}}]{Patil2016:CalibrationError}
{Patil}, A.~H., {Yatawatta}, S., {Zaroubi}, S., {et~al.} 2016, \mnras, 463,
  4317, \dodoi{10.1093/mnras/stw2277}

\bibitem[{{Paul} {et~al.}(2023){Paul}, {Santos}, {Chen}, \&
  {Wolz}}]{Paul2023:MeerKAT_Detection}
{Paul}, S., {Santos}, M.~G., {Chen}, Z., \& {Wolz}, L. 2023, arXiv e-prints,
  arXiv:2301.11943, \dodoi{10.48550/arXiv.2301.11943}

\bibitem[{{Pritchard} \& {Loeb}(2012)}]{Pritchard2012:Review}
{Pritchard}, J.~R., \& {Loeb}, A. 2012, Reports on Progress in Physics, 75,
  086901, \dodoi{10.1088/0034-4885/75/8/086901}

\bibitem[{{Rath} {et~al.}(2021){Rath}, {Dynes}, {Hewitt}, \&
  {Molewa}}]{HERA_Beam_Movement}
{Rath}, E., {Dynes}, S., {Hewitt}, J., \& {Molewa}, M. 2021, {Motion of HERA
  Antenna Feeds}, Tech. rep., HERA Collaboration.
\newblock
  \url{https://reionization.org/manual_uploads/HERA095_Motion_of_HERA_Antenna_Feeds.pdf}

\bibitem[{{Rath} {et~al.}(2024){Rath}, {Pascua}, {Josaitis}, {Ewall-Wice},
  {Fagnoni}, {de Lera Acedo}, {Martinot}, {Abdurashidova}, {Adams}, {Aguirre},
  {Baartman}, {Beardsley}, {Berkhout}, {Bernardi}, {Billings}, {Bowman},
  {Bull}, {Burba}, {Byrne}, {Carey}, {Chen}, {Choudhuri}, {Cox}, {DeBoer},
  {Dexter}, {Dillon}, {Dynes}, {Eksteen}, {Ely}, {Fritz}, {Furlanetto},
  {Gale-Sides}, {Garsden}, {Gehlot}, {Ghosh}, {Gorce}, {Gorthi}, {Halday},
  {Hazelton}, {Hewitt}, {Hickish}, {Huang}, {Jacobs}, {Kern}, {Kerrigan},
  {Kittiwisit}, {Kolopanis}, {Lanman}, {Liu}, {Ma}, {MacMahon}, {Malan},
  {Malgas}, {Malgas}, {Marero}, {McBride}, {Mesinger}, {Mohamed-Hinds},
  {Molewa}, {Morales}, {Murray}, {Nikolic}, {Nuwegeld}, {Parsons}, {Patra}, {La
  Plante}, {Qin}, {Razavi-Ghods}, {Riley}, {Robnett}, {Rosie}, {Santos},
  {Sims}, {Singh}, {Storer}, {Swarts}, {Tan}, {Wilensky}, {Williams}, {van
  Wyngaarden}, \& {Zheng}}]{Rath_Pascua2024:Mutual_Coupling}
{Rath}, E., {Pascua}, R., {Josaitis}, A.~T., {et~al.} 2024, arXiv e-prints,
  arXiv:2406.08549, \dodoi{10.48550/arXiv.2406.08549}

\bibitem[{{Roberts} {et~al.}(1987){Roberts}, {Lehar}, \&
  {Dreher}}]{Roberts1987:CLEAN_Time_Series}
{Roberts}, D.~H., {Lehar}, J., \& {Dreher}, J.~W. 1987, \aj, 93, 968,
  \dodoi{10.1086/114383}

\bibitem[{{Rybicki} \& {Press}(1992)}]{Rybicki1992:GPR}
{Rybicki}, G.~B., \& {Press}, W.~H. 1992, \apj, 398, 169,
  \dodoi{10.1086/171845}

\bibitem[{{Santos} {et~al.}(2015){Santos}, {Bull}, {Alonso}, {Camera},
  {Ferreira}, {Bernardi}, {Maartens}, {Viel}, {Villaescusa-Navarro}, {Abdalla},
  {Jarvis}, {Metcalf}, {Pourtsidou}, \& {Wolz}}]{SKA2015:HI_IM}
{Santos}, M., {Bull}, P., {Alonso}, D., {et~al.} 2015, in Advancing
  Astrophysics with the Square Kilometre Array (AASKA14), 19,
  \dodoi{10.22323/1.215.0019}

\bibitem[{{Santos} {et~al.}(2016){Santos}, {Bull}, {Camera}, {Chen}, {Fonseca},
  {Heywood}, {Hilton}, {Jarvis}, {Jozsa}, {Knowles}, {Leeuw}, {Maartens},
  {Malefahlo}, {McAlpine}, {Moodley}, {Patel}, {Pourtsidou}, {Prescott},
  {Spekkens}, {Taylor}, {Witzemann}, \& {Whittam}}]{MeerKAT2016:MeerKLASS}
{Santos}, M., {Bull}, P., {Camera}, S., {et~al.} 2016, in MeerKAT Science: On
  the Pathway to the SKA, 32, \dodoi{10.22323/1.277.0032}

\bibitem[{{Scargle}(1982)}]{Scargle1982}
{Scargle}, J.~D. 1982, \apj, 263, 835, \dodoi{10.1086/160554}

\bibitem[{{Sievers}(2017)}]{Sievers2017:Calibration}
{Sievers}, J.~L. 2017, arXiv e-prints, arXiv:1701.01860,
  \dodoi{10.48550/arXiv.1701.01860}

\bibitem[{{Sihlangu} {et~al.}(2020){Sihlangu}, {Oozeer}, \&
  {Bassett}}]{Sihlangu2020:MeerKAT_RFI}
{Sihlangu}, I., {Oozeer}, N., \& {Bassett}, B.~A. 2020, arXiv e-prints,
  arXiv:2008.08877, \dodoi{10.48550/arXiv.2008.08877}

\bibitem[{{Sims} {et~al.}(2022){Sims}, {Pober}, \&
  {Sievers}}]{Sims2022:Calibration}
{Sims}, P.~H., {Pober}, J.~C., \& {Sievers}, J.~L. 2022, \mnras, 517, 910,
  \dodoi{10.1093/mnras/stac1861}

\bibitem[{{Slepian}(1978)}]{Slepian1978:DPSS}
{Slepian}, D. 1978, AT T Technical Journal, 57, 1371

\bibitem[{{Sokolowski} {et~al.}(2016){Sokolowski}, {Wayth}, \&
  {Lewis}}]{Sokolowski2016:MWA_RFI}
{Sokolowski}, M., {Wayth}, R.~B., \& {Lewis}, M. 2016, arXiv e-prints,
  arXiv:1610.04696, \dodoi{10.48550/arXiv.1610.04696}

\bibitem[{{Starck} {et~al.}(2013){Starck}, {Fadili}, \&
  {Rassat}}]{Starck2013:CMB_Inpainting}
{Starck}, J.~L., {Fadili}, M.~J., \& {Rassat}, A. 2013, \aap, 550, A15,
  \dodoi{10.1051/0004-6361/201220332}

\bibitem[{{Sullivan} {et~al.}(2012){Sullivan}, {Morales}, {Hazelton}, {Arcus},
  {Barnes}, {Bernardi}, {Briggs}, {Bowman}, {Bunton}, {Cappallo}, {Corey},
  {Deshpande}, {deSouza}, {Emrich}, {Gaensler}, {Goeke}, {Greenhill}, {Herne},
  {Hewitt}, {Johnston-Hollitt}, {Kaplan}, {Kasper}, {Kincaid}, {Koenig},
  {Kratzenberg}, {Lonsdale}, {Lynch}, {McWhirter}, {Mitchell}, {Morgan},
  {Oberoi}, {Ord}, {Pathikulangara}, {Prabu}, {Remillard}, {Rogers}, {Roshi},
  {Salah}, {Sault}, {Udaya Shankar}, {Srivani}, {Stevens}, {Subrahmanyan},
  {Tingay}, {Wayth}, {Waterson}, {Webster}, {Whitney}, {Williams}, {Williams},
  \& {Wyithe}}]{Sullivan2012:Calibration}
{Sullivan}, I.~S., {Morales}, M.~F., {Hazelton}, B.~J., {et~al.} 2012, \apj,
  759, 17, \dodoi{10.1088/0004-637X/759/1/17}

\bibitem[{{Tan} {et~al.}(2021){Tan}, {Liu}, {Kern}, {Abdurashidova}, {Aguirre},
  {Alexander}, {Ali}, {Balfour}, {Beardsley}, {Bernardi}, {Billings}, {Bowman},
  {Bradley}, {Bull}, {Burba}, {Carey}, {Carilli}, {Cheng}, {DeBoer}, {Dexter},
  {de Lera Acedo}, {Dillon}, {Ely}, {Ewall-Wice}, {Fagnoni}, {Fritz},
  {Furlanetto}, {Gale-Sides}, {Glendenning}, {Gorthi}, {Greig}, {Grobbelaar},
  {Halday}, {Hazelton}, {Hewitt}, {Hickish}, {Jacobs}, {Julius}, {Kerrigan},
  {Kittiwisit}, {Kohn}, {Kolopanis}, {Lanman}, {La Plante}, {Lekalake},
  {MacMahon}, {Malan}, {Malgas}, {Maree}, {Martinot}, {Matsetela}, {Mesinger},
  {Molewa}, {Morales}, {Mosiane}, {Murray}, {Neben}, {Nikolic}, {Nunhokee},
  {Parsons}, {Patra}, {Pieterse}, {Pober}, {Razavi-Ghods}, {Ringuette},
  {Robnett}, {Rosie}, {Sims}, {Singh}, {Smith}, {Syce}, {Thyagarajan},
  {Williams}, \& {Zheng}}]{Tan2021:HERA_ErrorBar}
{Tan}, J., {Liu}, A., {Kern}, N.~S., {et~al.} 2021, \apjs, 255, 26,
  \dodoi{10.3847/1538-4365/ac0533}

\bibitem[{{Thyagarajan} {et~al.}(2013){Thyagarajan}, {Udaya Shankar},
  {Subrahmanyan}, {Arcus}, {Bernardi}, {Bowman}, {Briggs}, {Bunton},
  {Cappallo}, {Corey}, {deSouza}, {Emrich}, {Gaensler}, {Goeke}, {Greenhill},
  {Hazelton}, {Herne}, {Hewitt}, {Johnston-Hollitt}, {Kaplan}, {Kasper},
  {Kincaid}, {Koenig}, {Kratzenberg}, {Lonsdale}, {Lynch}, {McWhirter},
  {Mitchell}, {Morales}, {Morgan}, {Oberoi}, {Ord}, {Pathikulangara},
  {Remillard}, {Rogers}, {Anish Roshi}, {Salah}, {Sault}, {Srivani}, {Stevens},
  {Thiagaraj}, {Tingay}, {Wayth}, {Waterson}, {Webster}, {Whitney}, {Williams},
  {Williams}, \& {Wyithe}}]{Thyagarajan2013:wedge}
{Thyagarajan}, N., {Udaya Shankar}, N., {Subrahmanyan}, R., {et~al.} 2013,
  \apj, 776, 6, \dodoi{10.1088/0004-637X/776/1/6}

\bibitem[{{Tingay} {et~al.}(2013){Tingay}, {Goeke}, {Bowman}, {Emrich}, {Ord},
  {Mitchell}, {Morales}, {Booler}, {Crosse}, {Wayth}, {Lonsdale}, {Tremblay},
  {Pallot}, {Colegate}, {Wicenec}, {Kudryavtseva}, {Arcus}, {Barnes},
  {Bernardi}, {Briggs}, {Burns}, {Bunton}, {Cappallo}, {Corey}, {Deshpande},
  {Desouza}, {Gaensler}, {Greenhill}, {Hall}, {Hazelton}, {Herne}, {Hewitt},
  {Johnston-Hollitt}, {Kaplan}, {Kasper}, {Kincaid}, {Koenig}, {Kratzenberg},
  {Lynch}, {Mckinley}, {Mcwhirter}, {Morgan}, {Oberoi}, {Pathikulangara},
  {Prabu}, {Remillard}, {Rogers}, {Roshi}, {Salah}, {Sault}, {Udaya-Shankar},
  {Schlagenhaufer}, {Srivani}, {Stevens}, {Subrahmanyan}, {Waterson},
  {Webster}, {Whitney}, {Williams}, {Williams}, \&
  {Wyithe}}]{MWA2013:PhaseI_Overview}
{Tingay}, S.~J., {Goeke}, R., {Bowman}, J.~D., {et~al.} 2013, \pasa, 30, e007,
  \dodoi{10.1017/pasa.2012.007}

\bibitem[{{Trott} {et~al.}(2012){Trott}, {Wayth}, \&
  {Tingay}}]{Trott2012:wedge}
{Trott}, C.~M., {Wayth}, R.~B., \& {Tingay}, S.~J. 2012, \apj, 757, 101,
  \dodoi{10.1088/0004-637X/757/1/101}

\bibitem[{{Trott} {et~al.}(2016){Trott}, {Pindor}, {Procopio}, {Wayth},
  {Mitchell}, {McKinley}, {Tingay}, {Barry}, {Beardsley}, {Bernardi}, {Bowman},
  {Briggs}, {Cappallo}, {Carroll}, {de Oliveira-Costa}, {Dillon}, {Ewall-Wice},
  {Feng}, {Greenhill}, {Hazelton}, {Hewitt}, {Hurley-Walker},
  {Johnston-Hollitt}, {Jacobs}, {Kaplan}, {Kim}, {Lenc}, {Line}, {Loeb},
  {Lonsdale}, {Morales}, {Morgan}, {Neben}, {Thyagarajan}, {Oberoi},
  {Offringa}, {Ord}, {Paul}, {Pober}, {Prabu}, {Riding}, {Udaya Shankar},
  {Sethi}, {Srivani}, {Subrahmanyan}, {Sullivan}, {Tegmark}, {Webster},
  {Williams}, {Williams}, {Wu}, \& {Wyithe}}]{Trott2016:CHIPS}
{Trott}, C.~M., {Pindor}, B., {Procopio}, P., {et~al.} 2016, \apj, 818, 139,
  \dodoi{10.3847/0004-637X/818/2/139}

\bibitem[{{Trott} {et~al.}(2020){Trott}, {Jordan}, {Midgley}, {Barry}, {Greig},
  {Pindor}, {Cook}, {Sleap}, {Tingay}, {Ung}, {Hancock}, {Williams}, {Bowman},
  {Byrne}, {Chokshi}, {Hazelton}, {Hasegawa}, {Jacobs}, {Joseph}, {Li}, {Line},
  {Lynch}, {McKinley}, {Mitchell}, {Morales}, {Ouchi}, {Pober}, {Rahimi},
  {Takahashi}, {Wayth}, {Webster}, {Wilensky}, {Wyithe}, {Yoshiura}, {Zhang},
  \& {Zheng}}]{Trott2020:MWA_Limit}
{Trott}, C.~M., {Jordan}, C.~H., {Midgley}, S., {et~al.} 2020, \mnras,
  \dodoi{10.1093/mnras/staa414}

\bibitem[{{Ung} {et~al.}(2020){Ung}, {Sokolowski}, {Sutinjo}, \&
  {Davidson}}]{Ung2020:MWA_Coupling_Sim}
{Ung}, D. C.~X., {Sokolowski}, M., {Sutinjo}, A.~T., \& {Davidson}, D.~B. 2020,
  IEEE Transactions on Antennas and Propagation, 68, 5395,
  \dodoi{10.1109/TAP.2020.2980334}

\bibitem[{{van Haarlem} {et~al.}(2013){van Haarlem}, {Wise}, {Gunst}, {Heald},
  {McKean}, {Hessels}, {de Bruyn}, {Nijboer}, {Swinbank}, {Fallows},
  {Brentjens}, {Nelles}, {Beck}, {Falcke}, {Fender}, {H{\"o}randel},
  {Koopmans}, {Mann}, {Miley}, {R{\"o}ttgering}, {Stappers}, {Wijers},
  {Zaroubi}, {van den Akker}, {Alexov}, {Anderson}, {Anderson}, {van Ardenne},
  {Arts}, {Asgekar}, {Avruch}, {Batejat}, {B{\"a}hren}, {Bell}, {Bell}, {van
  Bemmel}, {Bennema}, {Bentum}, {Bernardi}, {Best}, {B{\^\i}rzan}, {Bonafede},
  {Boonstra}, {Braun}, {Bregman}, {Breitling}, {van de Brink}, {Broderick},
  {Broekema}, {Brouw}, {Br{\"u}ggen}, {Butcher}, {van Cappellen}, {Ciardi},
  {Coenen}, {Conway}, {Coolen}, {Corstanje}, {Damstra}, {Davies}, {Deller},
  {Dettmar}, {van Diepen}, {Dijkstra}, {Donker}, {Doorduin}, {Dromer}, {Drost},
  {van Duin}, {Eisl{\"o}ffel}, {van Enst}, {Ferrari}, {Frieswijk}, {Gankema},
  {Garrett}, {de Gasperin}, {Gerbers}, {de Geus}, {Grie{\ss}meier}, {Grit},
  {Gruppen}, {Hamaker}, {Hassall}, {Hoeft}, {Holties}, {Horneffer}, {van der
  Horst}, {van Houwelingen}, {Huijgen}, {Iacobelli}, {Intema}, {Jackson},
  {Jelic}, {de Jong}, {Juette}, {Kant}, {Karastergiou}, {Koers}, {Kollen},
  {Kondratiev}, {Kooistra}, {Koopman}, {Koster}, {Kuniyoshi}, {Kramer},
  {Kuper}, {Lambropoulos}, {Law}, {van Leeuwen}, {Lemaitre}, {Loose}, {Maat},
  {Macario}, {Markoff}, {Masters}, {McFadden}, {McKay-Bukowski}, {Meijering},
  {Meulman}, {Mevius}, {Middelberg}, {Millenaar}, {Miller-Jones}, {Mohan},
  {Mol}, {Morawietz}, {Morganti}, {Mulcahy}, {Mulder}, {Munk}, {Nieuwenhuis},
  {van Nieuwpoort}, {Noordam}, {Norden}, {Noutsos}, {Offringa}, {Olofsson},
  {Omar}, {Orr{\'u}}, {Overeem}, {Paas}, {Pandey-Pommier}, {Pandey}, {Pizzo},
  {Polatidis}, {Rafferty}, {Rawlings}, {Reich}, {de Reijer}, {Reitsma},
  {Renting}, {Riemers}, {Rol}, {Romein}, {Roosjen}, {Ruiter}, {Scaife}, {van
  der Schaaf}, {Scheers}, {Schellart}, {Schoenmakers}, {Schoonderbeek},
  {Serylak}, {Shulevski}, {Sluman}, {Smirnov}, {Sobey}, {Spreeuw}, {Steinmetz},
  {Sterks}, {Stiepel}, {Stuurwold}, {Tagger}, {Tang}, {Tasse}, {Thomas},
  {Thoudam}, {Toribio}, {van der Tol}, {Usov}, {van Veelen}, {van der Veen},
  {ter Veen}, {Verbiest}, {Vermeulen}, {Vermaas}, {Vocks}, {Vogt}, {de Vos},
  {van der Wal}, {van Weeren}, {Weggemans}, {Weltevrede}, {White}, {Wijnholds},
  {Wilhelmsson}, {Wucknitz}, {Yatawatta}, {Zarka}, {Zensus}, \& {van
  Zwieten}}]{LOFAR2013:Overview}
{van Haarlem}, M.~P., {Wise}, M.~W., {Gunst}, A.~W., {et~al.} 2013, \aap, 556,
  A2, \dodoi{10.1051/0004-6361/201220873}

\bibitem[{{Vanderlinde} {et~al.}(2019){Vanderlinde}, {Liu}, {Gaensler}, {Bond},
  {Hinshaw}, {Ng}, {Chiang}, {Stairs}, {Brown}, {Sievers}, {Mena}, {Smith},
  {Bandura}, {Masui}, {Spekkens}, {Belostotski}, {Dobbs}, {Turok}, {Boyle},
  {Rupen}, {Landecker}, {Pen}, \& {Kaspi}}]{CHORD2019:Overview}
{Vanderlinde}, K., {Liu}, A., {Gaensler}, B., {et~al.} 2019, in Canadian Long
  Range Plan for Astronomy and Astrophysics White Papers, Vol. 2020, 28,
  \dodoi{10.5281/zenodo.3765414}

\bibitem[{{Van{\'\i}{\v{c}}ek}(1969)}]{Vanivek1969:LSSA}
{Van{\'\i}{\v{c}}ek}, P. 1969, \apss, 4, 387, \dodoi{10.1007/BF00651344}

\bibitem[{{Van{\'\i}{\v{c}}ek}(1971)}]{Vanivek1971:LSSA}
---. 1971, \apss, 12, 10, \dodoi{10.1007/BF00656134}

\bibitem[{{Vedantham} {et~al.}(2012){Vedantham}, {Udaya Shankar}, \&
  {Subrahmanyan}}]{Vedantham2012:image_wedge}
{Vedantham}, H., {Udaya Shankar}, N., \& {Subrahmanyan}, R. 2012, \apj, 745,
  176, \dodoi{10.1088/0004-637X/745/2/176}

\bibitem[{{Virtanen} {et~al.}(2020){Virtanen}, {Gommers}, {Oliphant},
  {Haberland}, {Reddy}, {Cournapeau}, {Burovski}, {Peterson}, {Weckesser},
  {Bright}, {van der Walt}, {Brett}, {Wilson}, {Jarrod Millman}, {Mayorov},
  {Nelson}, {Jones}, {Kern}, {Larson}, {Carey}, {Polat}, {Feng}, {Moore}, {Vand
  erPlas}, {Laxalde}, {Perktold}, {Cimrman}, {Henriksen}, {Quintero}, {Harris},
  {Archibald}, {Ribeiro}, {Pedregosa}, {van Mulbregt}, \&
  {Contributors}}]{scipy}
{Virtanen}, P., {Gommers}, R., {Oliphant}, T.~E., {et~al.} 2020, Nature
  Methods, \dodoi{10.1038/s41592-019-0686-2}

\bibitem[{Waskom(2021)}]{seaborn}
Waskom, M.~L. 2021, Journal of Open Source Software, 6, 3021,
  \dodoi{10.21105/joss.03021}

\bibitem[{{Wayth} {et~al.}(2018){Wayth}, {Tingay}, {Trott}, {Emrich},
  {Johnston-Hollitt}, {McKinley}, {Gaensler}, {Beardsley}, {Booler}, {Crosse},
  {Franzen}, {Horsley}, {Kaplan}, {Kenney}, {Morales}, {Pallot}, {Sleap},
  {Steele}, {Walker}, {Williams}, {Wu}, {Cairns}, {Filipovic}, {Johnston},
  {Murphy}, {Quinn}, {Staveley-Smith}, {Webster}, \&
  {Wyithe}}]{MWA2018:PhaseII_Overview}
{Wayth}, R.~B., {Tingay}, S.~J., {Trott}, C.~M., {et~al.} 2018, \pasa, 35,
  e033, \dodoi{10.1017/pasa.2018.37}

\bibitem[{{Wilensky} {et~al.}(2022){Wilensky}, {Hazelton}, \&
  {Morales}}]{Wilensky2022:flag_systematics}
{Wilensky}, M.~J., {Hazelton}, B.~J., \& {Morales}, M.~F. 2022, \mnras, 510,
  5023, \dodoi{10.1093/mnras/stab3456}

\bibitem[{{Wilensky} {et~al.}(2019){Wilensky}, {Morales}, {Hazelton}, {Barry},
  {Byrne}, \& {Roy}}]{Wilensky2019:SSINS}
{Wilensky}, M.~J., {Morales}, M.~F., {Hazelton}, B.~J., {et~al.} 2019, \pasp,
  131, 114507, \dodoi{10.1088/1538-3873/ab3cad}

\bibitem[{{Wilensky} {et~al.}(2023){Wilensky}, {Morales}, {Hazelton}, {Star},
  {Barry}, {Byrne}, {Jordan}, {Jacobs}, {Pober}, \&
  {Trott}}]{Wilensky2023:MWA_Limit}
---. 2023, \apj, 957, 78, \dodoi{10.3847/1538-4357/acffbd}

\bibitem[{{Yatawatta} {et~al.}(2008){Yatawatta}, {Zaroubi}, {de Bruyn},
  {Koopmans}, \& {Noordam}}]{Yatawatta2008:SAGE_Calibration}
{Yatawatta}, S., {Zaroubi}, S., {de Bruyn}, G., {Koopmans}, L., \& {Noordam},
  J. 2008, arXiv e-prints, arXiv:0810.5751, \dodoi{10.48550/arXiv.0810.5751}

\bibitem[{{Yoshiura} {et~al.}(2021){Yoshiura}, {Pindor}, {Line}, {Barry},
  {Trott}, {Beardsley}, {Bowman}, {Byrne}, {Chokshi}, {Hazelton}, {Hasegawa},
  {Howard}, {Greig}, {Jacobs}, {Jordan}, {Joseph}, {Kolopanis}, {Lynch},
  {McKinley}, {Mitchell}, {Morales}, {Murray}, {Pober}, {Rahimi}, {Takahashi},
  {Tingay}, {Wayth}, {Webster}, {Wilensky}, {Wyithe}, {Zhang}, \&
  {Zheng}}]{Yoshiura2021:MWA_Limit}
{Yoshiura}, S., {Pindor}, B., {Line}, J.~L.~B., {et~al.} 2021, \mnras, 505,
  4775, \dodoi{10.1093/mnras/stab1560}

\bibitem[{{Zarka} {et~al.}(2012){Zarka}, {Girard}, {Tagger}, \&
  {Denis}}]{NenuFAR2012:Overview}
{Zarka}, P., {Girard}, J.~N., {Tagger}, M., \& {Denis}, L. 2012, in SF2A-2012:
  Proceedings of the Annual meeting of the French Society of Astronomy and
  Astrophysics, ed. S.~{Boissier}, P.~{de Laverny}, N.~{Nardetto}, R.~{Samadi},
  D.~{Valls-Gabaud}, \& H.~{Wozniak}, 687--694

\bibitem[{{Zheng} {et~al.}(2017){Zheng}, {Tegmark}, {Dillon}, {Kim}, {Liu},
  {Neben}, {Jonas}, {Reich}, \& {Reich}}]{Zheng2017:GSM}
{Zheng}, H., {Tegmark}, M., {Dillon}, J.~S., {et~al.} 2017, \mnras, 464, 3486,
  \dodoi{10.1093/mnras/stw2525}

\end{thebibliography}
\bibliographystyle{aasjournal}

\onecolumngrid
\appendix
\section{Probability distribution function of the inpainted visibility} \label{appendix:derivation}
In this appendix, we calculate the uncertainties in the inpainted visibility $\vect{v}'_\mathrm{inp}$ in the flagged channels. Namely, we wish to know the probability of the unobserved $\vect{v}'_\mathrm{inp}$ in the flagged channels given the observed visibility $\vect{v}_\mathrm{obs}$, the noise covariance $\vect{N}$, and the design matrix $\vect{A}$ that prescribes the shape of the underlying signal. Following Sec.\,\ref{subsec:inpaint_error}, we have

\begin{equation}
\label{eq:likelihood_integral}
    P(\mathbf{v}_\mathrm{inp}' | \mathbf{v}_\mathrm{obs}, \mathbf{N}, \mathbf{A}) = \int\mathrm{d}\mathbf{b}\,P(\mathbf{v}_\mathrm{inp}' | \mathbf{b}, \mathbf{N}_\mathrm{f}', \mathbf{A}) P(\mathbf{b} | \mathbf{v}_\mathrm{obs}, \mathbf{N}_\mathrm{u}, \mathbf{A})
\end{equation}
where 
\begin{equation}
    P(\mathbf{v}_\mathrm{inp}' | \mathbf{b}, \mathbf{N}_\mathrm{f}', \mathbf{A}) \propto \exp{\left[-(\mathbf{v}'_\mathrm{inp} - \mathbf{P}_\mathrm{f}\mathbf{A}\mathbf{b})^\dagger \mathbf{N'}_\mathrm{f}^{-1}(\mathbf{v}'_\mathrm{inp} - \mathbf{P}_\mathrm{f}\mathbf{A}\mathbf{b})\right]},
\end{equation}
and 
\begin{equation}
    P(\mathbf{b} | \mathbf{v}_\mathrm{obs}, \mathbf{N}_\mathrm{u}, \mathbf{A}) \propto P(\mathbf{v}_\mathrm{obs} | \mathbf{b}, \mathbf{N}_\mathrm{u}, \mathbf{A}) P(\mathbf{b}) \propto \exp{\left[-(\mathbf{v}_\mathrm{obs} - \mathbf{A}\mathbf{b})^\dagger \mathbf{N}_\mathrm{u}^{-1}(\mathbf{v}_\mathrm{obs} - \mathbf{A}\mathbf{b})\right]}
\end{equation}
assuming a flat prior on $\mathbf{b}$. By taking  $\partial P(\mathbf{b} | \mathbf{v}_\mathrm{obs}, \mathbf{N}_\mathrm{u}, \mathbf{A}) / \partial \mathbf{b} = 0$, we can solve for the mean of the Gaussian distribution $P(\mathbf{b} | \mathbf{v}_\mathrm{obs}, \mathbf{N}_\mathrm{u}, \mathbf{A})$ to be at
\begin{equation}
    {\mathbf{\hat{b}}} = (\mathbf{A}^\dagger\mathbf{N}_\mathrm{u}^{-1}\mathbf{A})^{-1} \mathbf{A}^\dagger\mathbf{N}_\mathrm{u}^{-1}\mathbf{v}_\mathrm{obs}\footnote{Formally speaking, the matrix $\mathbf{A}^\dagger\mathbf{N}_\mathrm{u}^{-1}\mathbf{A}$ might not be invertible as $\mathbf{N}_\mathrm{u}$ is not full rank. However, $\mathbf{A}^\dagger$ is a coordinate transformation that maps from a space of dimension $N_\mathrm{freq}$ to the space of DPSS coefficients. Since we have less DPSS modes than the number of frequency channels, here we will assume $\mathbf{A}^\dagger\mathbf{N}_\mathrm{u}^{-1}\mathbf{A}$ is invertible. This assumption holds if we only have a small amount of channels that is flagged.}.
\end{equation}
Meanwhile, by examining quadratic terms in the exponent, we know that the covariance must be $(\mathbf{A}^\dagger\mathbf{N}_\mathrm{u}^{-1}\mathbf{A})^{-1}$. Therefore, we can write
\begin{equation}
    P(\mathbf{b} | \mathbf{v}_\mathrm{obs}, \mathbf{N}_\mathrm{u}, \mathbf{A}) \propto \exp{\left[-(\mathbf{b} - \mathbf{\hat{b}})^\dagger(\mathbf{A}^\dagger\mathbf{N}_\mathrm{u}^{-1}\mathbf{A})(\mathbf{b} - \mathbf{\hat{b}})\right]}.
\end{equation}
The integration in Eq.\,\eqref{eq:likelihood_integral} can then be computed by ``completing the square''. We first note that we can re-write $(\mathbf{v}'_\mathrm{inp} - \mathbf{P}_\mathrm{f}\mathbf{A}\mathbf{b})^\dagger \mathbf{N}'^{-1}_\mathrm{f}(\mathbf{v}'_\mathrm{inp} - \mathbf{P}_\mathrm{f}\mathbf{A}\mathbf{b})$ into
\begin{equation}
    \begin{aligned}
        &(\mathbf{v}'_\mathrm{inp} - \mathbf{P}_\mathrm{f}\mathbf{A}\mathbf{b})^\dagger \mathbf{N}'^{-1}_\mathrm{f}(\mathbf{v}'_\mathrm{inp} - \mathbf{P}_\mathrm{f}\mathbf{A}\mathbf{b}) \\
        =& \mathbf{v}'^\dagger_\mathrm{inp}\mathbf{N}'^{-1}_\mathrm{f}\mathbf{v}'_\mathrm{inp} + \mathbf{b}^\dagger\mathbf{A}^\dagger\mathbf{P}_\mathrm{f}^\dagger\mathbf{N}'^{-1}_\mathrm{f}\mathbf{P}_\mathrm{f}\mathbf{A}\mathbf{b} + \mathbf{v}'^\dagger_\mathrm{inp}\mathbf{N}'^{-1}_\mathrm{f}\mathbf{P}_\mathrm{f}\mathbf{A}\mathbf{b} + \mathbf{b}^\dagger\mathbf{A}^\dagger\mathbf{P}_\mathrm{f}^\dagger\mathbf{N}'^{-1}_\mathrm{f}\mathbf{v}'_\mathrm{inp} \\
        =&\mathbf{v}'^\dagger_\mathrm{inp}\mathbf{N}'^{-1}_\mathrm{f}\mathbf{v}'_\mathrm{inp} + \mathbf{b}^\dagger\left(\mathbf{A}^\dagger\mathbf{P}_\mathrm{f}^\dagger\mathbf{N}'^{-1}_\mathrm{f}\mathbf{P}_\mathrm{f}\mathbf{A}\right)\mathbf{b} + \\
        &+\mathbf{b'}^\dagger\left(\mathbf{A}^\dagger\mathbf{P}_\mathrm{f}^\dagger\mathbf{N}'^{-1}_\mathrm{f}\mathbf{P}_\mathrm{f}\mathbf{A}\right)\mathbf{b} + \mathbf{b}^\dagger\left(\mathbf{A}^\dagger\mathbf{P}_\mathrm{f}^\dagger\mathbf{N}'^{-1}_\mathrm{f}\mathbf{P}_\mathrm{f}\mathbf{A}\right)\mathbf{b'} \\
        =&(\mathbf{b} - \mathbf{b'})^\dagger\left(\mathbf{A}^\dagger\mathbf{P}_\mathrm{f}^\dagger\mathbf{N}'^{-1}_\mathrm{f}\mathbf{P}_\mathrm{f}\mathbf{A}\right)(\mathbf{b} - \mathbf{b'}) + \mathbf{v}'^\dagger_\mathrm{inp}\mathbf{N}'^{-1}_\mathrm{f}\mathbf{v}'_\mathrm{inp} - \mathbf{b'}^\dagger\left(\mathbf{A}^\dagger\mathbf{P}_\mathrm{f}^\dagger\mathbf{N}'^{-1}_\mathrm{f}\mathbf{P}_\mathrm{f}\mathbf{A}\right)\mathbf{b'}
    \end{aligned}
\end{equation}
where we have defined $\mathbf{b'}\equiv\left(\mathbf{A}^\dagger\mathbf{P}_\mathrm{f}^\dagger\mathbf{N}'^{-1}_\mathrm{f}\mathbf{P}_\mathrm{f}\mathbf{A}\right)^{-1}\mathbf{A}^\dagger\mathbf{P}_\mathrm{f}^
\dagger\mathbf{N}'^{-1}_\mathrm{f}\mathbf{v}'_\mathrm{inp}$. Eq.\,\eqref{eq:likelihood_integral} thus becomes
\begin{equation}
\label{eq:full_integration}
    \begin{aligned}
        P(\mathbf{v}'_\mathrm{inp} | \mathbf{v}_\mathrm{obs}, \mathbf{N}, \mathbf{A}) \propto& \exp{\left[-\mathbf{v}'^\dagger_\mathrm{inp}\mathbf{N}'^{-1}_\mathrm{f}\mathbf{v}'_\mathrm{inp}+\mathbf{b'}^\dagger\left(\mathbf{A}^\dagger\mathbf{P}_\mathrm{f}^\dagger\mathbf{N}'^{-1}_\mathrm{f}\mathbf{P}_\mathrm{f}\mathbf{A}\right)\mathbf{b'}\right]}\\
        &\times\int\mathrm{d}\mathbf{b}\,\exp{\left[-(\mathbf{b} - \mathbf{b'})^\dagger\left(\mathbf{A}^\dagger\mathbf{P}_\mathrm{f}^\dagger\mathbf{N}'^{-1}_\mathrm{f}\mathbf{P}_\mathrm{f}\mathbf{A}\right)(\mathbf{b} - \mathbf{b'}) - (\mathbf{b} - \mathbf{\hat{b}})^\dagger(\mathbf{A}^\dagger\mathbf{N}_\mathrm{u}^{-1}\mathbf{A})(\mathbf{b} - \mathbf{\hat{b}})\right]} \\
        \propto& \exp\Big[-\mathbf{v}'^\dagger_\mathrm{inp}\mathbf{N}'^{-1}_\mathrm{f}\mathbf{v}'_\mathrm{inp}+\mathbf{b'}^\dagger\left(\mathbf{A}^\dagger\mathbf{P}_\mathrm{f}^\dagger\mathbf{N}'^{-1}_\mathrm{f}\mathbf{P}_\mathrm{f}\mathbf{A}\right)\mathbf{b'} \\
        &- (\mathbf{b'} - \mathbf{\hat{b}})^\dagger\left((\mathbf{A}^\dagger\mathbf{N}_\mathrm{u}^{-1}\mathbf{A})^{-1} + (\mathbf{A}^\dagger\mathbf{P}_\mathrm{f}\mathbf{N}'^{-1}_\mathrm{f}\mathbf{P}_\mathrm{f}\mathbf{A})^{-1}\right)^{-1}(\mathbf{b'} - \mathbf{\hat{b}}) \Big]
    \end{aligned}.
\end{equation}
To simplify the notation, we denote $\mathbf{C}_\mathrm{u}\equiv (\mathbf{A}^\dagger\mathbf{N}_\mathrm{u}^{-1}\mathbf{A})^{-1}$ and $\mathbf{C}_\mathrm{f}\equiv (\mathbf{A}^\dagger\mathbf{P}_\mathrm{f}^\dagger\mathbf{N}'^{-1}_\mathrm{f}\mathbf{P}_\mathrm{f}\mathbf{A})^{-1}$. Next, we will reduce Eq.\,\eqref{eq:full_integration} by ``completing the square'' again. First, we collect all the terms that are quadratic in $\mathbf{v}'_\mathrm{inp}$, noting that $\mathbf{b'} \equiv \mathbf{C}_\mathrm{f}\mathbf{A}^\dagger\mathbf{P}_\mathrm{f}^\dagger\mathbf{N}'^{-1}_\mathrm{f}\mathbf{v}'_\mathrm{inp}$,
\begin{equation}
\begin{aligned}
    &-\mathbf{v}'^\dagger_\mathrm{inp}\mathbf{N}'^{-1}_\mathrm{f}\mathbf{v}'_\mathrm{inp}
    +\mathbf{b'}^\dagger\mathbf{C}_\mathrm{f}^{-1}\mathbf{b'}
    -\mathbf{b'}^\dagger\left(\mathbf{C}_\mathrm{f} + \mathbf{C}_\mathrm{u}\right)^{-1}\mathbf{b'}\\
    = &-\mathbf{v}'^\dagger_\mathrm{inp}\left(\mathbf{N}'^{-1}_\mathrm{f} -(\mathbf{N}'^{-1}_\mathrm{f}\mathbf{A}\mathbf{P}_\mathrm{f}\mathbf{C}_\mathrm{f})(\mathbf{C}_\mathrm{f}^{-1} - (\mathbf{C}_\mathrm{f} + \mathbf{C}_\mathrm{u})^{-1})(\mathbf{C}_\mathrm{f}\mathbf{A}^\dagger\mathbf{P}_\mathrm{f}^\dagger\mathbf{N}'^{-1}_\mathrm{f})\right)\mathbf{v}'_\mathrm{inp}.
\end{aligned}
\end{equation}
We now show that
\begin{equation}
    \mathbf{C}^{-1} \equiv \left(\mathbf{N}'^{-1}_\mathrm{f} -(\mathbf{N}'^{-1}_\mathrm{f}\mathbf{P}_\mathrm{f}\mathbf{A}\mathbf{C}_\mathrm{f})(\mathbf{C}_\mathrm{f}^{-1} - (\mathbf{C}_\mathrm{f} + \mathbf{C}_\mathrm{u})^{-1})(\mathbf{C}_\mathrm{f}\mathbf{A}^\dagger\mathbf{P}_\mathrm{f}^\dagger\mathbf{N}'^{-1}_\mathrm{f}\right) = (\mathbf{N}'_\mathrm{f} + \mathcal{O'}_\mathrm{inp}\mathbf{N}_\mathrm{u}\mathcal{O'}_\mathrm{inp}^\dagger)^{-1}
\end{equation}
where $\mathcal{O'}_\mathrm{inp}$ is our proposed inpainting operator that maps observed data $\vect{v}_\mathrm{obs}$ into a model which we use to inpaint the flagged channels,
\begin{equation}
    \mathcal{O'}_\mathrm{inp}\equiv\mathbf{P}_\mathrm{f}\mathbf{A}(\mathbf{A}^\dagger\mathbf{N}_\mathrm{u}^{-1}\mathbf{A})^{-1} \mathbf{A}^\dagger\mathbf{N}_\mathrm{u}^{-1}.
\end{equation}
To show this relation, we note that by the Woodbury matrix identity,
\begin{equation}
    (\mathbf{C}_\mathrm{f} + \mathbf{C}_\mathrm{u})^{-1} = \mathbf{C}_\mathrm{f}^{-1} - \mathbf{C}_\mathrm{f}^{-1}(\mathbf{C}_\mathrm{f}^{-1} + \mathbf{C}_\mathrm{u}^{-1})\mathbf{C}_\mathrm{f}^{-1}.
\end{equation}
Thus,
\begin{equation}
      (\mathbf{C}_\mathrm{f}^{-1} - (\mathbf{C}_\mathrm{f} + \mathbf{C}_\mathrm{u})^{-1}) = \mathbf{C}_\mathrm{f}^{-1}(\mathbf{C}_\mathrm{f}^{-1} + \mathbf{C}_\mathrm{u}^{-1})\mathbf{C}_\mathrm{f}^{-1}.
\end{equation}
This gives us 
\begin{equation}
(\mathbf{N}'^{-1}_\mathrm{f}\mathbf{A}\mathbf{P}_\mathrm{f}\mathbf{C}_\mathrm{f})(\mathbf{C}_\mathrm{f}^{-1} - (\mathbf{C}_\mathrm{f} + \mathbf{C}_\mathrm{u})^{-1})(\mathbf{C}_\mathrm{f}\mathbf{A}^\dagger\mathbf{P}_\mathrm{f}^\dagger\mathbf{N}'^{-1}_\mathrm{f}) = \mathbf{N}'^{-1}_\mathrm{f}\mathbf{A}\mathbf{P}_\mathrm{f}(\mathbf{C}_\mathrm{f}^{-1} + \mathbf{C}_\mathrm{u}^{-1})\mathbf{A}^\dagger\mathbf{P}_\mathrm{f}^\dagger\mathbf{N}'_\mathrm{f}.
\end{equation}
Hence
\begin{equation}
\label{eq:covariance_formula}
    \mathbf{C}^{-1} = \mathbf{N}'^{-1}_\mathrm{f} - \mathbf{N}'^{-1}_\mathrm{f}\mathbf{P}_\mathrm{f}\mathbf{A}(\mathbf{C}_\mathrm{f}^{-1} + \mathbf{C}_\mathrm{u}^{-1})\mathbf{A}^\dagger\mathbf{P}_\mathrm{f}^\dagger\mathbf{N}'^{-1}_\mathrm{f}.
\end{equation}
On the other hand, the Woodbury matrix identity also gives us
\begin{equation}
\begin{aligned}
    (\mathbf{N}'_\mathrm{f} + \mathcal{O'}_\mathrm{inp}\mathbf{N}_\mathrm{u}\mathcal{O'}_\mathrm{inp}^\dagger)^{-1} &= \left(\mathbf{N}'_\mathrm{f} + \mathbf{P}_\mathrm{f}\mathbf{A}(\mathbf{A}^\dagger\mathbf{N}_\mathrm{u}^{-1}\mathbf{A})^{-1} \mathbf{A}^\dagger\mathbf{P}_\mathrm{f}^\dagger\right)^{-1} \\
    &= \mathbf{N}'^{-1}_\mathrm{f} - \mathbf{N}'^{-1}_\mathrm{f}\mathbf{P}_\mathrm{f}\mathbf{A}(\mathbf{C}_\mathrm{u}^{-1} + \mathbf{A}^\dagger\mathbf{P}_\mathrm{f}^\dagger \mathbf{N}'^{-1}_\mathrm{f}\mathbf{P}_\mathrm{f}\mathbf{A})\mathbf{A}^\dagger\mathbf{P}_\mathrm{f}^\dagger\mathbf{N}'^{-1}_\mathrm{f} \\
    &= \mathbf{N}'^{-1}_\mathrm{f} - \mathbf{N}'^{-1}_\mathrm{f}\mathbf{P}_\mathrm{f}\mathbf{A}(\mathbf{C}_\mathrm{u}^{-1} + \mathbf{C}_\mathrm{f}^{-1})\mathbf{A}^\dagger\mathbf{P}_\mathrm{f}^\dagger\mathbf{N}'^{-1}_\mathrm{f},
\end{aligned}
\end{equation}
which agrees with Eq.\,\eqref{eq:covariance_formula}. Thus we have shown that 
\begin{equation}
    \mathbf{C} = \mathbf{N}'_\mathrm{f} + \mathcal{O'}_\mathrm{inp}\mathbf{N}_\mathrm{u}\mathcal{O'}_\mathrm{inp}^\dagger.
\end{equation}
To examine rest of the terms in Eq.\,\eqref{eq:full_integration}, we note that the term linear in $\vect{v}'_\mathrm{inp}$ can be written as 
\begin{equation}
\begin{aligned}
    &\mathbf{b'}^\dagger\left(\mathbf{C}_\mathrm{f} + \mathbf{C}_\mathrm{u}\right)\mathbf{\hat{b}} \\
    =& \vect{v}'^\dagger_\mathrm{inp}\mathbf{N}'^{-1}_\mathrm{f}\mathbf{P}_\mathrm{f}\mathbf{A}\mathbf{C}_\mathrm{f}\left(\mathbf{C}_\mathrm{f} + \mathbf{C}_\mathrm{u}\right)\mathbf{\hat{b}} \\
    =&\vect{v}'^\dagger_\mathrm{inp}\mathbf{N}'^{-1}_\mathrm{f}\mathbf{P}_\mathrm{f}\mathbf{A}\mathbf{C}_\mathrm{f}\left(\mathbf{C}_\mathrm{f}^{-1} - \mathbf{C}_\mathrm{f}^{-1}(\mathbf{C}_\mathrm{f}^{-1} + \mathbf{C}_\mathrm{u}^{-1})\mathbf{C}_\mathrm{f}^{-1}\right)\mathbf{\hat{b}}\\
    =&\vect{v}'^\dagger_\mathrm{inp}\left(\mathbf{N}'^{-1}_\mathrm{f}\mathbf{P}_\mathrm{f}\mathbf{A} - \mathbf{N}'^{-1}_\mathrm{f}\mathbf{P}_\mathrm{f}\mathbf{A}(\mathbf{C}_\mathrm{f}^{-1} + \mathbf{C}_\mathrm{u}^{-1})\mathbf{A}^\dagger\mathbf{P}_\mathrm{f}^\dagger\mathbf{N}'^{-1}_\mathrm{f}\mathbf{P}_\mathrm{f}\mathbf{A}\right)\mathbf{\hat{b}} \\
    =&\vect{v}'^\dagger_\mathrm{inp}\left(\mathbf{N}'^{-1}_\mathrm{f} - \mathbf{N}'^{-1}_\mathrm{f}\mathbf{P}_\mathrm{f}\mathbf{A}(\mathbf{C}_\mathrm{f}^{-1} + \mathbf{C}_\mathrm{u}^{-1})\mathbf{A}^\dagger\mathbf{P}_\mathrm{f}^\dagger\mathbf{N}'^{-1}_\mathrm{f}\right)\mathbf{P}_\mathrm{f}\mathbf{A}\mathbf{\hat{b}}\\
    =&\vect{v}'^\dagger_\mathrm{inp}\mathbf{C}^{-1}\mathbf{P}_\mathrm{f}\mathbf{A}\mathbf{\hat{b}}.
\end{aligned}
\end{equation}
Combining all these results, we finally arrive at
\begin{equation}
    P(\mathbf{v}'_\mathrm{inp} | \mathbf{v}_\mathrm{obs}, \mathbf{N}, \mathbf{A}) \propto \exp\left[-(\mathbf{v}_\mathrm{inp}' - \mathbf{P}_\mathrm{f}\mathbf{A}\mathbf{\hat{b}})^\dagger\mathbf{C}^{-1}(\mathbf{v}_\mathrm{inp}' - \mathbf{P}_\mathrm{f}\mathbf{A}\mathbf{\hat{b}})\right]
\end{equation}
where $\mathbf{C} = \mathbf{N}'_\mathrm{f} + \mathcal{O'}_\mathrm{inp}\mathbf{N}_\mathrm{u}\mathcal{O'}_\mathrm{inp}^\dagger$, as promised in Eq.\,\eqref{eq:inp_likelihood}. We note that while the derivation is quite complicated here, since everything is Gaussian in Eq.\,\eqref{eq:likelihood_integral}, it should not be surprising that we have arrived in a Gaussian distribution after performing this integration. 

\listofchanges

\end{document}